\documentclass[a4paper,12pt]{article}
\usepackage[dvipdfmx]{graphicx}
\usepackage{cite}
\usepackage{amsmath}
\usepackage{amsfonts}
\usepackage{amssymb}
\usepackage{graphicx, rotating}
\usepackage{epsfig}
\usepackage{latexsym}
\usepackage{graphicx}
\usepackage{color}
\usepackage{amsmath,bm,amssymb}
\usepackage{ulem}
\usepackage{setspace}
\usepackage{lmodern}
\usepackage{microtype}
\usepackage{mathrsfs}
\usepackage[table]{xcolor} 
\usepackage{braket}
\usepackage{mleftright} %\mleftright
\usepackage{comment}
\usepackage{array}
\usepackage{booktabs}
\usepackage{slashed}
\usepackage[english]{babel} 
\usepackage{caption} 
\usepackage{subcaption} %for subfigures
\usepackage{bm} %for making symbols bold by \bm command   
\usepackage{multirow} %for drawing table
\usepackage{calc} 
\usepackage{makecell} %for adjusting table /hline thickness
\usepackage{physics}
\usepackage{url}
\usepackage{hyperref}
\usepackage{cleveref}
\usepackage{ragged2e}     % command \RaggedRight
\usepackage{verbatim} 
\usepackage[symbol]{footmisc} %symbol instead of number for footnote
\usepackage{textcomp} %for \textrightarrow
\renewcommand{\thefootnote}{\fnsymbol{footnote}}
\definecolor{blizzardblue}{rgb}{0.93, 0.93, 0.93} % light gray
\newcolumntype{?}{!{\vrule width 0.8pt}} %makes vertical lines in tables with thickness 0.8pt
\mathchardef\mhyphen="2D %small hyphen in math mode
\newcommand{\RN}[1]{ %command for writing roman numbers 
  \textup{\uppercase\expandafter{\romannumeral#1}}%  
\renewcommand\thesubfigure{(\alph{subfigure})} %for referencing to subfigure as 2(a) instead of 2a
\captionsetup[sub]{labelformat=simple} 
}
\newcommand{\ra}{ %command for writing roman numbers 
\rightarrow
}
\mathchardef\mhyphen="2D

\newcommand{\Slash}[1]{{\ooalign{\hfil#1\hfil\crcr\raise.167ex\hbox{/}}}}
\newcommand{\beq}{\begin{equation}}  \newcommand{\eeq}{\end{equation}}
\newcommand{\bef}{\begin{figure}}  \newcommand{\eef}{\end{figure}}
\newcommand{\bec}{\begin{center}}  \newcommand{\eec}{\end{center}}
  
\newcommand{\laq}[1]{\label{eq:#1}}  
\newcommand{\Eq}[1]{Eq.~(\ref{eq:#1})}

\newcommand{\lac}[1]{\label{chap:#1}}
\newcommand{\SU}[1]{{\rm SU{#1} } }
\newcommand{\SO}[1]{{\rm SO{#1}} }
\def\({\left(}
\def\){\right)}

\def\O{\mathcal{O}}
\def\U{\mathop{\rm U}}

\def\tr{\mathop{\rm tr}}

\newcommand{\OR}{~{\rm or}~}
\newcommand{\AND}{~{\rm and}~}

\newcommand{\KEV}{ {\rm \, keV} }
\newcommand{\MEV}{ {\rm \, MeV} }
\newcommand{\GEV}{ {\rm \, GeV} }
\newcommand{\TEV}{ {\rm \, TeV} }

\def\a{\alpha}

\def\d{\delta}
\def\e{\epsilon}
\def\f{\phi}
\def\g{\gamma}

\def\k{\kappa}
\def\l{\lambda}
\def\m{\mu}
\def\n{\nu}
\def\p{\psi}

\def\r{\rho}

\def\D{\Delta}
\def\G{\Gamma}
\def\L{\Lambda}
\def\F{\Phi}

\def\tl{\tilde}
\def\*{\dagger}

\begin{document}

\begin{flushright}
TU-1169
\end{flushright}

\begin{center}

\vspace{0.1cm}

{
\begin{singlespace}
\Large\bf Probing a light dark sector at future lepton colliders via invisible decays of the SM-like and \\dark Higgs bosons \\ 
\end{singlespace}
}

\vspace{.8cm}
\renewcommand{\thefootnote}{\fnsymbol{footnote}}
{\bf Gholamhossein Haghighat$^{1}$},\footnote{E-mail: \texttt{h.haghighat@ipm.ir}} 
{\bf Mojtaba Mohammadi Najafabadi$^{1}$},\footnote{E-mail: \texttt{mojtaba@ipm.ir}} 
{\bf  Kodai Sakurai$^{2}$\footnote{E-mail:  \texttt{kodai.sakurai.e3@tohoku.ac.jp}}}
and {\bf Wen Yin$^{2}$\footnote{E-mail:  \texttt{yin.wen.b3@tohoku.ac.jp}}}

\vspace{5pt}
\vspace{0.3cm}
{\em 
$^{1}${School of Particles and Accelerators, Institute for Research in Fundamental Sciences (IPM)}, 
{P.O. Box 19395-5531, Tehran, Iran}\\
$^{2}${Department of Physics, Tohoku University, Sendai, Miyagi 980-8578, Japan} 
}

\vspace{0.3cm}
\abstract{
A renormalizable UV model for Axion-Like Particles (ALPs) or hidden photons, that may explain the dark matter usually involves a dark Higgs field which is a singlet under the standard model (SM) gauge group. 
The dark sector can couple to the SM particles via the portal coupling between the SM-like Higgs and dark Higgs fields. 
Through this coupling, the dark sector particles can be produced in either the early universe or the collider experiments. Interestingly, not only the SM-like Higgs boson can decay into the light dark bosons, but also a light dark Higgs boson may be produced and decay into the dark bosons in a collider. 
In this paper, we perform the first collider search for invisible decays by taking both the Higgs bosons into account. We use a multivariate technique to best discriminate the signal from the background. We find that a large parameter region can be {probed} at the International Linear Collider (ILC) operating at the center-of-mass energy of 250 GeV. {In particular, even when the SM-like Higgs invisible decay is a few orders of magnitude below the planned sensitivity reaches of the ILC and the high luminosity LHC (HL-LHC), the scenario can be {probed} by the invisible decay of the dark Higgs boson produced via a similar diagram.} Measuring {the dark Higgs boson} decay into the dark sector will be a smoking gun signal of the light dark sector. 
{{A} similar search of the dark sector would be expected in, e.g., Cool Copper Collider (C$^3$), Circular Electron Positron Collider (CEPC), Compact Linear Collider (CLIC) and {Future Circular electron-positron Collider} (FCC-ee). }
}

\end{center}
\clearpage

\setcounter{footnote}{0}

%\tableofcontents

%%%%%%%%%%%%%%%%%%%%%%%%%%%%%%%%%%%%%%
\section{Introduction}
The existence of a dark sector is plausible due to the evidence for dark matter (DM). In particular, a light dark sector may be reasonable since DM stability is easily guaranteed. ALPs and hidden photons are widely studied and considered dark matter candidates. In a large class of models, they arise from the spontaneous breaking of a global (gauged) U(1) symmetry by the vacuum expectation value (VEV) of a dark Higgs field whose (would-be) Nambu-Goldstone boson is ALP (hidden photon) (see Refs.\,\cite{Jaeckel:2010ni,Ringwald:2012hr,Arias:2012az,Graham:2015ouw,Marsh:2015xka,Irastorza:2018dyq,DiLuzio:2020wdo} for reviews). Given the portal coupling~\cite{Silveira:1985rk, Burgess:2000yq, Barger:2008jx, Barger:2010yn, Gonderinger:2012rd} between the dark and {SM-like Higgs doublet} fields, which may not be too small due to the 't Hooft naturalness argument~\cite{tHooft:1979rat}, and given that they both get VEVs, mixing between the dark and SM-like Higgs bosons is induced (see appendix \ref{app:1}). Thanks to this mixing, light dark matter can be successfully produced in the early universe (see appendix \ref{app:2}). 
This mixing leads to a universal prediction for a large class of models\footnote{{Our analysis can even be applied to probe the case that the pseudo-Nambu-Goldstone boson is the WIMP dark matter whose mass is around the half of the Higgs boson mass \cite{Ishiwata:2018sdi, Cline:2019okt, Grzadkowski:2020frj, Abe:2020iph, Abe:2021nih, Abe:2021jcz}. Although, in this case, the mass effect is not negligible, we expect that we can still have a very nice opportunity to search for the {dark Higgs} decay into DM at future lepton colliders. To distinguish the WIMP models from light-dark sector models, we focused on, after we detect the {dark Higgs} invisible decay, we may check the SM-like Higgs boson decay rate into the massive DM pair and, further, consider the indirect/direct detection experiments.}  }, which is the invisible decay of the SM-like Higgs boson into the light dark sector particles, {\it and} the existence of the dark Higgs boson. 
The dark Higgs boson and the SM-like Higgs boson may be produced and decay into the light dark matter particles in the early universe or collider experiments. In a collider, especially in a Higgs factory, the SM-like Higgs invisible decay and dark Higgs invisible decay, which have similar diagrams, may be both probed. 
The former has been studied widely. {The current upper limit on the invisible branching fraction of the SM-like Higgs boson is 0.11 (0.18) at $95\%$ Confidence Level (CL) from the ATLAS (CMS) searches \cite{ATLAS:2020kdi,CMS:2022qva}. The projected upper limit at the high luminosity LHC (HL-LHC) is also found to be 0.038 \cite{Cepeda:2019klc} (also see the studies \cite{Kato:2020pyl,Steinhebel:2021tep}).} {For the latter, excluded regions on a model parameter space by the dark Higgs invisible decay with LEP data is studied in~\cite{Sakurai:2022hwh}.} 
{In addition, in the case the SM-like Higgs boson does not decay invisibly, collider searches for dark Higgs boson {production} at the ILC with the center-of-mass energies 250 GeV and 500 GeV were studied in Refs \cite{Wang:2018awp}  (irrelevant to the dark Higgs boson decay products) and \cite{Kamon:2017yfx}, respectively. The invisible decay with either Higgs boson at CLIC was studied in~\cite{Mekala:2020zys} (see also other relevant papers~\cite{Shrock:1982kd, Chen:2019ebq, Grzadkowski:2020frj, Abe:2021nih, Bhattacherjee:2021rml, Sakurai:2021ipp}).} {However, the collider simulation involving productions and decays of both Higgs bosons has not been studied in detail so far. Since the productions, as well as decays, of the two Higgs bosons are tightly linked in a wide class of light dark sector models, in a large parameter region, they cannot be separately discussed. When they both contribute to the events, the collider signals would be affected statistically, and thus, as we will see, either enhance or suppress the resulting significance depending on the observables and parameter regions.
%By taking into the two Higgs bosons decaying invisibly, we will get more sensitive reaches than previously studied. 
%  that (a) both invisible decays of the two Higgs bosons are predictions in the wide class of the light-dark sector models, (b) neither of the processes can be simply decoupled if the dark Higgs boson is not too heavy and not have very close mass to $125$\,GeV, since they are from the similar diagrams, e.g., decoupling the SM-like Higgs boson decay while keeping the mixing parameter would lead to the dark Higgs boson mainly decay visibly (c) by considering the two Higgs boson invisible decays depending on the parameter region the reaches can be suppressed or (significantly) enhanced compared to the previous studies with single Higgs boson and simplified assumptions.
  }
%  \footnote{\WY{
%  The authors of \cite{Mekala:2020zys} considered the test of the very similar model at $380\GEV$ and $1.5\TEV$ CLIC, but they perform the collider analysis by focusing on Higgs  constraints are from the two separate collider studies by focusing on SM Higgs boson invisible decay and dark Higgs boson invisible decay. Although we have claimed that This may be justified since they focus on the dark Higgs boson mass to be very heavy than }}
%  )}

In this paper, we study the invisible decays of the two Higgs bosons at the ILC \cite{Asner:2013psa} as an approach to probe the light dark sector. 
Choosing to work at a lepton collider for such a search is mainly motivated by the fact that lepton colliders provide a clean environment with less background compared with the hadron colliders and that there is no ambiguity about the energies of the colliding particles at lepton colliders. Assuming the ILC operates at the center-of-mass energy of 250 GeV, we search for the missing energy signature from the {light dark particles} produced in the decays of the Higgs bosons. The production of a SM-like Higgs boson $h$ or a dark Higgs boson $s$ in association with a $Z$ boson, $e^-e^+ \rightarrow h/s  + Z$, with subsequent decay of the produced Higgs boson into {light dark particles} is assumed as the signal process in this study. We estimate the backgrounds relevant to the signal process and use a multivariate technique to separate the signal from the background. We pay careful attention to all decay modes of the produced $Z$ boson. Different signal regions are defined so that all possible final states are covered, and the analysis is performed for each signal region independently. It is seen that the backgrounds are well under control using a number of carefully defined discriminating variables. {We find that even if the SM-like Higgs boson invisible decay rate is a few orders of magnitude smaller than the reaches of the HL-LHC and ILC, the light dark sector models can still be probed by the invisible decay of the produced dark Higgs boson. It is also shown that a similar improvement with respect to the indirect limits on the invisible decay of the SM-like Higgs boson derived from Higgs signal strength measurements \cite{Biekotter:2022ckj,ATLAS:2019nkf, CMS:2020gsy} can be achieved.} It is also seen that the SM-like and dark Higgs bosons can be reconstructed and can even be distinguished if they have different masses. This affords the opportunity to detect the dark Higgs boson and measure its physical properties, e.g., its mass, which can provide a smoking gun signal of the light dark sector. 
{We choose ILC to perform the analysis for illustrative purposes. Our conclusions and numerical results can be directly applied to lepton colliders like CEPC, CLIC, FCC-ee~\cite{dEnterria:2016sca,Abramowicz:2016zbo,CEPC-SPPCStudyGroup:2015csa} as well as the recently proposed C$^3$~\cite{Bai:2021rdg} if the detector effects do not differ. }

This paper is organized as follows. In section \ref{model}, we discuss the model for the renormalizable light natural dark sector, its cosmology, and the prediction of the Higgs boson invisible decay {(more precisely in appendices \ref{app:1} and \ref{app:2})}. Section \ref{analysis} is dedicated to the search for Higgs invisible decays at the ILC. Section \ref{signal} discusses the signal process under consideration and relevant background processes. Monte Carlo (MC) simulation method, used to generate signal and background samples, is described in section \ref{MC}. In section \ref{selection}, the employed object identification and event selection procedure is discussed. Section \ref{Analysis} provides a detailed discussion of the multivariate method used for signal-background discrimination, and in section \ref{limits}, the constraints on the model parameters obtained in the analysis are provided.

\section{Renormalizable model for a generic light dark sector} 
\label{model}
\lac{1}

\subsection{Model}
Let us consider a dark sector with one dark Higgs field, $\F$,  which spontaneously breaks a continuous $G_{\rm dark}$ symmetry by its VEV, 
\beq 
\vev{\F}=v_\F. 
\eeq
In this minimal setup, the only renormalizable interaction between the SM and dark sector is the portal coupling between the SM{-like} and dark Higgs fields. We consider the most general Higgs potential, given by
\beq
V=-m^2_\F|\F|^2+{\lambda} |\F|^4 +\l_P |H|^2 |\F|^2 + \lambda_H |H|^4-\mu_H^2 |H|^2, \label{V}
\eeq 
where $\F$ ($H$) is the dark (SM{-like}) Higgs singlet (doublet) field which breaks the $G_{\rm dark}$ ($\SU(2)_L\times \U(1)_Y$) symmetry, $\lambda_P, \l (>0) \AND \l_H (>0)$ are constants,  $\m_H^2 \simeq (125\GEV)^2/2$ is the bare Higgs mass term in the SM, and $m^2_\F (>0)$ is the dark Higgs mass squared parameter. 

Via the symmetry breaking, we obtain, 
\begin{align}
H=
\begin{pmatrix}
G^{+} \\
v +\frac{1}{\sqrt{2}}(\phi_r+iG^0)
\end{pmatrix},\quad
\Phi=v_{\F}+\frac{1}{\sqrt{2}}(\rho+ia^\prime),
\end{align}
where $v$ denotes the VEV of the {SM-like} Higgs field{, and $G^{\pm},\  G^{0}$ are the Nambu Goldstone bosons (NGBs) absorbed as a longitudinal mode of the weak gauge bosons}. Here, for simplicity of discussion, we assume $G_{\rm dark}$ is a global $\U(1)$ symmetry. {In this case, the NGB associated with the spontaneous breaking of $G_{\rm dark}$ is an ALP. We assume the ALP, $a'$, acquires its mass via explicit symmetry breaking with a generic renormalizable potential controlled by a real order parameter $\k$,
\beq\label{Vexp}
\d V=  \kappa \(\sum_{j=1}^4 c_j  m_\F^{4-j} \Phi^{j}+ \sum_{j=1}^2 (\tl c_j^H m_\F^{2-j} \Phi^{j}  |H|^2+\tl c_j^\F m_\F^{2-j} \Phi^{j}  |\F|^2) \) +{\rm h.c.}.
\eeq
Here, $c_j, \tl c^x_j$ $( x= H, \Phi)$ are complex dimensionless coefficients which involve generic CP phases. 
At $\k\to 0$, the $\U(1)$ symmetry is exact, and {we obtain a naturally small ALP mass scaling as}~\cite{tHooft:1979rat}
\beq
m_a \sim \sqrt{\k} v_\F.
\label{makappa}
\eeq
Interestingly, if $c_j, \tl c^x_j$ have sizes of $\O(1)$ with generic $\O(1)$ CP phases, $a$ behaves as a {\it CP-even} ALP~\cite{Sakurai:2021ipp}. 
This can be easily found that in any singlet scalar extension, the generic renormalizable potential has the form $V_{\rm full}=V_{\rm full}(|H|^2, \rho, a').$  
Once we assign that $\rho, a'$ are CP-even, we find the spatially integrated potential CP conserving. 
With the generic CP phases residing in $c_j, \tl c^x_j$, this CP-even ALP can mix with the SM-like and dark CP-even Higgs bosons in the low-energy effective theory and decay into the light components in the SM, like photons. 
The mixing between the SM-like Higgs boson and ALP can be estimated as~\cite{Sakurai:2021ipp}
\beq
\laq{Defc_h}
{\theta_{ah}= c_h \frac{m_a^2}{m_h m_\F} },
\eeq
with $c_h$ being $\O(1)$ given that $c_j, \tl c^H_j, \tl c^\F_j$ are $\O(1)$. The model predicts a consistent CP-even ALP DM in the regime of $10\KEV-\MEV$ with a thermal production with a reheating temperature of a few GeV thanks to the non-vanishing portal coupling $\lambda_P$.

Assuming this model as the theoretical framework, we perform a collider search to probe the dark sector. As we have emphasized, our analysis can apply to a wider class of models, like hidden photons and many pairs of (would-be) NGBs (see appendix \ref{app:2}). If the (would-be) NGB mass and the irrelevant couplings are not too large, our numerical results will be the same quantitatively with certain parameter redefinition (due to the equivalence theorem).}

\subsection{Prediction of the Higgs boson invisible decay}
\label{prediction}
We consider a scenario in which the CP-even ALP has a mass in the sub-MeV range so that (based on the discussion in appendix \ref{app:2}) it can be a viable dark matter. To have an ALP in this mass range, the order parameter for the breaking of the $\U(1)$ symmetry should be small, $\kappa\ll1$, according to Eq.\,\eqref{makappa}. We define the physical states in the mass basis as
\begin{align}\label{eq:basis}
\begin{pmatrix}
h \\
s \\
a
\end{pmatrix}=
R_{\alpha}
\begin{pmatrix}
\phi_{r} \\
\rho \\
a^{\prime}
\end{pmatrix},\ \ \
\end{align}
where the rotation matrix $R_\alpha$ is given by
\begin{equation}
          {R}_{\alpha} = \begin{pmatrix}
                     \cos \alpha & \sin \alpha & 0 \\
            -\sin \alpha & \cos \alpha & 0 \\
            0 & 0 & 1 \end{pmatrix},
\end{equation}
in the limit $\kappa\rightarrow0$. The states $h$, $s$, and $a$ are, respectively, identified as the SM-like Higgs boson, the dark Higgs boson, and the CP-even ALP. Working on the physical basis, the model input parameters are $\alpha$, $v_\Phi$, the dark Higgs mass $m_s$, the CP-even ALP mass $m_a${, the mass of the {SM-like} Higgs boson $m_{h}$ and the EW VEV $v$\footnote{{While the complex dimensionless coefficients $c_{j}, \tilde{c}^{x}_{j}$ are also model input parameters, their effects are negligible in the regime of $\kappa\ll1$. }}. The last two parameters are fixed as $m_{h}=125.25$ GeV and $v=246$ GeV}. Here, we assume $m_s >m_{h}/2$ {for simplicity of the analysis. 
%Here, $m_{h}$ is the {SM-like} Higgs boson mass, and in this range, the {SM-like} Higgs boson cannot decay into the on-shell dark Higgs boson pair.\footnote{{In this case, the SM-like Higgs invisible decay rate would be enhanced by $\O(1)$ due to the new channel. Still, the production {of} $s$ followed by the invisible decay should be a good probe of the scenario in the parameter region.}}}
In this range, the {SM-like} Higgs boson cannot decay into the on-shell dark Higgs boson pair.\footnote{{In this case, the SM-like Higgs invisible decay rate would be enhanced by $\O(1)$ due to the new channel. Still, the production {of} $s$ followed by the invisible decay should be a good probe of the scenario in the parameter region.}}}
{We also assume the range of the mixing angle to be $0<\alpha<\pi/4$ because predicted SM-like Higgs boson couplings considerably deviate from those of the SM in the case of $\pi/4< \alpha<\pi/2$, which obviously does not fit the properties of the discovered 125 GeV Higgs boson. }

\begin{figure*}[!t]
  \centering  
    \begin{subfigure}[b]{0.482\textwidth} 
    \centering
    \includegraphics[width=\textwidth]{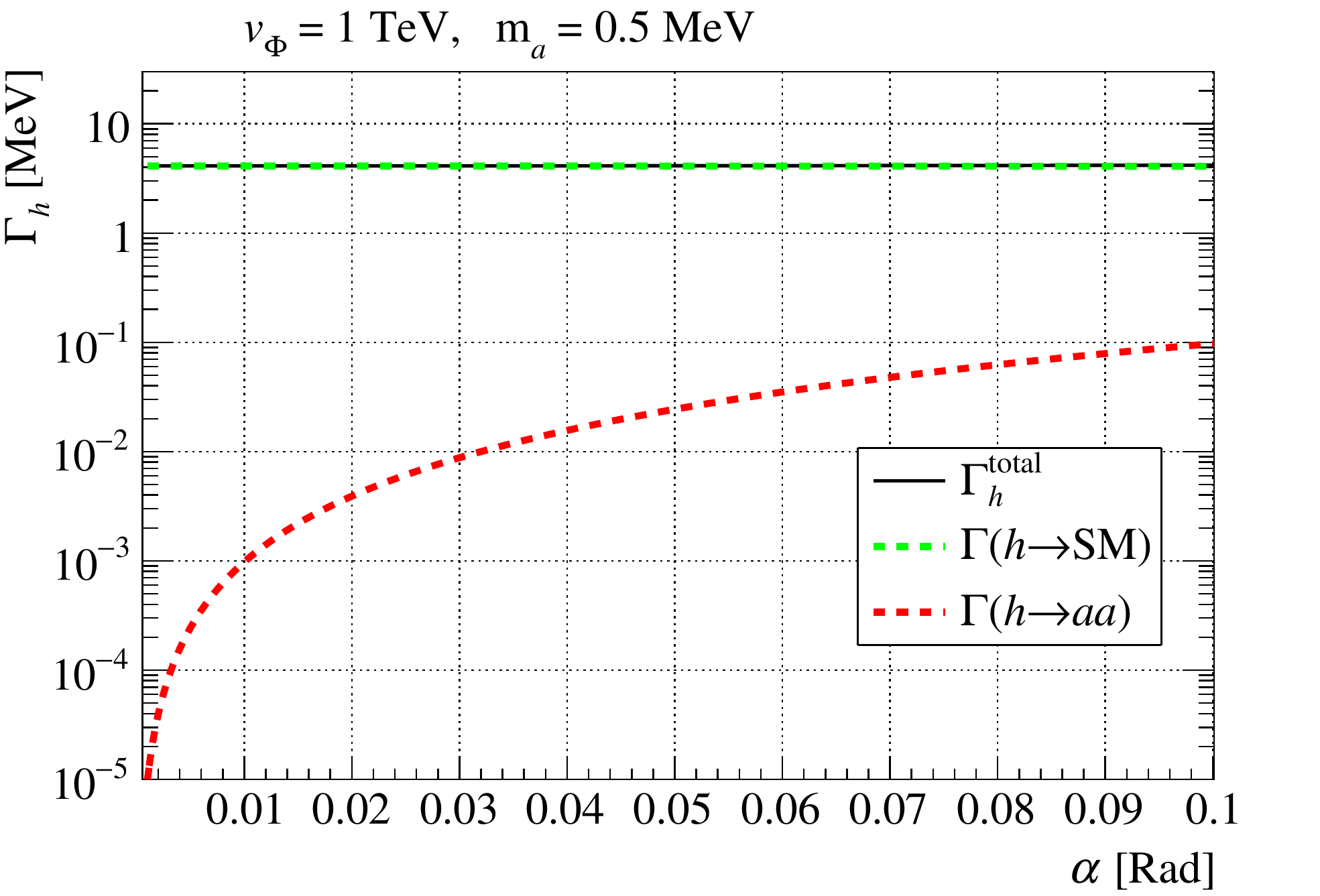}
    \caption{}
    \label{hWidth-Alpha-All}
    \end{subfigure} 
\vspace{0.25cm}
    \begin{subfigure}[b]{0.482\textwidth} 
    \centering
    \includegraphics[width=\textwidth]{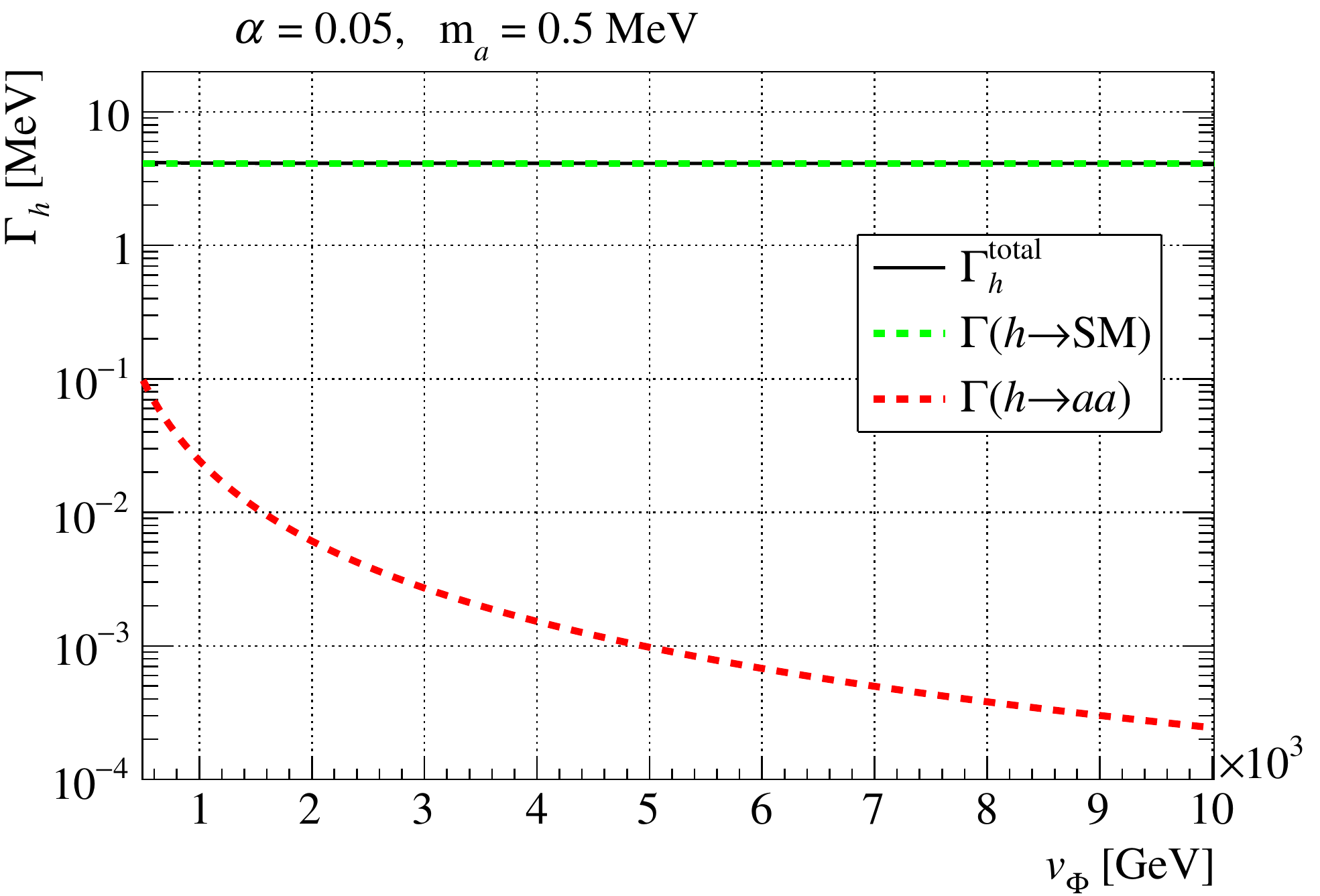}
    \caption{}
    \label{hWidth-vPhi-All}            
    \end{subfigure} 
    \begin{subfigure}[b]{0.482\textwidth} 
    \centering
    \includegraphics[width=\textwidth]{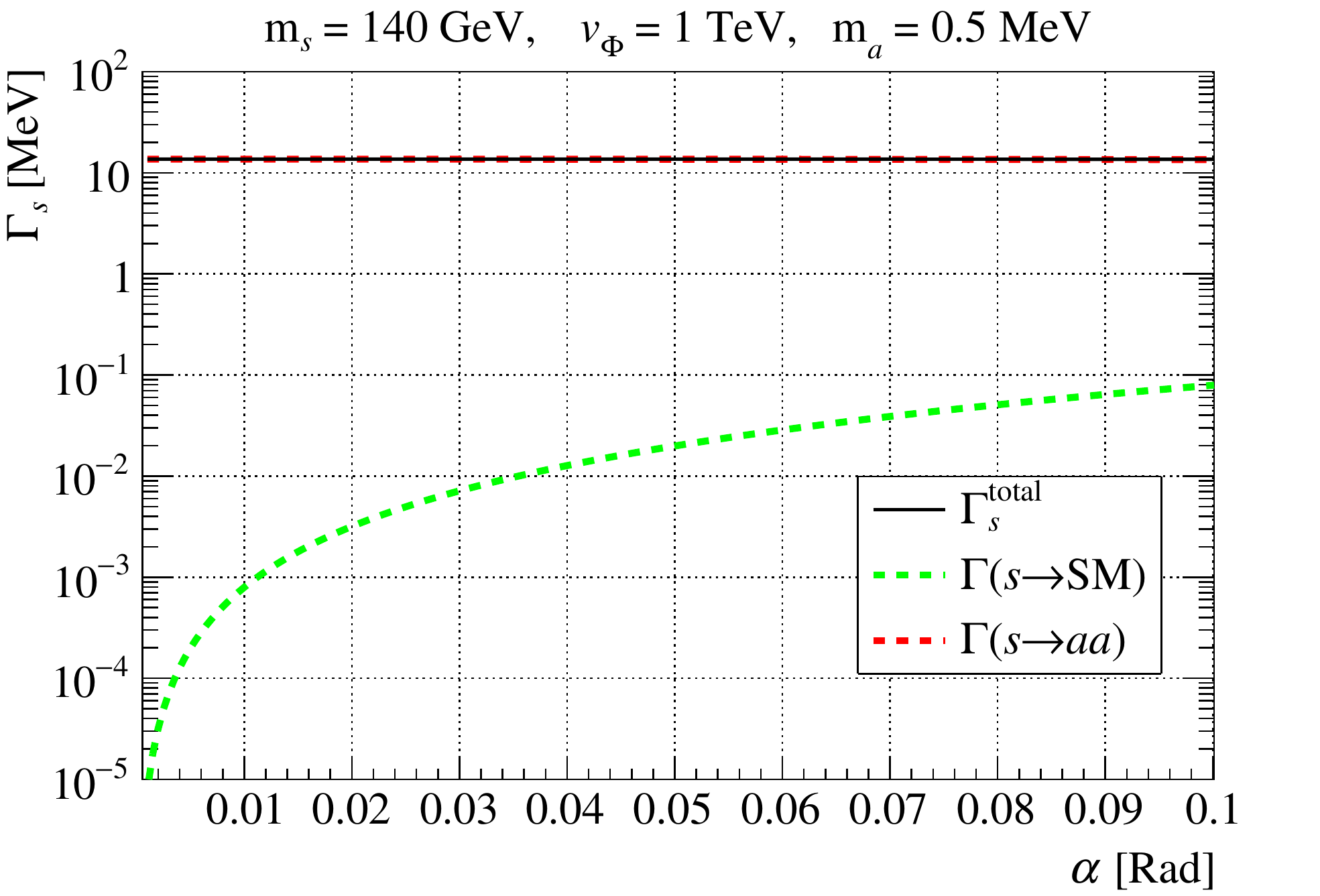}
    \caption{}
    \label{sWidth-alpha-All}
    \end{subfigure} 
\vspace{0.1cm}
    \begin{subfigure}[b]{0.482\textwidth}
    \centering
    \includegraphics[width=\textwidth]{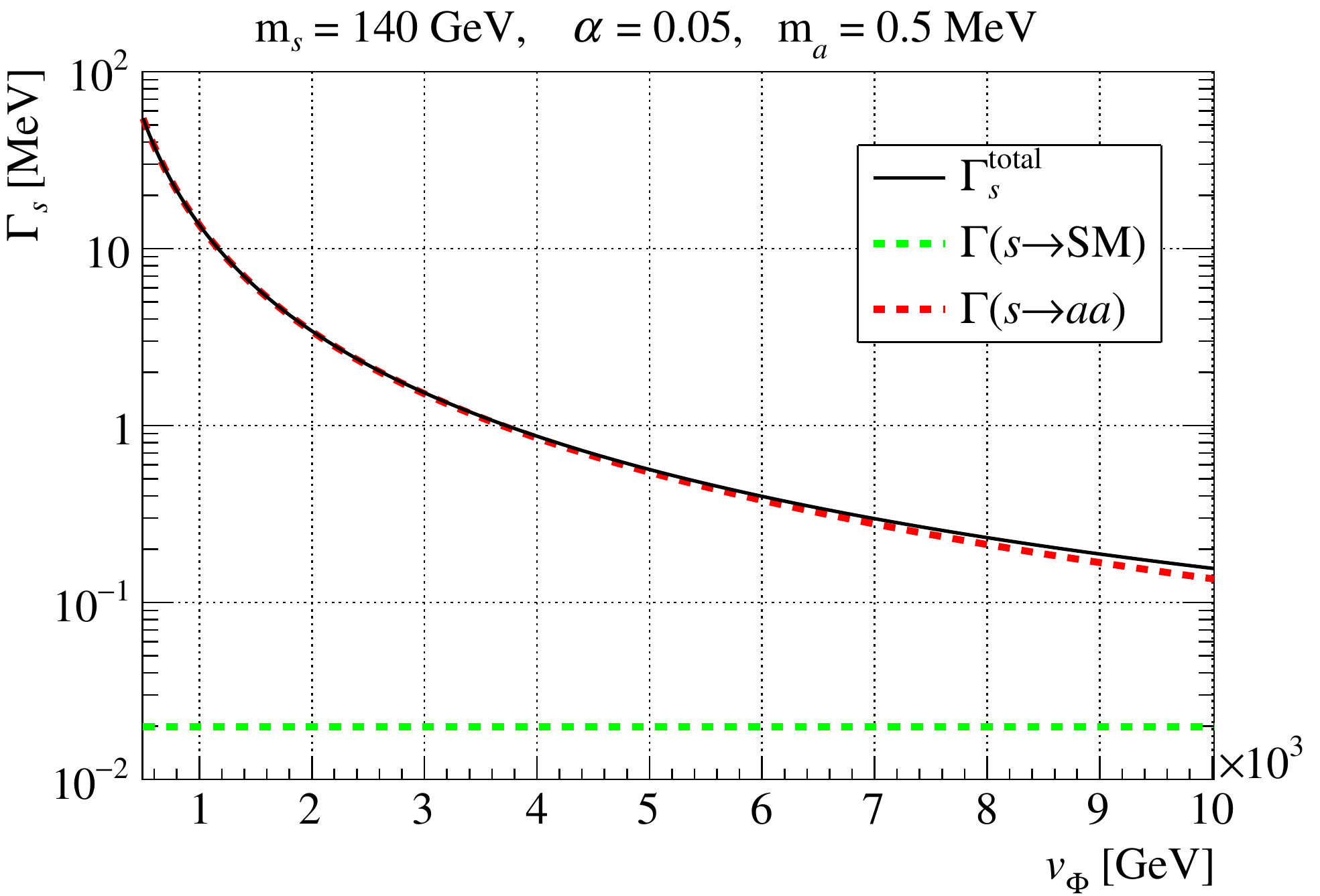}
    \caption{}
    \label{sWidth-vPhi-All}   
    \end{subfigure}
    \begin{subfigure}[b]{0.482\textwidth}
    \centering
    \includegraphics[width=\textwidth]{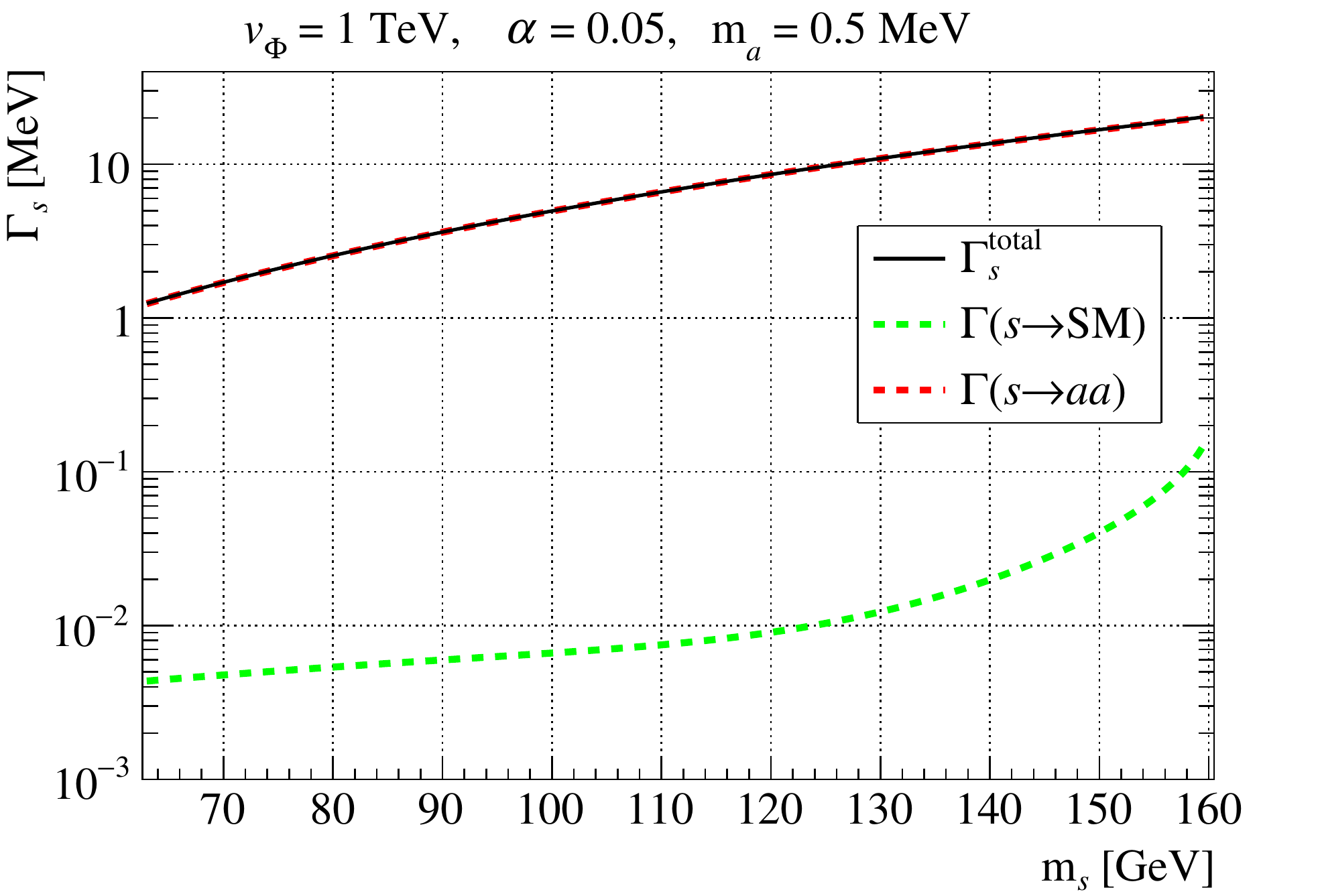}
    \caption{}
    \label{sWidth-Ms-All}
    \end{subfigure}
    \caption{Decay widths of the {SM-like} Higgs and dark Higgs bosons as a function of the model parameters $\alpha$, $v_{\Phi}$ and $m_s$. $\Gamma(h/s\ra \mathrm{SM})$ and $\Gamma(h/s\ra aa)$ are respectively the $h/s$ decay width into SM particles and ALPs, and $\Gamma^\mathrm{\,total}_{h/s}$ denotes the $h/s$ total decay width.}
  \label{hswidth}
\end{figure*}

It can be shown that the couplings of the SM/dark Higgs boson to the SM fermions $g_{h/sf\bar{f}}$ and SM gauge bosons $g_{h/sVV}$ ($V=Z,W^\pm$) are given by 
\begin{equation}
g_{h/sf\bar{f}} = \kappa^{h/s} g_{hf\bar{f}}^{\mathrm{SM}},\ \ \ 
g_{h/sVV} = \kappa^{h/s} g_{hVV}^{\mathrm{SM}},
\end{equation}
where $g^{\mathrm{SM}}_{hf\bar{f}}$ ($g^{\mathrm{SM}}_{hVV}$) denotes the SM prediction for the SM-like coupling to the SM fermions (gauge bosons), and the modification factors $\kappa^{h/s}$ are given by
\begin{equation}
\kappa^h = \cos\alpha,\ \ \ 
\kappa^s = \sin\alpha.
\end{equation}
The SM{-like} and dark Higgs bosons decay into ALPs through the couplings 
\begin{equation}
g_{haa} = \frac{m^2_h}{\sqrt{2}v_\Phi} \sin\alpha,\ \ \
g_{saa} = -\frac{m^2_s}{\sqrt{2}v_\Phi} \cos\alpha,
\end{equation}
respectively. The model also predicts the three-body decays $h\to aa s$ for the {SM-like} Higgs boson {and $s\to aah$ for the dark Higgs boson} {if kinematically allowed}. The {widths} of these decay modes {are} shown to be highly suppressed, and thus {these decays are} neglected in this study~\footnote{{In the limit of $\alpha\ll1$, the analytic expression for {the width of the decay} $h\to aa s$ is given by $\Gamma_{h\to aa s}\simeq m_{s}^{4} \sin^{2}\alpha/(2048\pi^{3}m_{h}^{3}v_{\Phi}^{4})\left[m_{h}^{4}-m_{s}^{4}-m_{h}^{2}m_{s}^{2}\log(m_{s}^{2}/m_{h}^{2})\right]$. Compared with two-body decays of $h$, this width is suppressed by the phase space factor, the mixing angle $\alpha$ and the factor $m_{h}^{4}/v_{\Phi}^{4}$. The width for $s\to aa h$ is obtained by exchanging $m_{s}\leftrightarrow m_{h}$. }}. The prediction of the model for the decay widths of the SM and dark Higgs bosons into ALPs and SM particles are shown in Fig. \ref{hswidth}. 
The decay widths are provided as functions of the model parameters. {For the SM-like Higgs boson, the partial width for the decays into SM particles is expressed by {$\Gamma(h\to \mathrm{SM})=(\kappa^{h})^{2}\,\Gamma^{\rm SM}(h\to \mathrm{SM})$}, where {$\Gamma^{\rm SM}(h\to \mathrm{SM})$} denotes the decay width into the SM (The analytical formulae can be found in, e.g.,~Refs.\cite{Djouadi:2005gi,Spira:2016ztx}.). The width for $h\to aa$ is given by ${\Gamma(h\to aa)}=g_{haa}^{2}/(32\pi m_{h})\sqrt{1-4m_{a}^{2}/m_{h}^{2}}$.}
In computing the decay width of the dark Higgs boson, all decay modes available for the considered dark Higgs masses, i.e., the decays into charged leptons, quarks, gluons, $WW^*$, $ZZ^*$, $Z\gamma$, $\gamma\gamma$ and ALPs, have been taken into account. 
{Analytic formula for each decay width of the dark Higgs boson is obtained from the one for the SM-like Higgs boson by the replacement $(m_{h},\kappa^{h},g_{haa})\to (m_{s},\kappa^{s},g_{saa})$.}
 We have also considered the NNLO QCD corrections for the decays into quarks~{\cite{Chetyrkin:1995pd,Larin:1995sq}} and gluons{~\cite{Chetyrkin:1997iv}}. 

Although the ALP interactions with light SM particles are highly suppressed by the positive powers of the ALP mass, the ALP interaction with the {SM-like} Higgs boson is not suppressed. This feature makes the search for invisible decays of the SM{-like} Higgs boson a helpful approach to probe the dark sector. However, the invisible decay of the SM{-like} Higgs boson can be probed down to branching fractions of $\O(0.1)\%$ at the ILC 
{\cite{Asner:2013psa,Kato:2020pyl,Steinhebel:2021tep} ({cf.}~\cite{Cepeda:2019klc,dEnterria:2016sca,Abramowicz:2016zbo,CEPC-SPPCStudyGroup:2015csa} for sensitivity of other future colliders)}, which corresponds to a {specific} parameter region (low $v_\Phi$$\mhyphen$high $\alpha$) for the SM{-like} Higgs boson decay into ALPs, as can be seen from Figs. \ref{hWidth-Alpha-All},\ref{hWidth-vPhi-All} {(With the above future sensitivity of the branching fraction, the decay width {$\Gamma(h\to aa)$} can be probed {down} to the order of $10^{-3}$ MeV)}. 
Invisible decays of the SM{-like} Higgs boson, therefore, only allow for probing the dark sector to a very limited extent. As mentioned before, the dark Higgs boson can also be produced at the ILC. Interestingly, the diagrams through which the dark Higgs boson is produced are similar to those of the SM-like boson. Moreover, as seen in Figs. \ref{sWidth-alpha-All},\ref{sWidth-vPhi-All},\ref{sWidth-Ms-All}, the invisible decay is the dominant decay mode of the dark Higgs boson for the whole parameter region considered here.  
{This is because the visible decay of the dark Higgs boson in the mass range is not only suppressed by the mixing but also by the small Higgs-Yukawa coupling or loop factor. 
Therefore, a hadron collider is difficult to probe the parameter region for $v_\Phi\lesssim 10\TEV$ \footnote{{That said, for $v_\F\gtrsim 10\TEV$ and $m_s\gtrsim 150\GEV$, the visible decay of the dark Higgs gets more important, which will be discussed elsewhere. For a search for visible decays of a heavy scalar, with a mass in the range $0.25\mhyphen\mhyphen1$ TeV, see e.g. {Refs.\, \cite{Robens:2015gla,Robens:2016xkb,No:2018fev}.}} }}. %~\cite{Aad:2019mbh, CMS:2020gsy}
A very beneficial approach to probe the dark sector is, therefore, to consider both the dark and SM-like bosons in the search for invisible decays. 
{As described below}, taking account of both the Higgs bosons offers a golden opportunity to extend the probe of the dark sector to regions below the reach of the ILC for invisible decays of the SM{-like} Higgs boson.

It is worth mentioning that, for the dark Higgs boson, the decay into SM particles is suppressed by $\sin^2\alpha$ and the invisible decay, which depends on $\cos^2\alpha/v_{\Phi}^2$, is dominant unless for large enough $v_{\Phi}$ values. For $v_{\Phi}$ values significantly larger than those considered in this study, the dark Higgs decay into the visible sector can become dominant and, therefore, of interest in the search for the dark sector.

\section{Probing light dark sector via (dark) Higgs invisible decay at the ILC}
\label{analysis}
Assuming ILC operates at the center-of-mass energy of 250 GeV, we perform a search for the missing energy signature {due to the ALP production}. This search takes advantage of a signal to which both the SM-like and dark Higgs bosons contribute. The ALP is produced in the decays of the SM and dark Higgs bosons. In this search, the model parameters are constrained and the resulting bounds, which correspond to $v_{\Phi}=0.5$, 1 and 10 TeV, will be presented in the $\sin\alpha\mhyphen m_s$ plane.

%***********************************************************
\subsection{Signal and background processes}
\label{signal}
\begin{figure}[t]
  \centering  
    \begin{subfigure}[b]{0.9\textwidth} 
    \centering
    \includegraphics[width=\textwidth]{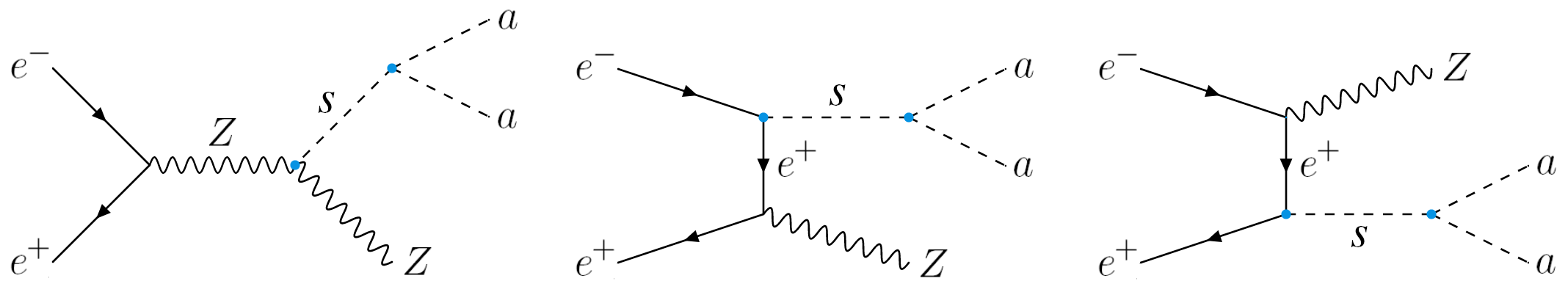}
    \caption{}
    \label{feyn1}
    \end{subfigure} 
    \begin{subfigure}[b]{0.9\textwidth}
    \centering
    \includegraphics[width=\textwidth]{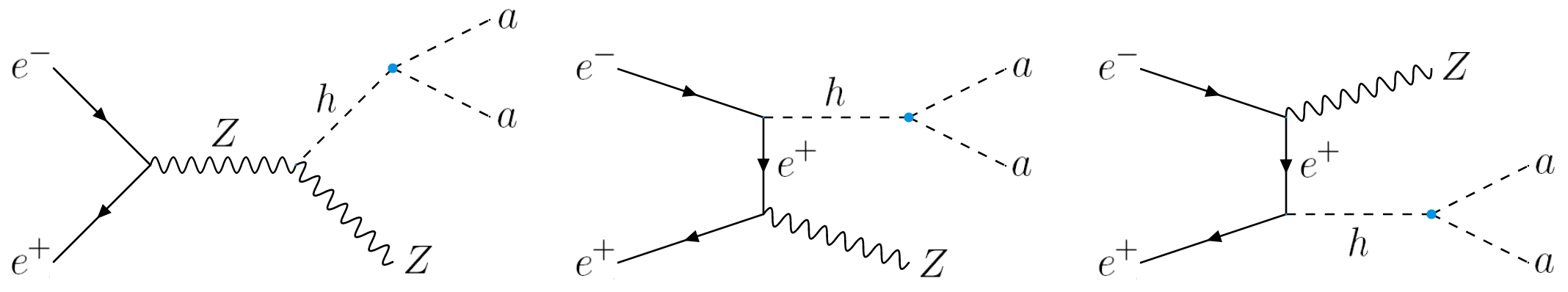}
    \caption{}
    \label{feyn2}
    \end{subfigure}
    \caption{Leading order Feynman diagrams contributing to the signal process $e^-e^+\ra aaZ$. The blue dots show the beyond SM interactions.}
  \label{diagrams}
\end{figure}
We search for the signal process $e^-e^+ \rightarrow h/s  + Z  \rightarrow aaZ$, in which the $Z$ boson is produced in association with the SM/dark Higgs boson, and the produced Higgs boson subsequently decays into ALPs. The mass of the ALP is assumed to be in the sub-MeV range, which corresponds to a lifetime {much longer} than the age of the universe for the ALP. The produced ALPs remain invisible in the detector and manifest themselves as missing energy. For the parameter space considered in this study, the width of the dark Higgs boson into ALPs is at least of $\O(10^{-6})$ GeV, which corresponds to a lifetime of $\O(10^{-19})$ s for the most energetic dark Higgs bosons produced at the ILC with the center-of-mass energy of 250 GeV. The dark Higgs boson, therefore, promptly decays into ALPs after production. The leading order Feynman diagrams contributing to the signal process are shown in Fig. \ref{diagrams}. The diagrams in Fig. \ref{feyn1} involve the dark Higgs boson, and the diagrams in Fig. \ref{feyn2} involve the SM-like boson. 

\begin{figure}[t]
  \centering  
    \begin{subfigure}[b]{0.49\textwidth} 
    \centering
    \includegraphics[width=\textwidth]{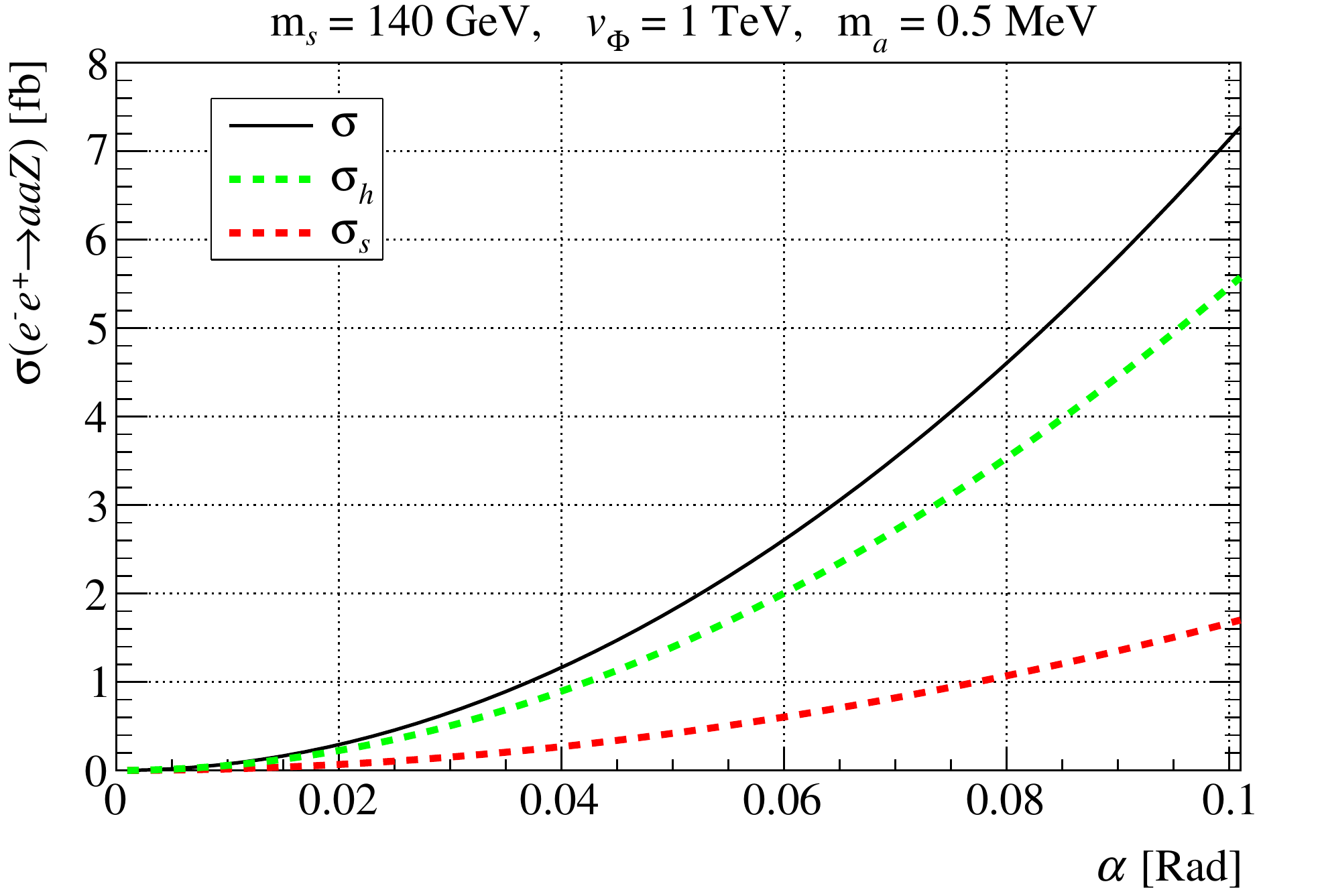}
    \caption{}
    \label{xsec-alpha-All}
    \end{subfigure} 
\vspace{0.3cm}
    \begin{subfigure}[b]{0.49\textwidth}
    \centering
    \includegraphics[width=\textwidth]{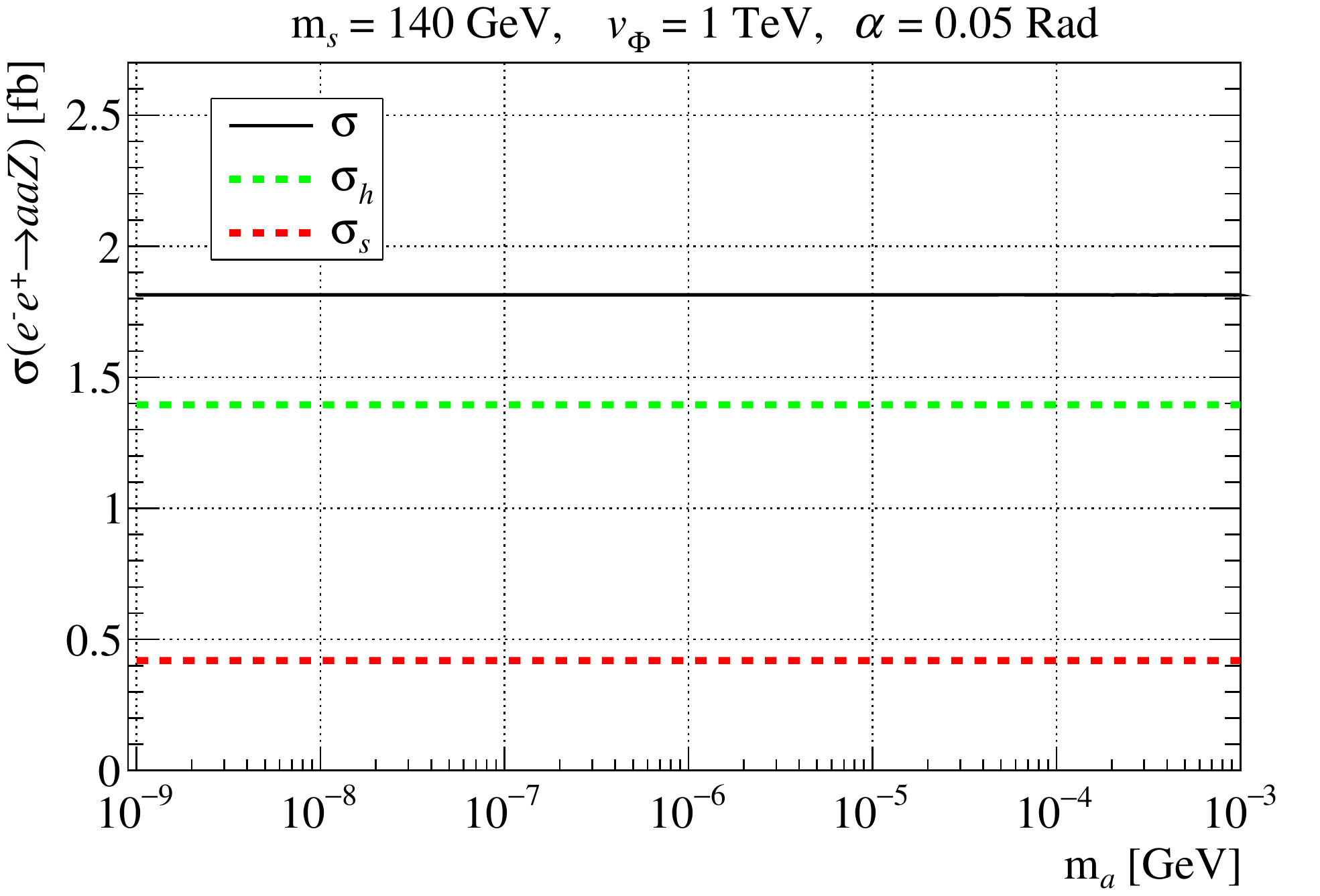}
    \caption{}
    \label{xsec-Ma-All}
    \end{subfigure}
    \begin{subfigure}[b]{0.49\textwidth}
    \centering
    \includegraphics[width=\textwidth]{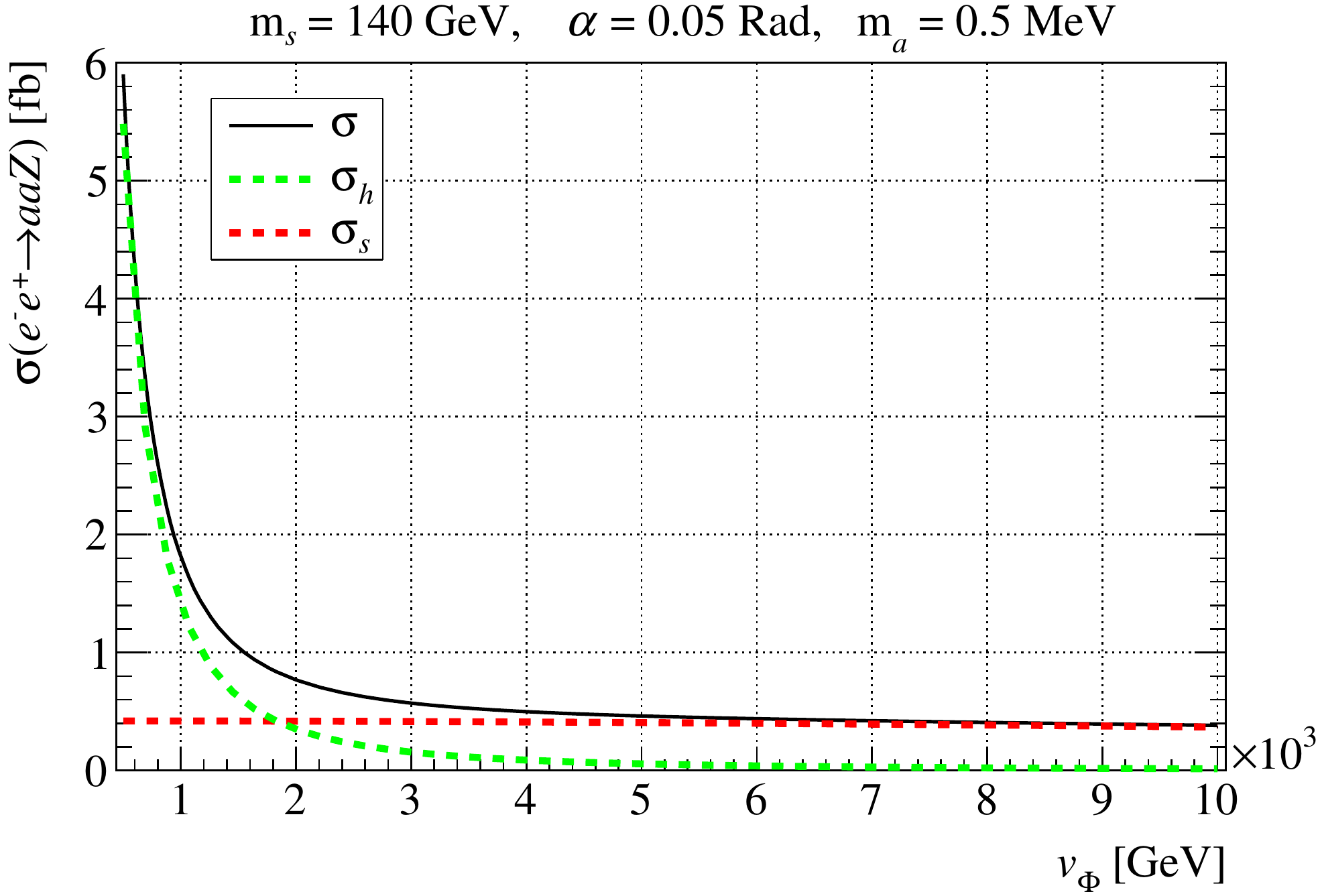}
    \caption{}
    \label{xsec-vPhi-All}
    \end{subfigure}
    \begin{subfigure}[b]{0.49\textwidth}
    \centering
    \includegraphics[width=\textwidth]{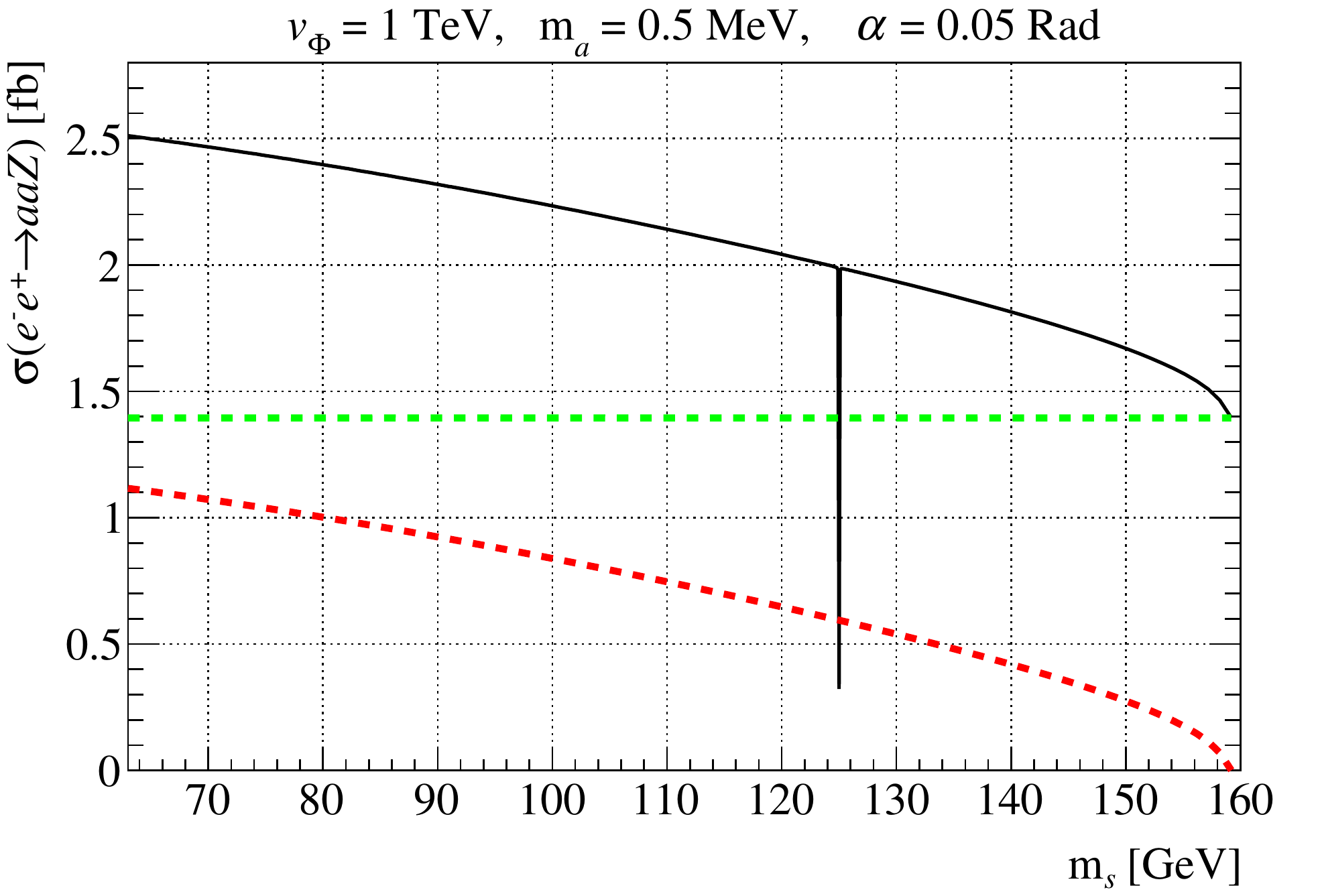}
    \caption{}
    \label{xsec-Ms-All}
    \end{subfigure}
    \caption{Cross section of the signal process $e^-e^+\ra aaZ$ versus the model parameters a) $\alpha$, b) ALP mass $m_a$, c) $v_{\Phi}$ and d) dark Higgs mass $m_s$. The signal cross section $\sigma$ is shown by the black solid line and $\sigma_{h}$ ($\sigma_{s}$), which denotes the contribution of the $h$-involved ($s$-involved) processes to the signal cross section, is shown by the green (red) dashed line.}
  \label{xsec}
\end{figure}

The signal cross section $\sigma$ can be written as $\sigma_h+\sigma_s+\sigma_{inter.}$, where $\sigma_h$ and $\sigma_s$ are respectively the contributions of the $h$-involved and $s$-involved processes, and $\sigma_{inter.}$ denotes the interference between the $h$-involved and $s$-involved processes{, which {is} important when $m_s \approx m_h.$} Fig. \ref{xsec} shows the signal cross section as a function of the model parameters. The presented cross sections are obtained with the use of \texttt{MadGraph5\_aMC@NLO} \cite{Alwall:2011uj,Alwall:2014hca,Frederix:2018nkq}.
The diagrams in Fig. \ref{diagrams} depend on $\sin2\alpha$ and thus the signal cross section increases with $\alpha$ as seen in Fig. \ref{xsec-alpha-All}. The signal cross section does not show a significant dependence on the ALP mass, as seen in Fig. \ref{xsec-Ma-All}. This results from the smallness of the ALP mass in the sub-MeV range when compared with the masses of the SM and dark Higgs bosons and the energy of the colliding particles. The analysis in this work is independent of the ALP mass {in the {considered} range,} 
and thus is only performed for one mass value. Throughout this paper, it is assumed that $m_a=0.5$ MeV. According to Fig. \ref{xsec-vPhi-All}, which shows the signal cross section scan over $v_{\Phi}$, the contribution of the $h$-involved processes to the signal cross section is dominant at low $v_{\Phi}$ values. As $v_{\Phi}$ increases, $\sigma_h$ decreases rapidly while $\sigma_s$ 
{does not change much. This is because the $h$-involved cross section is proportional to the SM-like Higgs invisible decay rate scaling roughly as $\propto \a^2 m_h^3/v_\F^2$, {but the $s$-involved cross section is only determined by the $s$ production cross section which is independent of $v_\F$ {at the leading order} as long as the $s$ invisible decay is faster than the visible decay suppressed by the mixing} \footnote{{In this case, the branching fraction for $s\to aa$ is almost independent of $v_{\Phi}$. }}.  
Thus, }the $s$-involved processes become dominant soon {by increasing $v_\F$.} 
As we will see, this behavior has significant effects on the limits obtained at different $v_{\Phi}$ values. Fig. \ref{xsec-Ms-All} shows the signal cross section as a function of the dark Higgs boson mass. As seen, the cross section decreases as the dark Higgs mass increases since the available phase space {gets kinematically suppressed.} For dark Higgs masses above the threshold $\sqrt{s}-m_Z\approx 158.8$ GeV, the contribution of the $s$-involved processes to the signal is highly suppressed as the dark Higgs boson can only be produced off-shell. This study concentrates on the dark Higgs masses below this threshold. As seen, the contribution of the interference term to the signal cross section is negligible for all dark Higgs masses except for $m_s\approx m_h$. At masses in the vicinity of the SM-like boson mass, a large destructive interference causes a rapid decrease in the signal cross section {(See Ref.\,\cite{Sakurai:2022cki} for a detailed theoretical study in this regime. The suppression can be interpreted as a quantum Zeno effect.)}

The $Z$ boson produced in the signal process decays via different hadronic and leptonic decay modes. These decay modes and their corresponding branching fractions are shown in Tab. \ref{ZBR}. As seen, the highest branching fraction ($\approx 70 \%$) belongs to the decay into a di-quark. The signal is, therefore, dominated by events with a final state di-jet. The leptonic decay $Z\rightarrow \ell^-\ell^+$ ($\ell=e,\mu,\tau$) has a branching fraction of $3.4\%$. In case of the decay into a tau pair, the produced tau leptons promptly decay through the hadronic, semileptonic, and leptonic modes. The lowest branching fraction corresponds to the decay into a tau pair with subsequent leptonic decays of the tau leptons.
\begin{table}[t]
\normalsize
    \begin{center}
         \begin{tabular}{cc}
Decay mode & BR \\ \Xhline{1\arrayrulewidth}
\multicolumn{1}{r}{$q\bar{q}$}  & 0.699  \\ 
\multicolumn{1}{r}{$ e^-e^+$}  & 0.034  \\ 
\multicolumn{1}{r}{$ \mu^-\mu^+$}  & 0.034  \\ 
\multicolumn{1}{r}{$ \tau^-\tau^+ (\mathrm{hadronic}) $}  & 0.014 \\ %3.3696×64.7261×64.7261      0.014116831
\multicolumn{1}{r}{$ \tau^-\tau^+ (\mathrm{semileptonic}) $}  & 0.015  \\ % 3.3696×(64.7261×17.3937+64.7261×17.8175)×2   0.015359201
\multicolumn{1}{r}{$ \tau^-\tau^+ (\mathrm{leptonic})$}  & 0.004  \\ % 3.3696×(17.3937×17.3937+2×17.3937×17.8175+17.8175×17.8175)    0.004177726
\end{tabular}
\caption{Decay modes of the $Z$ boson with corresponding branching fractions based on data taken from \cite{Workman:2022ynf}.}
\label{ZBR}
  \end{center}
\end{table}

The dominant background processes relevant to the assumed signal process are $\ell\bar{\ell}$ ($\ell=e,\mu,\tau$), $W^-W^+$, $ZZ$, $hZ$, $Z\gamma$ and multijet production. These background processes are considered in this study, and their contributions are estimated. {There are also some subdominant background processes producing the signal signature. These processes have a small contribution to the total background, and therefore, are not included in the list of backgrounds. The production of $ZZ^*\rightarrow Z \nu_\ell \bar{\nu_\ell}$ and $W^\pm W^{\mp*}\rightarrow W^\pm \ell \bar{\nu_\ell} (\bar{\ell} \nu_\ell)$ is an example of these processes. 
{The effects of such processes will be irrelevant and be neglected compared with the systematic uncertainty of the event selection efficiencies in sec \ref{limits}}.}

\subsection{Monte Carlo simulation}
\label{MC}
The model described in section \ref{model} is implemented into \texttt{FeynRules} \cite{Alloul:2013bka} to generate the Universal FeynRules Output (UFO) model. \texttt{MadGraph5\_aMC@NLO} uses the generated UFO file and generates hard events. \texttt{Pythia 8.2.43} \cite{Sjostrand:2006za} is used to perform parton showering, hadronization and decays of unstable particles. The simulation of detector effects is performed by \texttt{Delphes 3.4.2}~\cite{deFavereau:2013fsa} using the ILC card based on the proposed International Large Detector (ILD) \cite{ILD:2019kmq}. The center-of-mass energy of the colliding electron-positron beams is set to 250 GeV and the beams are assumed to be unpolarized. Event generation is independently performed for different $v_\Phi,m_s,\alpha$ parameter sets. $v_\Phi$ has the values 0.5, 1 and 10 TeV and the dark Higgs mass ranges from $m_h/2\approx 63$ GeV to the on-shell dark Higgs production threshold $\sqrt{s}-m_Z\approx 158.8$ GeV. In all generations, the mass of the ALP is assumed to be 0.5 MeV. 

\subsection{Object identification and event selection}
\label{selection}
Jets in the events are reconstructed by the anti-$k_t$ algorithm~\cite{Cacciari:2008gp} inside \texttt{FastJet 3.3.2} \cite{Cacciari:2011ma} assuming the jet cone size of 0.5. The reconstructed jets are required to satisfy the transverse momentum and pseudorapidity conditions $p_T>30$ GeV and $\vert \eta \vert < 2.5$. Jets are tagged using a tau tagging algorithm with $40\%$ efficiency for tau-jets and $0.1\%$ mistag rate for light jets. This tagger uses the distance criterion $\Delta R<0.5$, where $\Delta R = \sqrt{\Delta\eta^2+\Delta\phi^2}$, with $\eta$ and $\phi$ being the pseudorapidity and azimuth angle, respectively. Isolated objects are identified using the relative isolation variable $I_{rel}$, which is evaluated for each candidate particle. The relative isolation variable is defined as $I_{rel}=\Sigma \, p_T^{\,i}/p_T^{\,\mathrm{P}}$, where P is the candidate particle, and $i$ is the summation index running over all particles (excluding the particle P) within a cone with size 0.5 centered on the candidate particle. The minimum transverse momentum required for particles to be taken into account in the summation is 0.5 GeV. An object is identified as an isolated object if $I_{rel}<I_{rel}^{\,\mathrm{max}}$. $I_{rel}^{\,\mathrm{max}}$ is set to 0.12 for electrons and photons, and is assumed to be 0.25 for muons. Isolated electrons, photons and muons are required to satisfy the conditions $p_T>10$ GeV and $\vert \eta \vert < 2.5$. 

Seven signal regions are defined based on different signal final states the decay modes of the $Z$ boson produce. The criteria defining these signal regions are provided in Tab. \ref{SRs}.
\begin{table}[!t]
\normalsize
    \begin{center}
         \begin{tabular}{cccccc}
Signal region & $N_{jet}$ & $N_{\tau\mhyphen jet}$ & $N_{\mu}$ & $N_{e}$ & \\ \Xhline{1\arrayrulewidth}
SR1 & $2$ & $0$ & $0$ & $0$ &\\ 
SR2 & $0$ & $0$ & $2$ & $0$ & \multicolumn{1}{l}{($N_{\mu^-}=1$, $N_{\mu^+}=1$)}  \\ 
SR3 & $0$ & $0$ & $0$ & $2$ & \multicolumn{1}{l}{($N_{e^-}=1$, $N_{e^+}=1$)} \\ 
SR4 & $0$ & $2$ & $0$ & $0$ & \\ 
SR5 & $0$ & $1$ & $1$ & $0$ & \\ 
SR6 & $0$ & $1$ & $0$ & $1$ & \\  
SR7 & $0$ & $0$ & $1$ & $1$ & \multicolumn{1}{l}{($N_{e^-}=1$, $N_{\mu^+}=1$ or $N_{e^+}=1$, $N_{\mu^-}=1$)} \\ 
  \end{tabular}
\caption{Signal regions and corresponding defining criteria. $N_{jet}$, $N_{\tau\mhyphen jet}$, $N_{\mu},$ and $N_{e},$ respectively, denote the number of reconstructed jets which are not tau-tagged, number of tau-tagged jets, number of muons, and number of electrons in an event.}
\label{SRs}
  \end{center}
\end{table}
The signal regions SR1, SR2, and SR3 respectively correspond to the di-jet (where jets are not tau-tagged), di-muon and di-electron final states. The signal region SR4 is defined by requiring exactly two tau-tagged jets. The signal region SR5 (SR6) requires exactly one tau-tagged jet and one muon (electron). The last signal region, SR7, requires exactly one electron and one muon. The leptons required in these signal regions should pass the isolation criterion. Moreover, in signal regions requiring two leptons, the leptons should have opposite charges. 

\subsection{Analysis}
\label{Analysis}
Applying the object identification conditions and signal regions criteria, events are selected and categorized into the seven assumed signal regions. Different signal regions are analyzed independently. Tab. \ref{eff} shows the obtained signal and background event selection efficiencies corresponding to different signal regions for an assumed set of model parameters. As seen, the highest signal selection efficiency is obtained for the signal region SR1. This signal region benefits from the dominance of the hadronic decays of the $Z$ boson. The signal regions SR2 and SR3 have the second and third highest signal selection efficiencies. The efficiency obtained for SR2 is slightly better than that of SR3 due to the better efficiency for the muon reconstruction in the detector compared with the electron. The lowest signal selection efficiencies correspond to the signal regions SR4 to SR7, which are dominated by the $Z$ decay into a tau pair. In particular, SR4, which requires a pair of tau-tagged jets, has the lowest signal selection efficiency. Difficulties in reconstructing jets and the low tau-tagging efficiency are the main reasons leading to such a low event selection efficiency for this signal region.

\begin{figure}[!t]
    \centering
    \includegraphics[width=.60\textwidth]{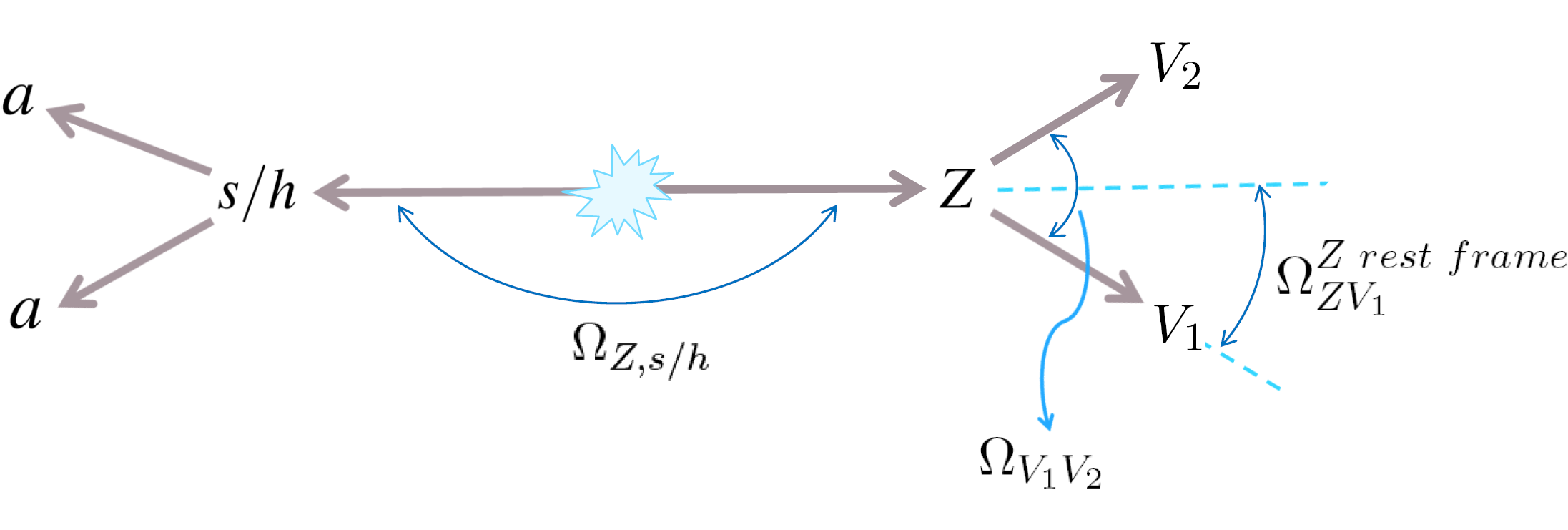}
    \caption{Cartoon showing the associated production of a SM or dark Higgs boson and a $Z$ boson with subsequent decay of the Higgs boson into ALPs and decay of the $Z$ boson into $V_1$ and $V_2$. $V_1$ and $V_2$ are visible decay products of the $Z$ boson, which can be a jet, a tau-jet, a muon, or an electron. Neutrinos produced in association with $V_1$ and $V_2$ in some decay modes are not shown. The angles $\Omega_{V_1 V_2}$, $\Omega^{Z\,\,rest\,\,frame}_{ZV_1}$ and $\Omega_{Z,s/h}$ are used as discriminating variables and are defined in the main text.}
    \label{signalCartoon}	
\end{figure}

\begin{figure}[!t]
  \centering  
    \begin{subfigure}[b]{0.495\textwidth} 
    \centering
    \includegraphics[width=\textwidth]{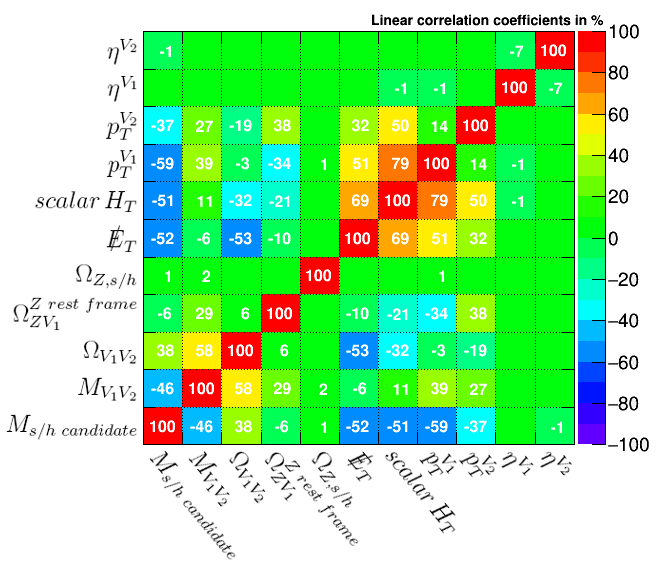}
    \caption{}
    \label{corrtable-SR1-vPhi1000Ms80}
    \end{subfigure} 
    \begin{subfigure}[b]{0.495\textwidth}
    \centering
    \includegraphics[width=\textwidth]{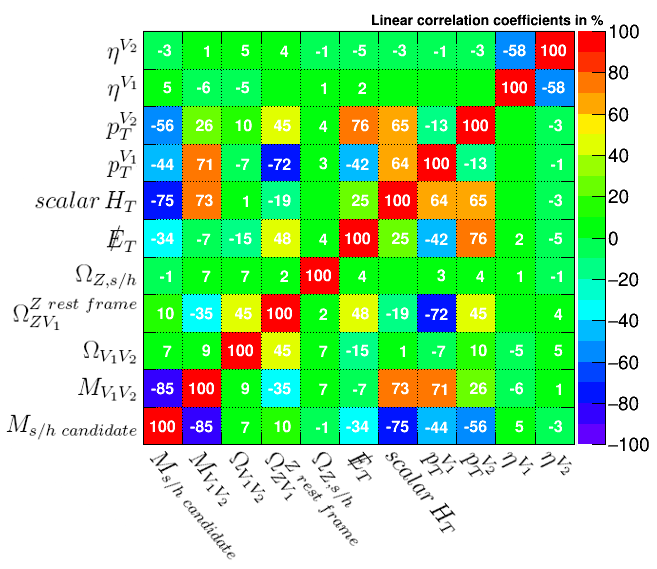}
    \caption{}
    \label{corrtable-SR6-Ma.0005vPhi10000Ms150}
    \end{subfigure}
%    \begin{subfigure}[b]{0.495\textwidth}
%    \centering
%    \includegraphics[width=\textwidth]{corrtable-SR7-vPhi1000Ms150}
%    \caption{}
%    \label{corrtable-SR7-Ma.0005vPhi1000Ms150}
%    \end{subfigure}
    \begin{subfigure}[b]{0.495\textwidth}
    \centering
    \includegraphics[width=\textwidth]{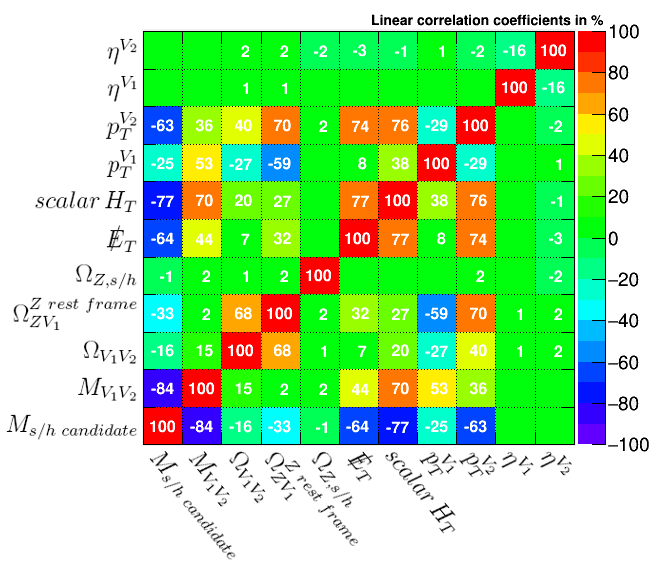}
    \caption{}
    \label{corrtable-SR7-Ma.0005vPhi10000Ms90}
    \end{subfigure}
    \caption{{Correlation matrices for the used discriminating variables corresponding to a) the signal region SR1 assuming $v_\Phi=1$ TeV and $m_s=80$ GeV, b) the signal region SR6 assuming $v_\Phi=10$ TeV and $m_s=150$ GeV, and c) the signal region SR7 assuming $v_\Phi=10$ TeV and $m_s=90$ GeV. In all the cases, it is assumed that $\alpha=0.05$ Rad and $m_a = 0.5$ MeV.}}
  \label{corrtable}
\end{figure}

Preselected events are analyzed to separate the signal from the background using a number of discriminating variables. Fig. \ref{signalCartoon} provides a cartoon showing the signal process $e^-e^+\ra aaZ$ with subsequent decay of the $Z$ boson into $V_1$ and $V_2$, where $V_1$ and $V_2$ are visible products of the $Z$ boson decay and $V_1$ is assumed to have a larger transverse momentum than $V_2$. The list of the discriminating variables used in this analysis is provided below. Some of the variables are also shown in Fig. \ref{signalCartoon}. The discriminating variables are assumed to be measured in the laboratory frame unless stated otherwise.
\begin{table}[!t]
\normalsize
    \begin{center}
         \begin{tabular}{ccccccccc}
& signal & $hZ$  & $\tau^+\tau^-$ & $W^+W^-$ & $Z\gamma$ & $ZZ$ & $\ell\bar{\ell}$ & multijet \\ \Xhline{1\arrayrulewidth}
SR1 & 0.17712 & 0.31771 & 0.10622 & 0.26258 & 0.14977 & 0.30364 & 0.00021 & 0.64717 \\ 
SR2 & 0.02722 & 0.00333 & 0.01709 & 0.01108 & 0.00714 & 0.01430 & 0.00076  & 0 \\
SR3 & 0.02442 & 0.00280 & 0.01657 & 0.01001 & 0.00734 & 0.01281 & 0.21458 & 1.3e-06  \\
SR4 &  0.00023 & 0.00060 & 0.04445 & 9.2e-05 & 0.00022 & 0.00023 & 0 & 1.7e-05 \\
SR5 & 0.00054 & 0.00097 & 0.05365 & 0.00206 & 0.00050 & 0.00046 & 0 & 1.8e-06 \\
SR6 & 0.00051 & 0.00097 & 0.05389 & 0.00192 & 0.00051 & 0.00046 & 9.3e-06 & 0 \\
SR7 & 0.00056 & 0.00162 & 0.03376 & 0.02078 & 0.00053 & 0.00037 & 0 & 0 \\
\end{tabular}
\caption{Signal and background event selection efficiencies obtained for different signal regions corresponding to $v_{\Phi}=1$ TeV, $m_s=140$ GeV, $\alpha=0.05$ and $m_a$ = 0.5 MeV.}
\label{eff}
  \end{center}
\end{table}
\begin{itemize} 
\item
$M_{s/h\,\,candidate}$: reconstructed mass of the dark/SM-like boson candidate.
\item
$M_{V_1 V_2}$: invariant mass of the $V_1$ and $V_2$ objects.
\item
$\Omega_{V_1 V_2}$: angle between the momentum vectors of the $V_1$ and $V_2$ objects. 
\item
$\Omega^{Z\,\,rest\,\,frame}_{ZV_1}$: angle between the momentum vector of the reconstructed $Z$ boson in the laboratory frame and the momentum vector of the object $V_1$ as measured in the rest frame of the reconstructed $Z$ boson. 
\item
$\Omega_{Z,s/h}$: angle between the momentum vectors of the reconstructed $Z$ boson and the reconstructed dark/SM-like Higgs boson.
\item
$\slashed{E}_T$: missing transverse energy.
\item
$scalar\,\,H_T$: scalar sum of transverse momenta of all objects reconstructed in the detector.
\item
$p_{T}^{V_1},\,\,p_{T}^{V_2}$: transverse momenta of the $V_1$ and $V_2$ objects. 
\item
$\eta^{V_1},\,\,\eta^{V_2}$: pseudorapidities of the $V_1$ and $V_2$ objects. 
\end{itemize}
The four-momenta of the $Z$ boson  $p_\mu^Z$ and the dark/SM-like boson $p_\mu^{s/h}$ are reconstructed using the relations $p_\mu^Z=p_\mu^{V_1}+p_\mu^{V_2}$ and $p_\mu^{s/h}=(\sqrt{s}-\sum\limits_{i}E^i,-\sum\limits_{i}\vec{p}^{\,i})$, where $p_\mu^X$, $E^X$ and $\vec{p}^{\,X}$ respectively denote the measured four-momentum, energy and three-momentum of the object $X$, and the summation index $i$ runs over the objects $V_1$ and $V_2$ and all reconstructed photons. {At a lepton collider, there is no ambiguity in the energies of the colliding particles compared with hadron colliders. The variables $M_{s/h\,\,candidate}$ and $\Omega_{Z,s/h}$, defined above, greatly benefit from this feature of lepton colliders.
The same set of variables is used for all the signal regions and parameter space points under study in this analysis. Fig. \ref{corrtable} shows several examples of the correlation matrix for the discriminating variables corresponding to different signal regions. The total missing energy $\slashed{E}$, is a well-motivated variable for searches performed at lepton colliders. In this analysis, however, we use the missing transverse energy $\slashed{E}_T$, instead, because $\slashed{E}$ is highly correlated with the variable $M_{s/h\,\,candidate}$.} 
Fig. \ref{dist} shows the distributions obtained for the discriminating variables corresponding to the signal region SR2 for an assumed set of model parameters as an example. The main contribution of the signal events to the signal region SR2 comes from events with the $Z$ boson decay into a di-muon. The objects $V_1$ and $V_2$ in Fig. \ref{signalCartoon} are, therefore, the reconstructed muons, $\mu_1$ and $\mu_2$, in this case. As seen in Fig. \ref{dist}, the signal distribution has two sharp peaks for several variables, i.e. $M_{s/h\,\,candidate}$, $\Omega_{\mu_1\mu_2}$ and $\slashed{E}_T$. These peaks correspond to the $h$-involved and $s$-involved signal processes. These processes have comparable cross sections for the $v_{\Phi}$ value assumed in Fig. \ref{dist}.  
\begin{figure*}[!ht]
  \centering  
    \begin{subfigure}[b]{0.485\textwidth} 
    \centering
    \includegraphics[width=\textwidth]{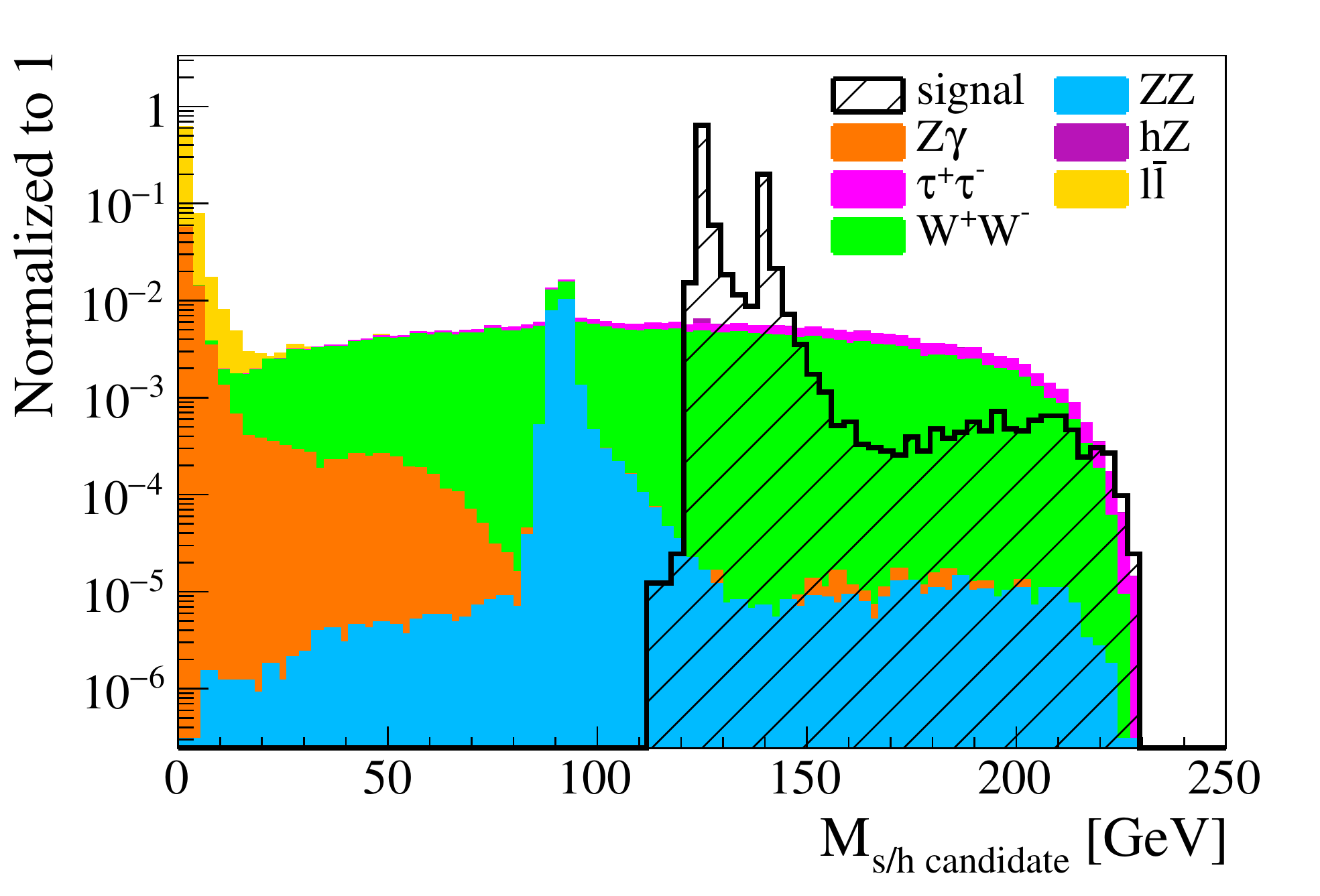}
    \caption{}
    \label{dist-SHmass}
    \end{subfigure} 
    \begin{subfigure}[b]{0.485\textwidth}
    \centering
    \includegraphics[width=\textwidth]{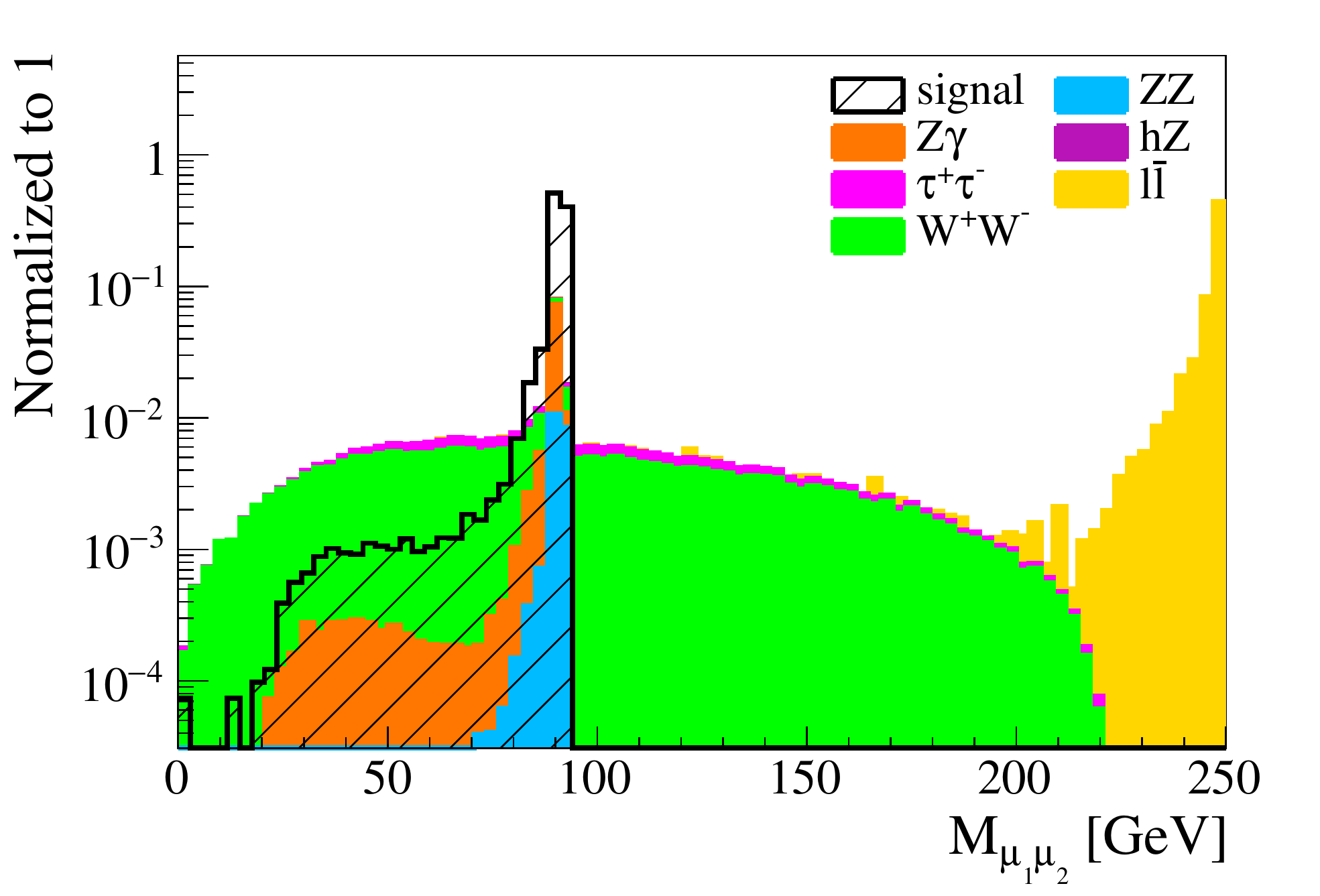}
    \caption{}
    \label{dist-Zmass}
    \end{subfigure} 
    \begin{subfigure}[b]{0.485\textwidth} 
    \centering
    \includegraphics[width=\textwidth]{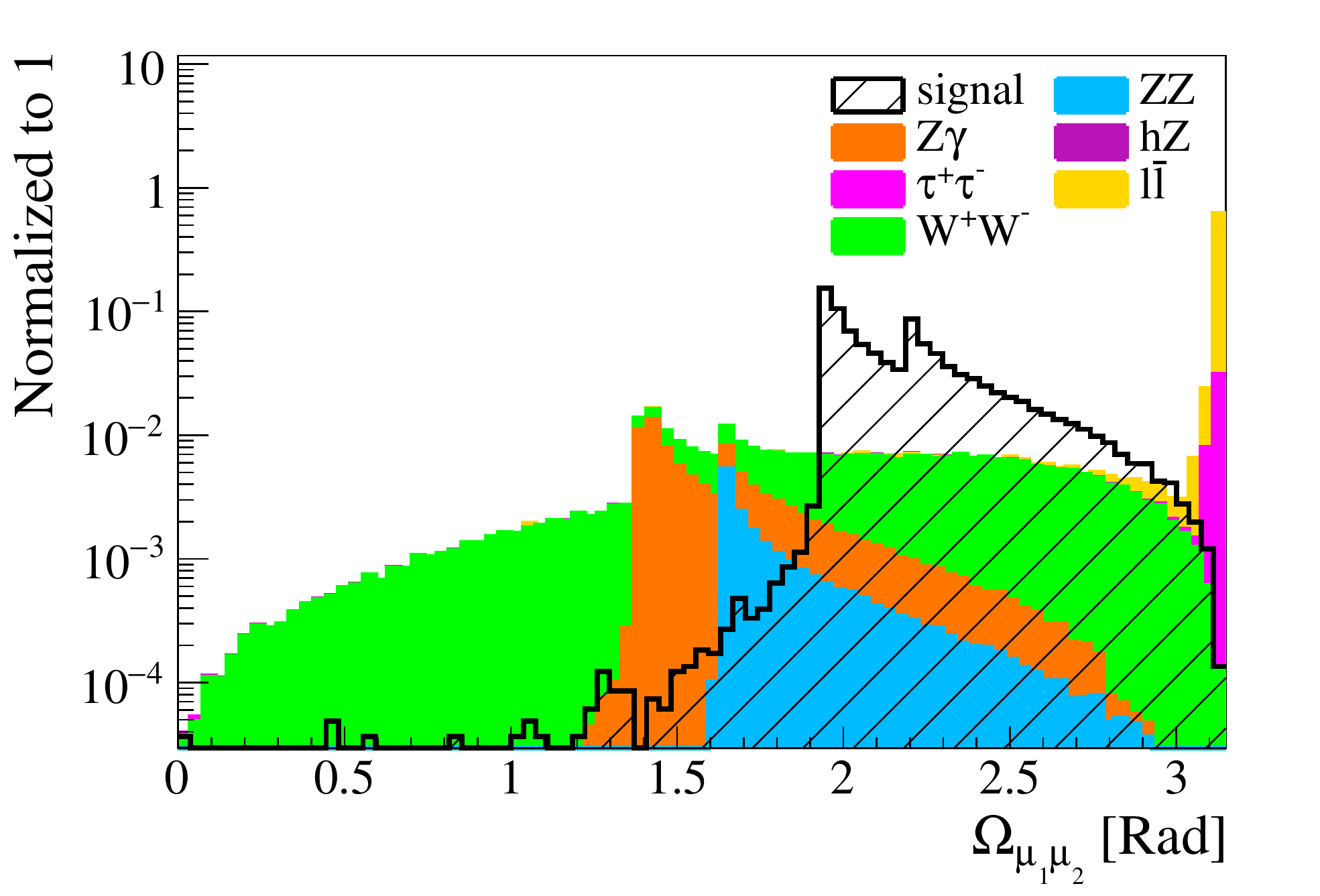}
    \caption{}
    \label{dist-ZobjectsAngle}
    \end{subfigure}
    \begin{subfigure}[b]{0.485\textwidth} 
    \centering
    \includegraphics[width=\textwidth]{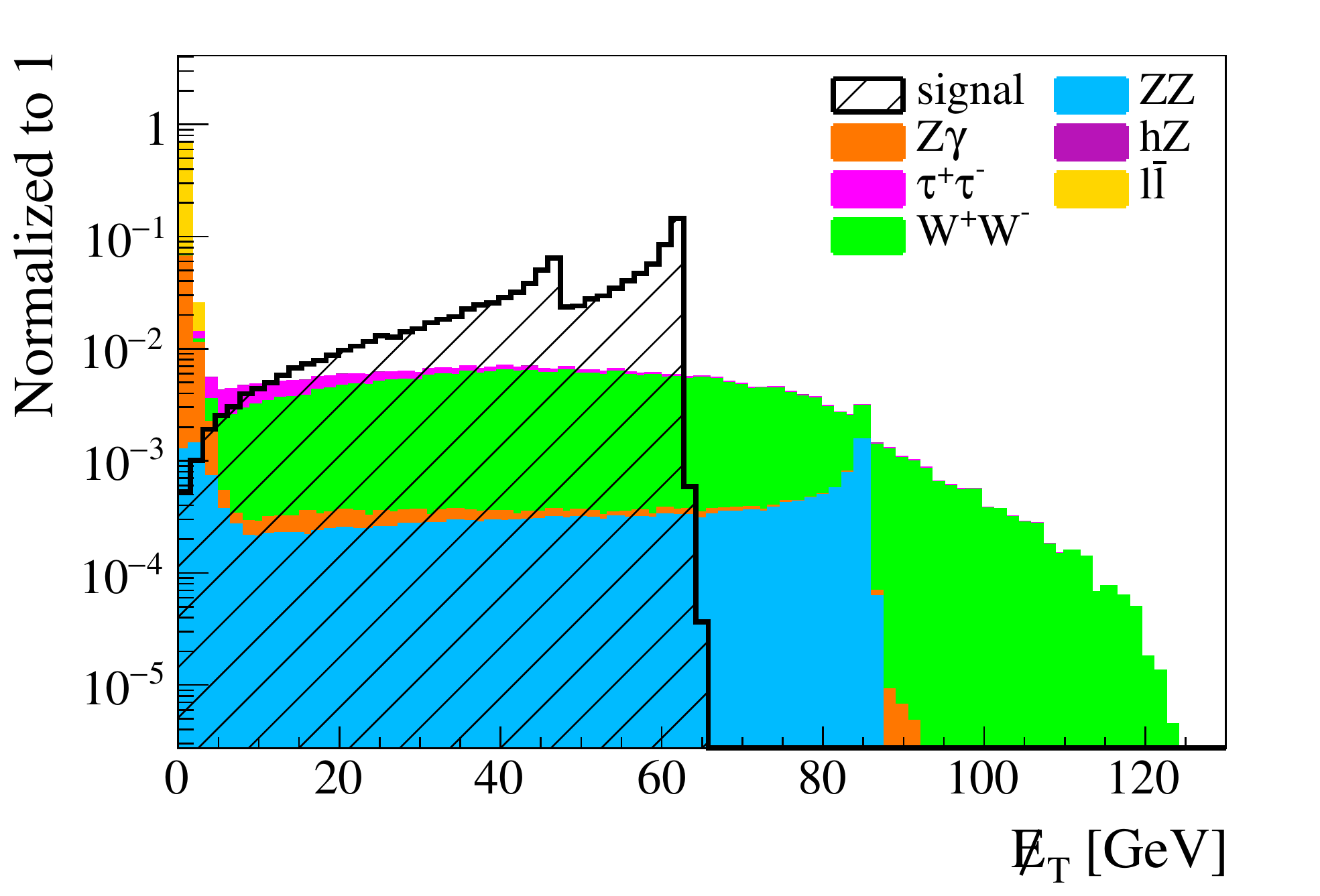}
    \caption{}
    \label{dist-met}
    \end{subfigure} 
    \begin{subfigure}[b]{0.485\textwidth}
    \centering
    \includegraphics[width=\textwidth]{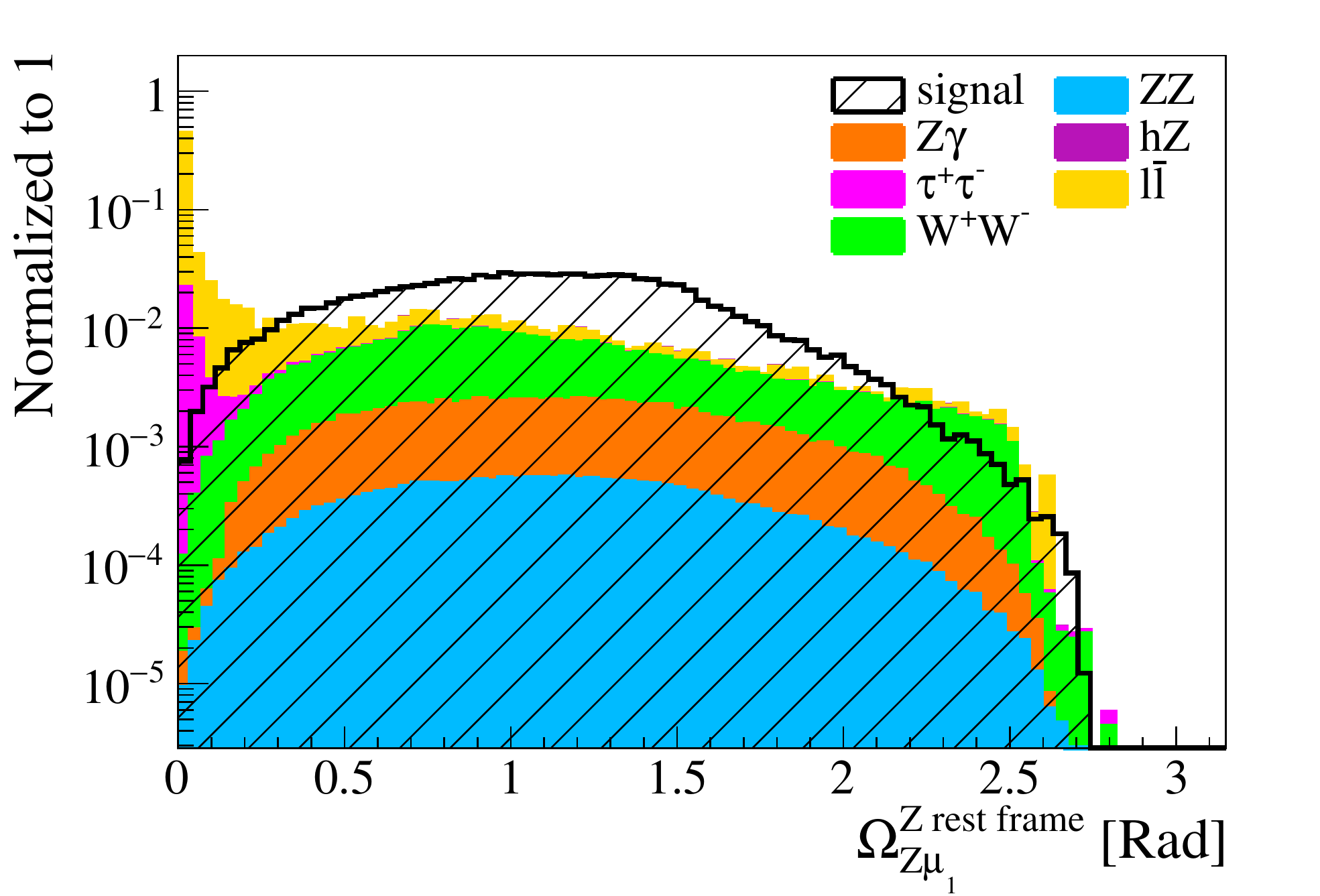}
    \caption{}
    \label{dist-DeviAngleRF1}
    \end{subfigure}
\begin{subfigure}[b]{0.485\textwidth}
    \centering
    \includegraphics[width=\textwidth]{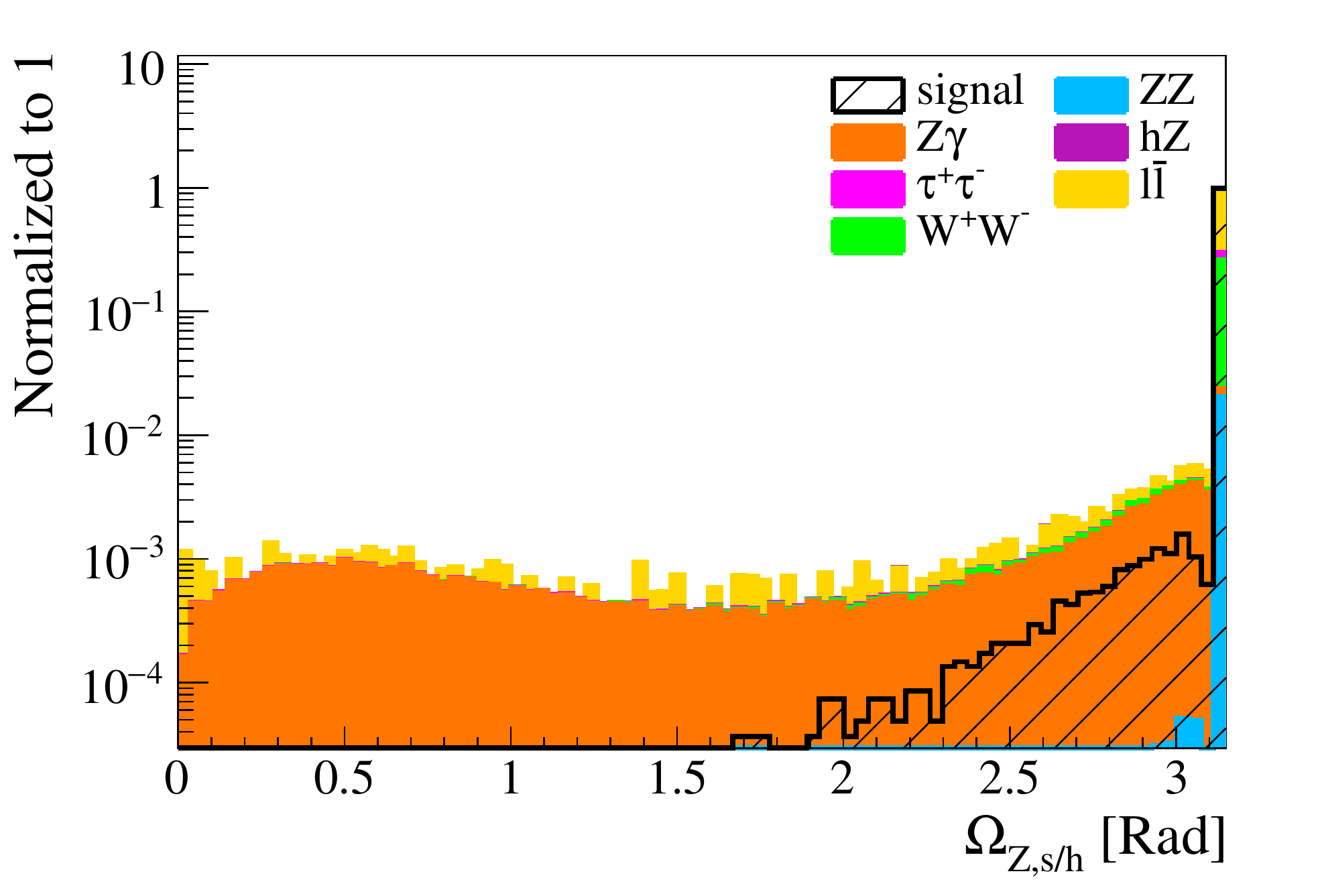}
    \caption{}
    \label{dist-ZSHangle}
    \end{subfigure}
\caption{Distributions obtained for the discriminating variables corresponding to the signal region SR2 and the model parameters $v_{\Phi}=1$ TeV, $m_s=140$ GeV, $\alpha=0.05$ Rad and $m_a = 0.5$ MeV. The distributions of the signal and the total background are separately normalized to unity.}
\label{dist}
\end{figure*}
\begin{figure*}[!h]\ContinuedFloat
    \centering
    \begin{subfigure}[b]{0.485\textwidth}
    \centering
    \includegraphics[width=\textwidth]{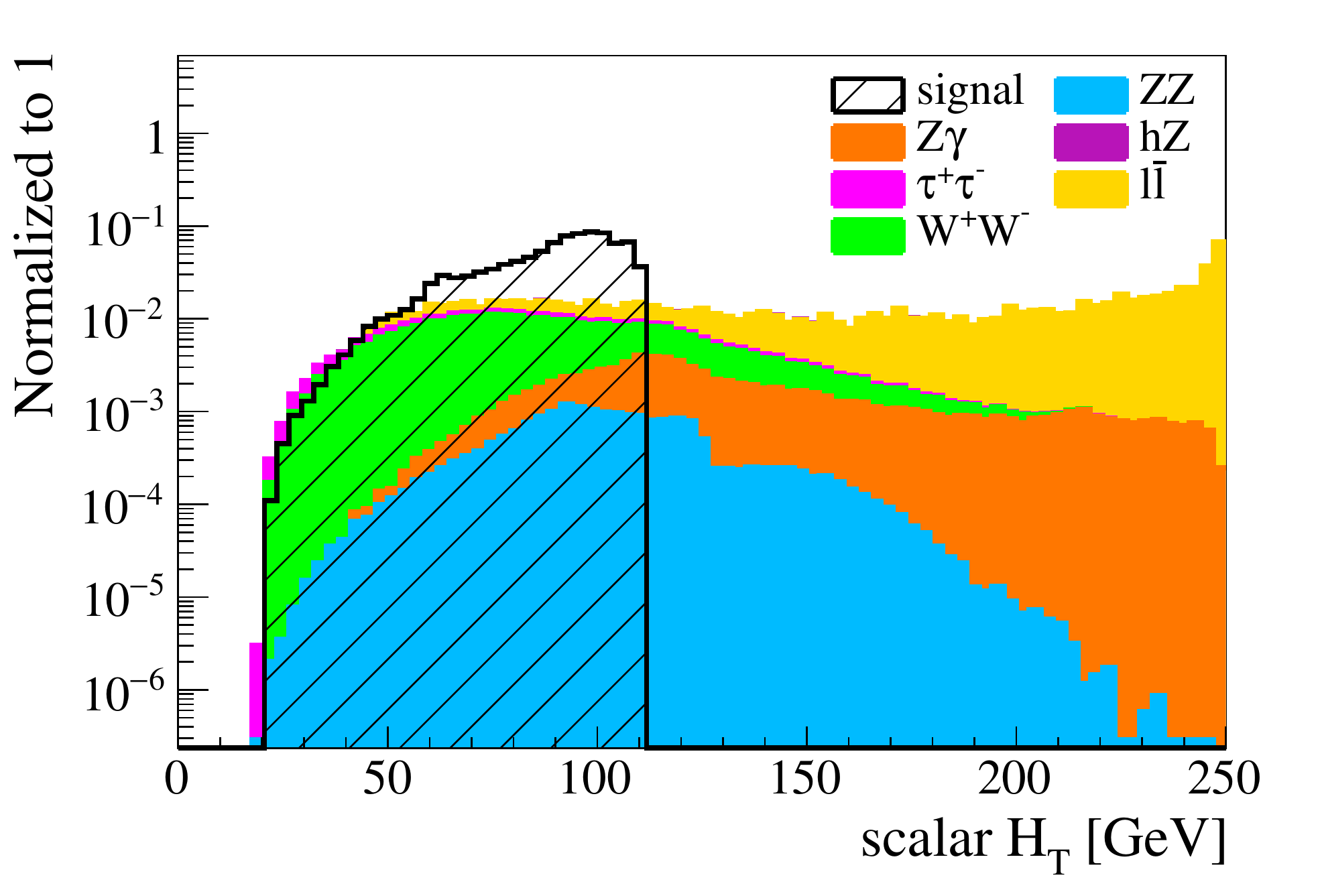}
    \caption{}
    \label{dist-scalarHT}
    \end{subfigure} 
    \begin{subfigure}[b]{0.485\textwidth}
    \centering
    \includegraphics[width=\textwidth]{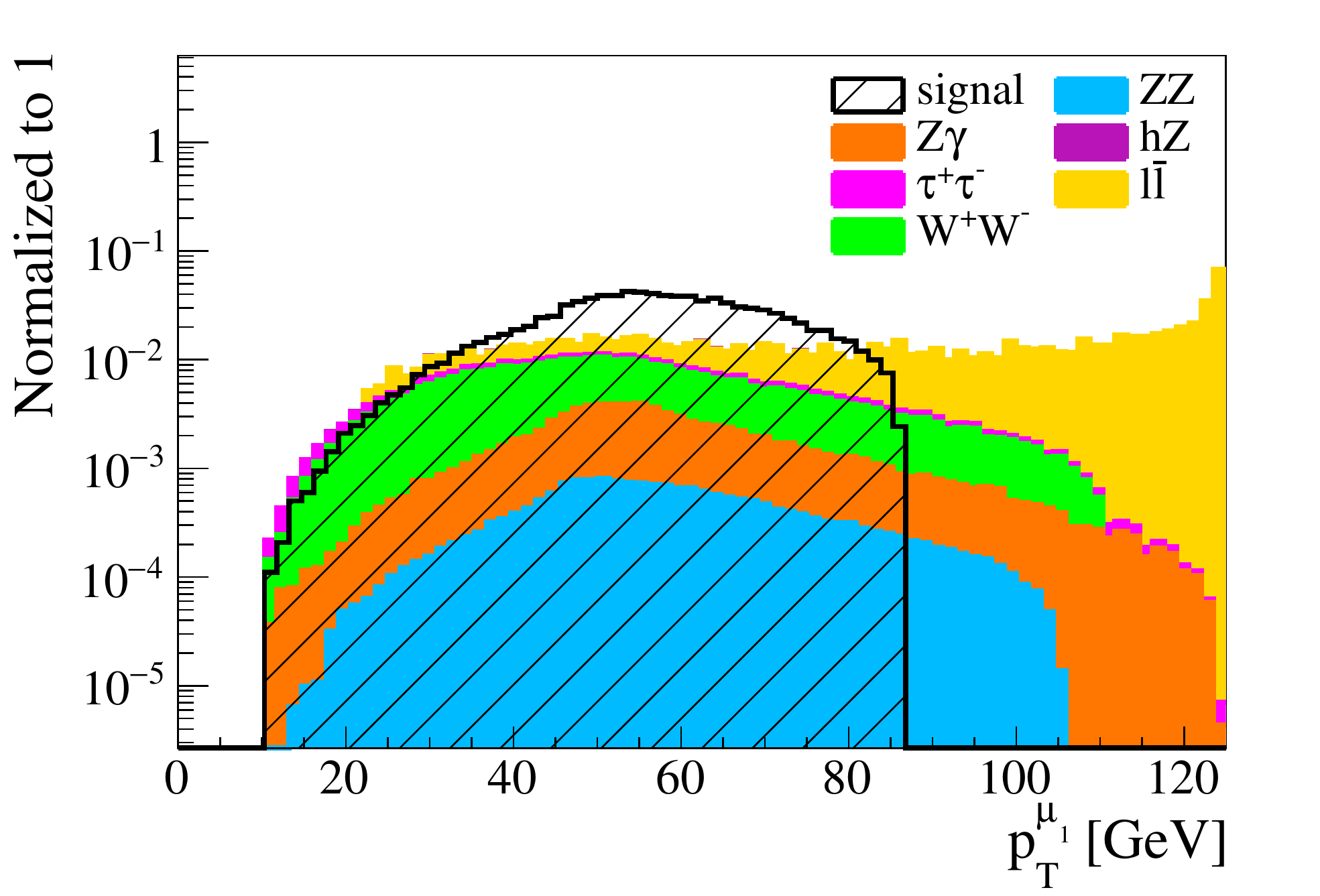}
    \caption{}
    \label{dist-ZobjectsPt1}
    \end{subfigure}
    \begin{subfigure}[b]{0.485\textwidth}
    \centering
    \includegraphics[width=\textwidth]{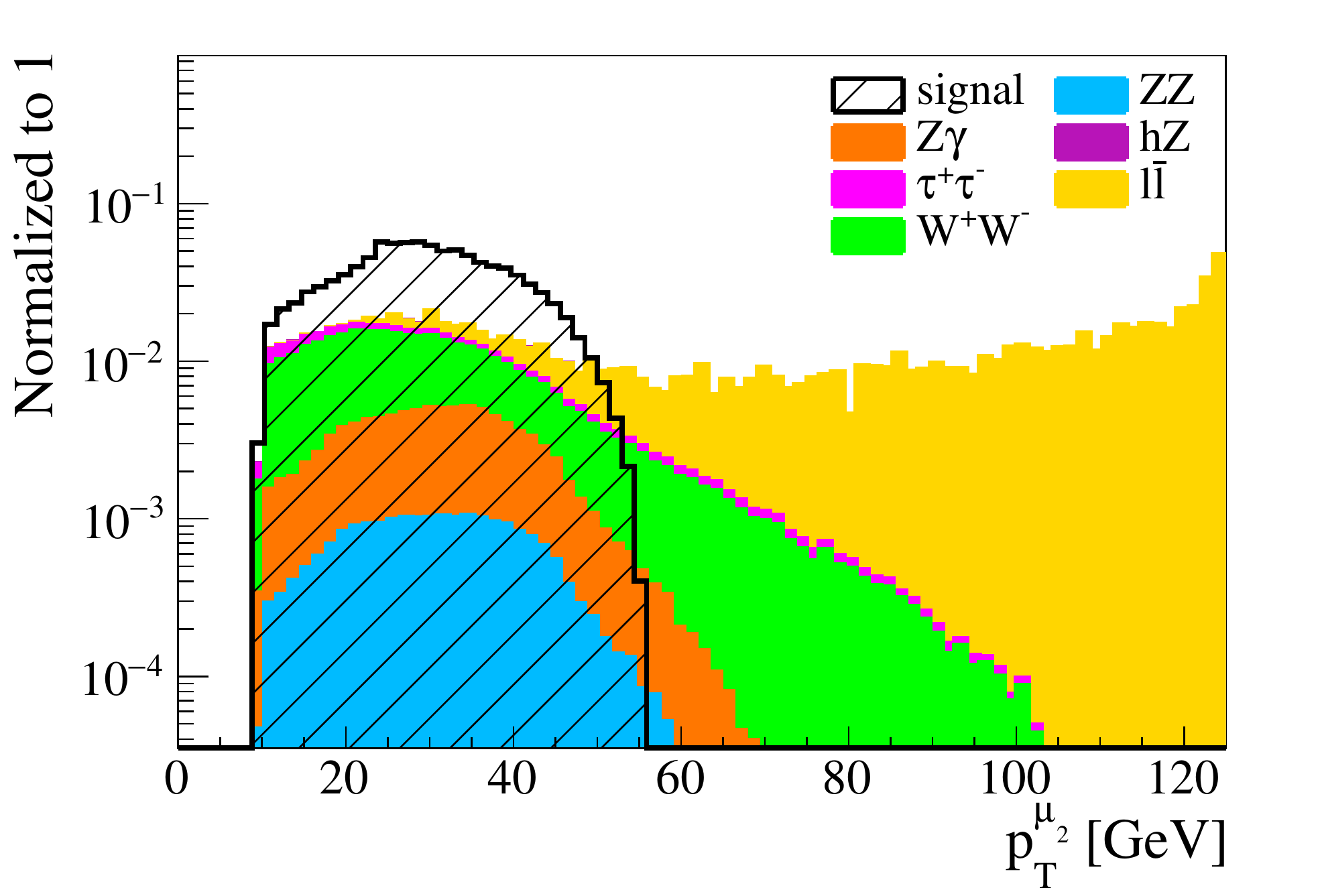}
    \caption{}
    \label{dist-ZobjectsPt2}
    \end{subfigure}
    \begin{subfigure}[b]{0.485\textwidth}
    \centering
    \includegraphics[width=\textwidth]{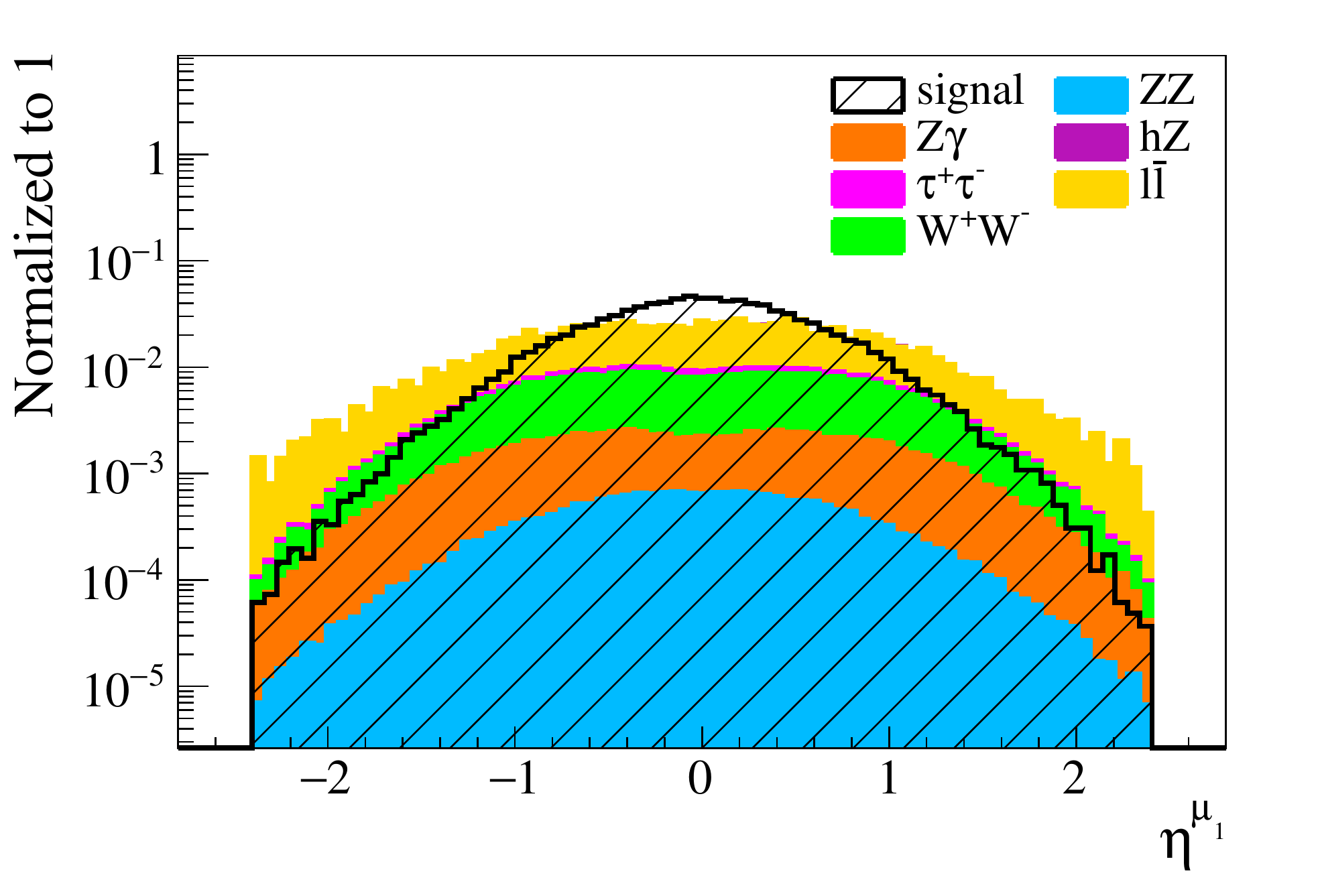}
    \caption{}
    \label{dist-ZobjectsEta1}
    \end{subfigure}
    \begin{subfigure}[b]{.485\textwidth}
   % \centering
%    \hspace{4.3cm}
    \includegraphics[width=\textwidth]{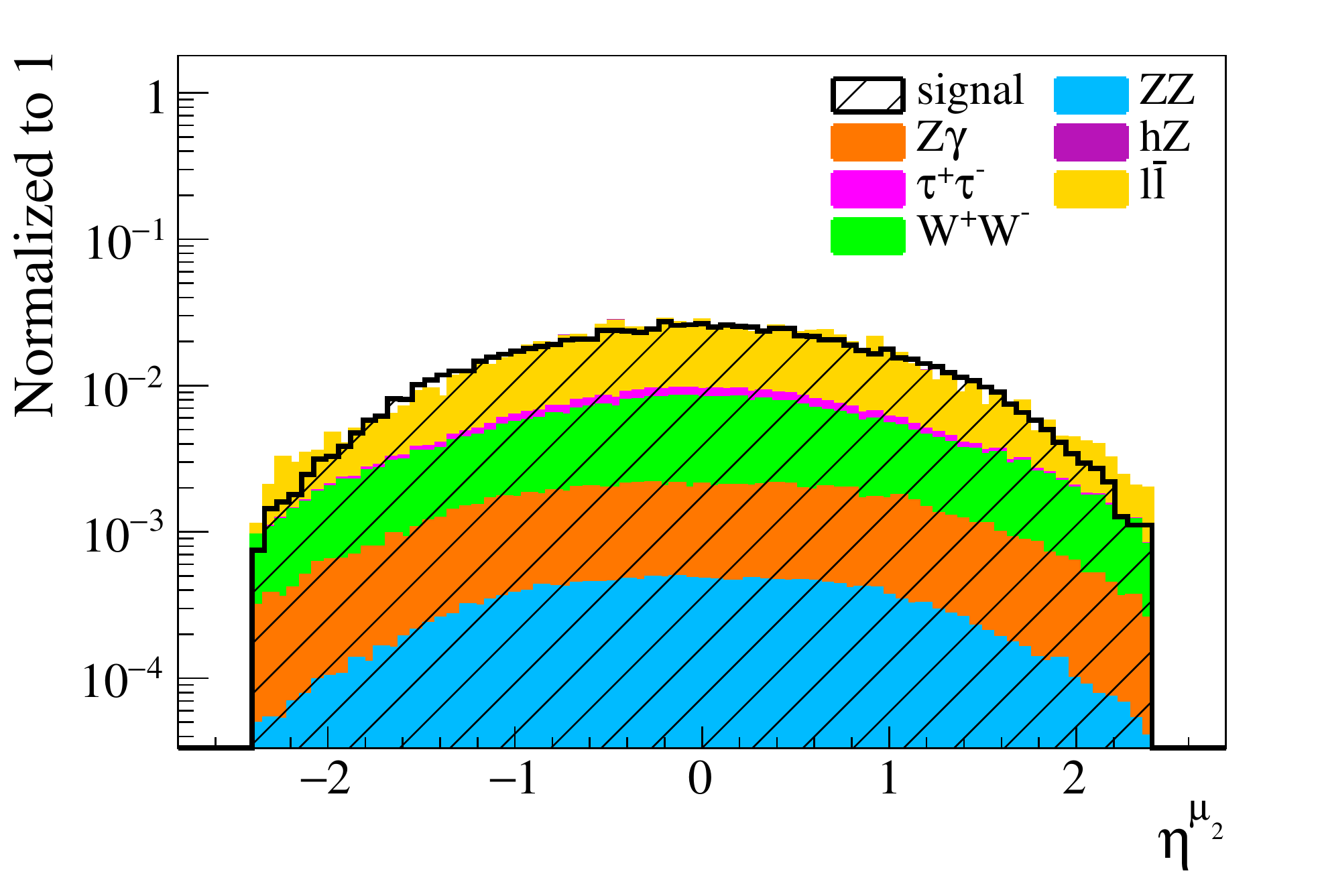}
    \caption{}
    \label{dist-ZobjectsEta2}
    \end{subfigure}
    \caption{Distributions obtained for the discriminating variables corresponding to the signal region SR2 and the model parameters $v_{\Phi}=1$ TeV, $m_s=140$ GeV, $\alpha=0.05$ Rad and $m_a = 0.5$ MeV. The distributions of the signal and the total background are separately normalized to unity.}
  \label{dist}
\end{figure*}

\begin{figure}[t]
    \centering
    \includegraphics[width=.60\textwidth]{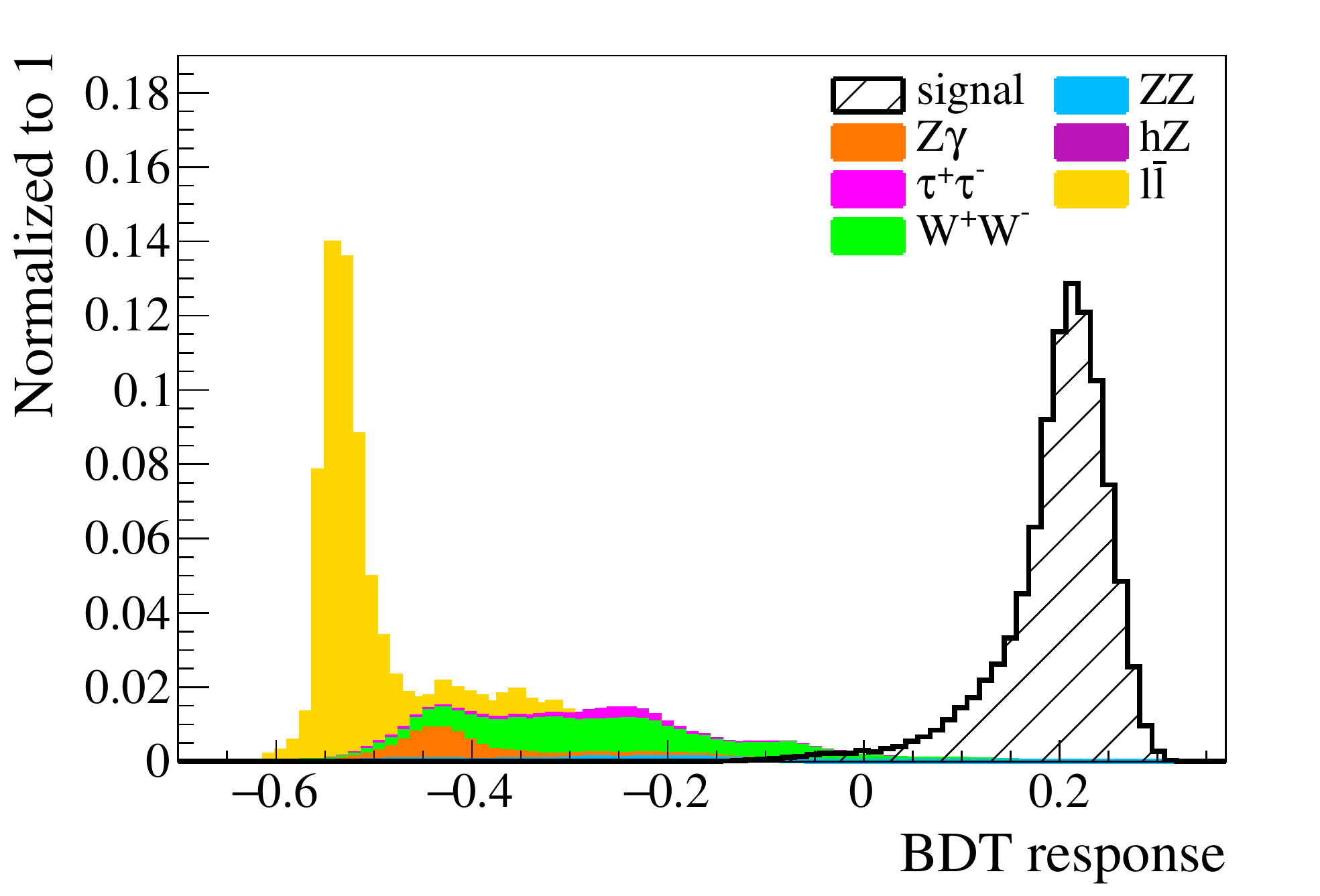}
    \caption{BDT response obtained for the signal and background processes using the distributions in Fig. \ref{dist} corresponding to the signal region SR2. The assumed model parameters are $v_{\Phi}=1$ TeV, $m_s=140$ GeV, $\alpha=0.05$ Rad and $m_a = 0.5$ MeV. The signal and the total background distributions are separately normalized to unity.}
    \label{dist-cls}	
\end{figure}

{To take full advantage of the defined variables and achieve the greatest possible sensitivity, a multivariate technique using the TMVA package \cite{Hocker:2007ht} is employed to discriminate the signal from the background rather than using a cut-based method.} All algorithms available in the TMVA package are examined in terms of the discrimination power using the receiver operating characteristic (ROC) curve. The distributions obtained for the discriminating variables are passed to the algorithms as input and training is performed considering all the background processes according to their respective weights. A comparison of the obtained ROC curves shows that the Boosted Decision Trees (BDT) algorithm has the best performance in signal-background discrimination. The BDT algorithm is, therefore, used in this analysis to separate the signal from the background. Fig. \ref{dist-cls} shows the BDT response when the distributions in Fig. \ref{dist} are passed to the BDT as input. As seen, the signal is well separated from the background. Among the background processes, the $ZZ$ and $W^+W^-$ production processes lead to the most severe, but well under control, backgrounds. To ensure that overtraining does not occur in the multivariate analysis, the TMVA overtraining check is performed. The BDT response for the test and training samples of the signal and background events are compared, and it is ensured that they are consistent.

According to the TMVA output, the discriminating variables $M_{s/h\,\,candidate}$, $\Omega_{\mu_1\mu_2}$ and $\slashed{E}_T$ are the most powerful discriminants. The discrimination power of the variables depends on the point of parameter space that is considered. Among the model parameters, $v_{\Phi}$ and the dark Higgs mass are the parameters with major effects on the signal-background discrimination. The contributions of the $s$-involved and $h$-involved processes to the signal change for different $v_{\Phi}$ values. Moreover, the kinematics of the signal $s$-involved events strongly depend on the dark Higgs mass. Consequently, the kinematics of the signal and thus the discrimination power of the variables can be strongly sensitive to $v_{\Phi}$ and the dark Higgs mass. {It can be inferred} from Figs. \ref{dist-SHmass}, and \ref{dist-ZobjectsAngle} that when the dark Higgs mass changes from 140 GeV to around the $Z$ boson mass, the signal peak corresponding to the $s$-involved processes coincides with the $ZZ$ background peak, giving rise to a degradation in the signal-background discrimination. The signal-background discrimination also depends on the signal region under consideration. In signal regions involving jets, the signal-background discrimination can be significantly degraded when compared with the pure leptonic signal regions. Technical difficulties and uncertainties in the reconstruction of jets, and also the invisible neutrinos produced as a by-product in the hadronization process, are the main reasons behind this degradation. 

It is worth mentioning that the distribution obtained for the Higgs candidate mass $M_{s/h\,\,candidate}$ in this analysis can be used to measure the mass of the dark Higgs boson. It is outside the scope of this paper, {but we can measure the mass of the dark Higgs boson by setting a proper fit function, e.g., a Gaussian function, to fit to the signal Higgs candidate mass distribution}. The dark Higgs mass can then be read from the fitted value of the ``mean'' parameter of the Gaussian function. 
The signal regions SR2 and SR3 are especially useful for this purpose because low uncertainties in these signal regions allow for accurate mass measurements. {If measured, it would be a smoking-gun signal of the light dark sector.}

\subsection{Prospects for constraints on the model parameters}
\label{limits}
\begin{figure}[!h]
  \centering  
    \begin{subfigure}[b]{0.555\textwidth} 
    \centering
    \includegraphics[width=\textwidth]{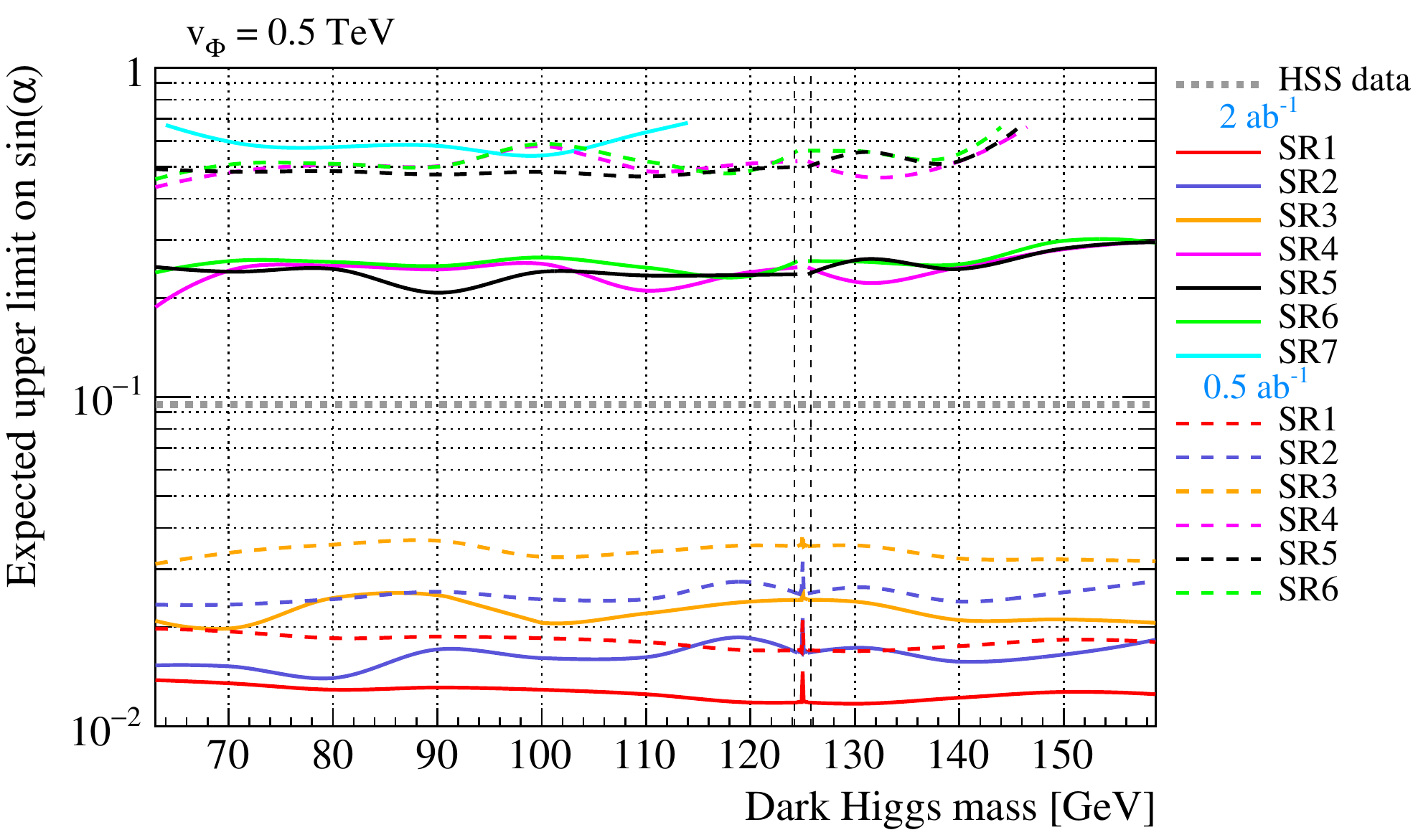}
    \caption{}
    \label{limit-vPhi500}
    \end{subfigure} 
    \begin{subfigure}[b]{0.555\textwidth}
    \centering
    \includegraphics[width=\textwidth]{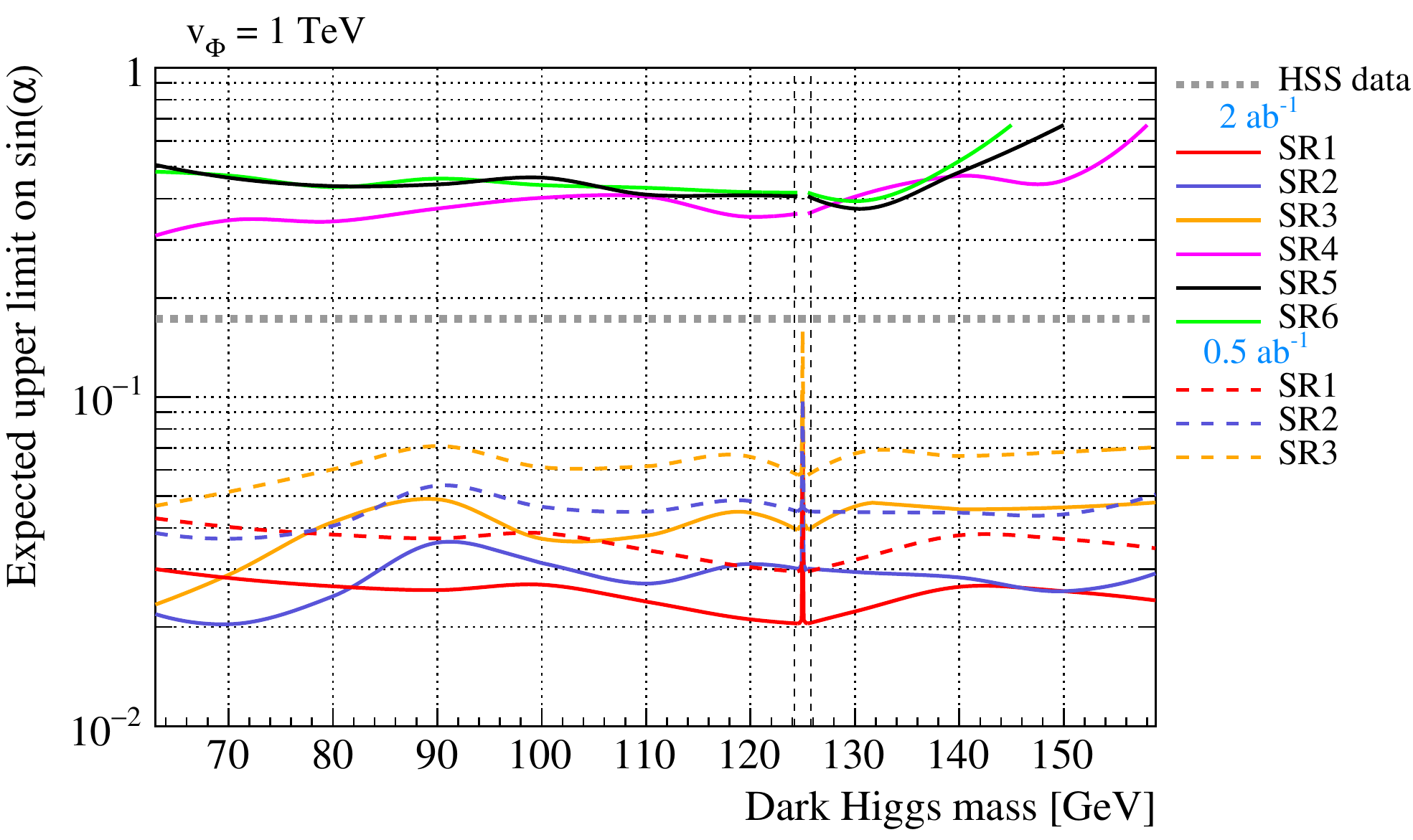}
    \caption{}
    \label{limit-vPhi1000}
    \end{subfigure}
    \begin{subfigure}[b]{0.555\textwidth}
    \centering
    \includegraphics[width=\textwidth]{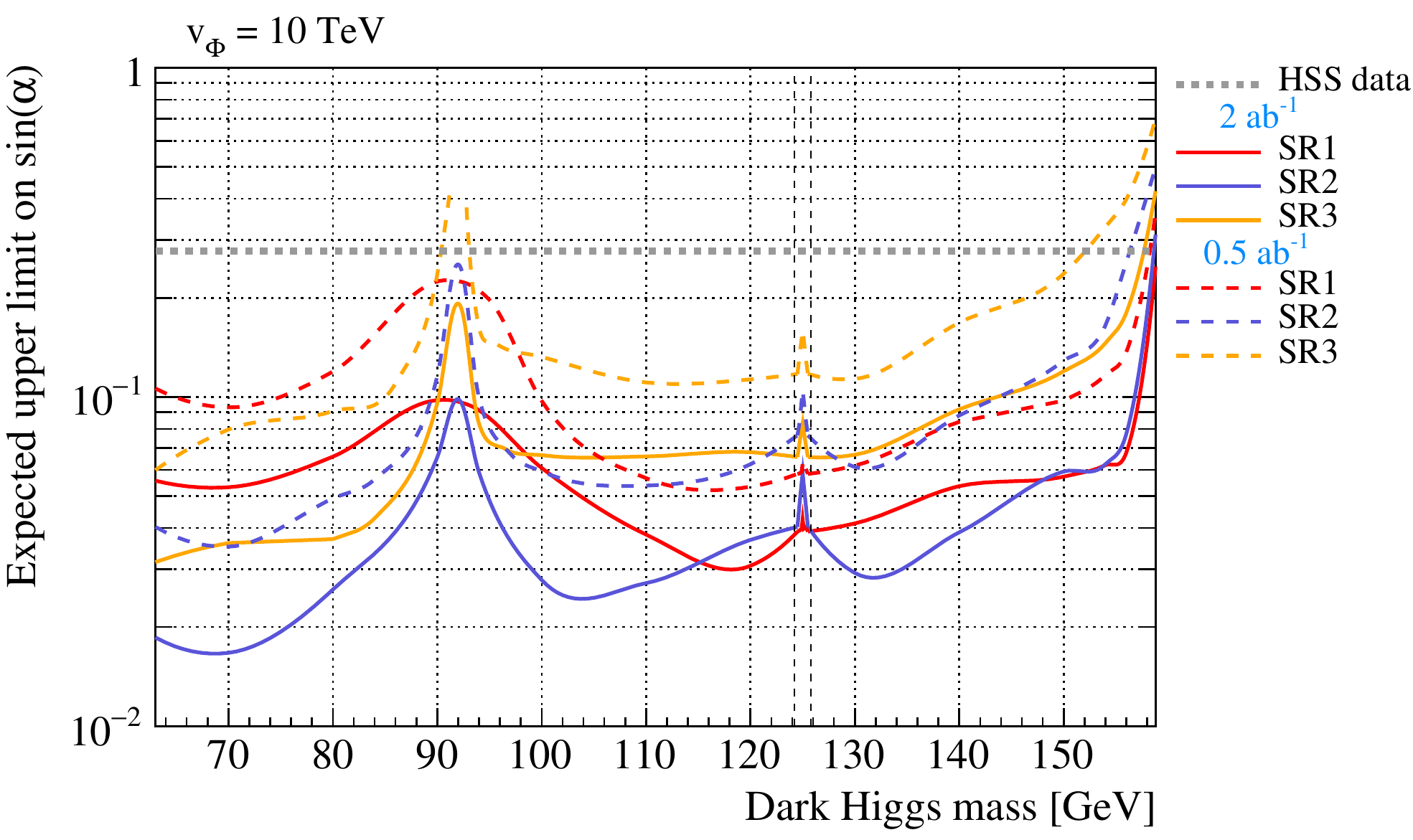}
    \caption{}
    \label{limit-vPhi10000}
    \end{subfigure}
    \caption{Expected $95\%$ CL upper limits on $\sin\alpha$ corresponding to different signal regions obtained for the $v_\Phi$ values a) 0.5 TeV, b) 1 TeV and c) 10 TeV and the integrated luminosities of 0.5 and 2 $\mathrm{ab}^{-1}$. {Upper limits at $95\%$ CL derived from Higgs signal strength data taken from Ref. \cite{Biekotter:2022ckj} are also shown.}}
  \label{limit}
\end{figure}

\begin{figure}[!h]
  \centering  
    \begin{subfigure}[b]{0.555\textwidth} 
    \centering
    \includegraphics[width=\textwidth]{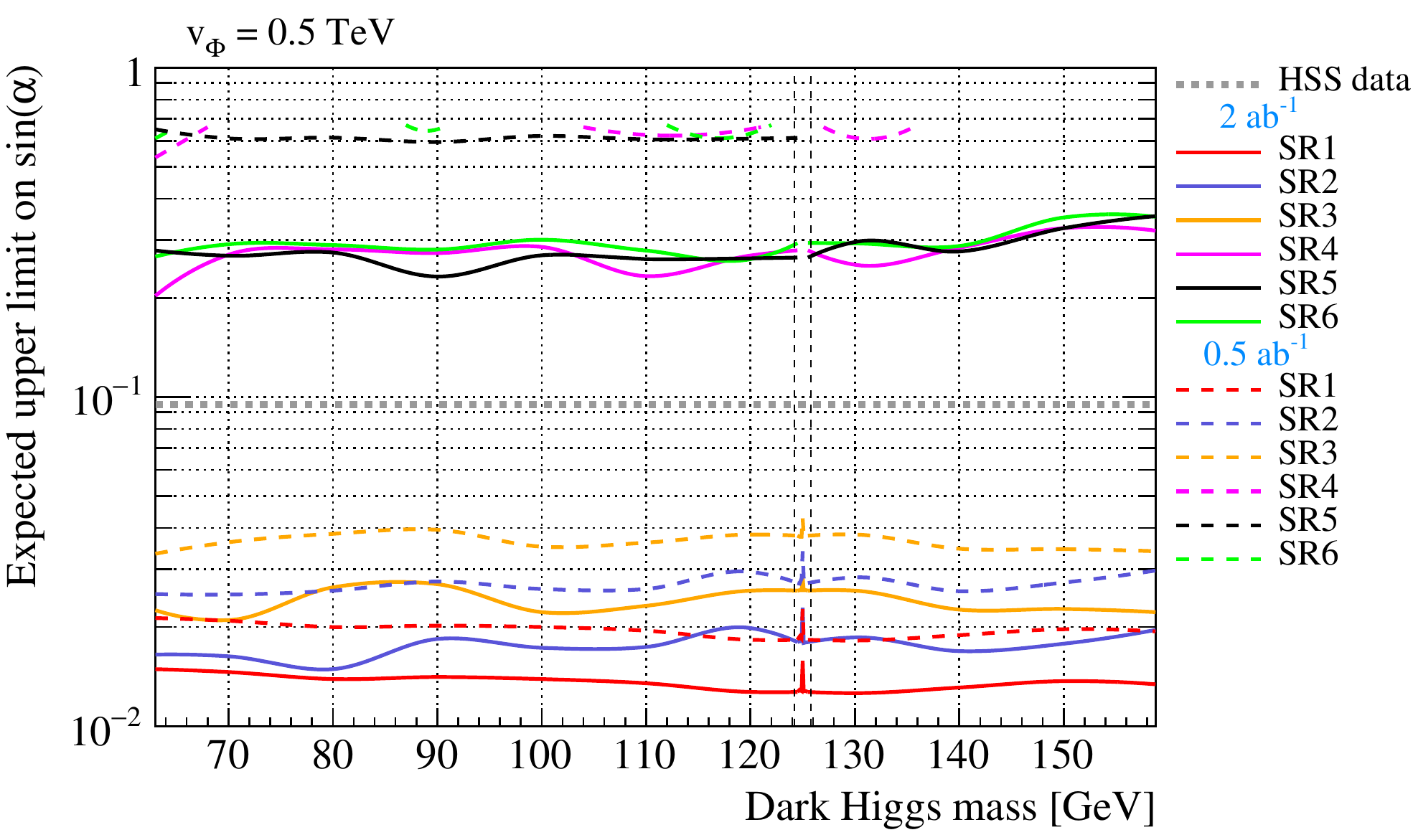}
    \caption{}
    \label{limitSys-vPhi500}
    \end{subfigure} 
    \begin{subfigure}[b]{0.555\textwidth}
    \centering
    \includegraphics[width=\textwidth]{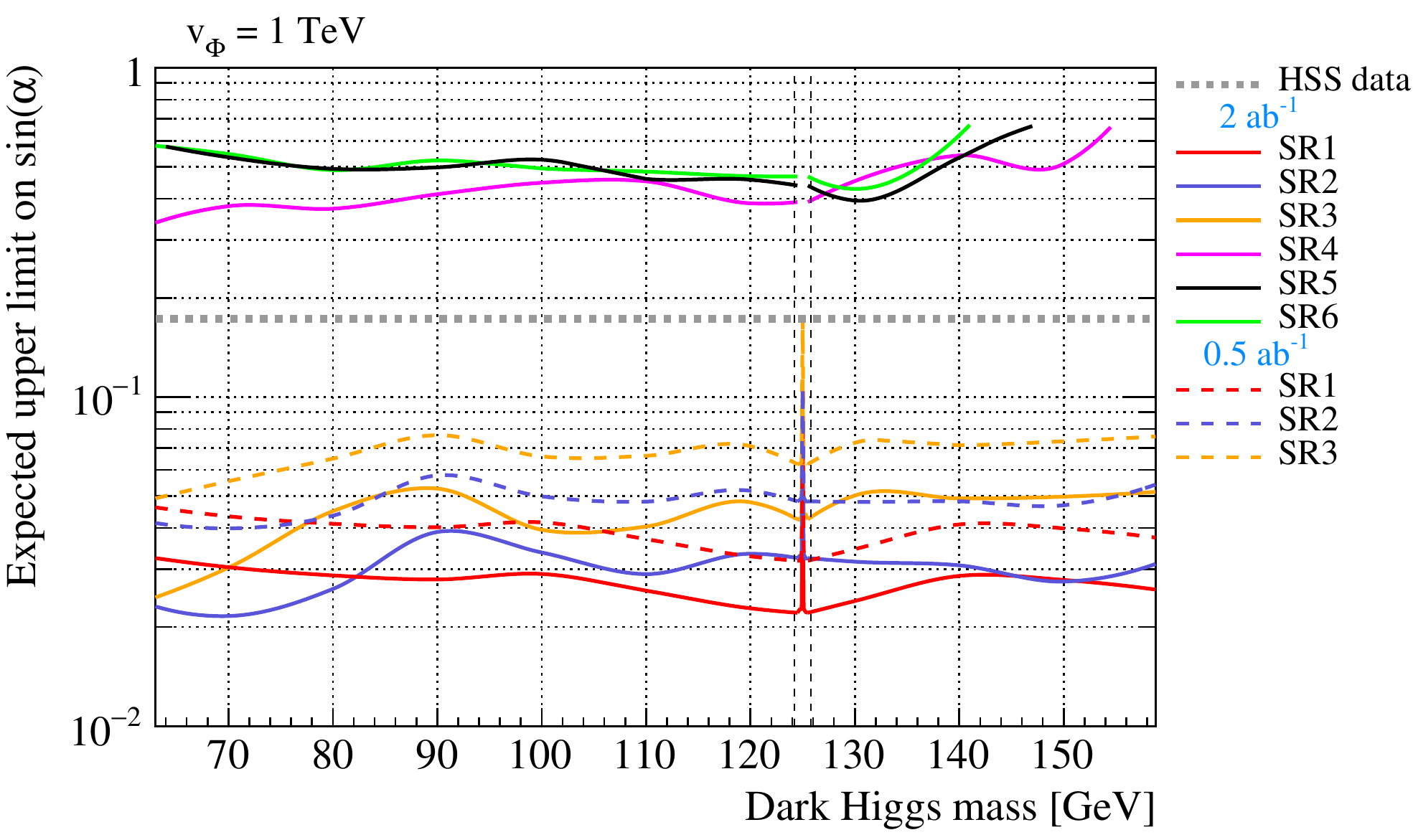}
    \caption{}
    \label{limitSys-vPhi1000}
    \end{subfigure}
    \begin{subfigure}[b]{0.555\textwidth}
    \centering
    \includegraphics[width=\textwidth]{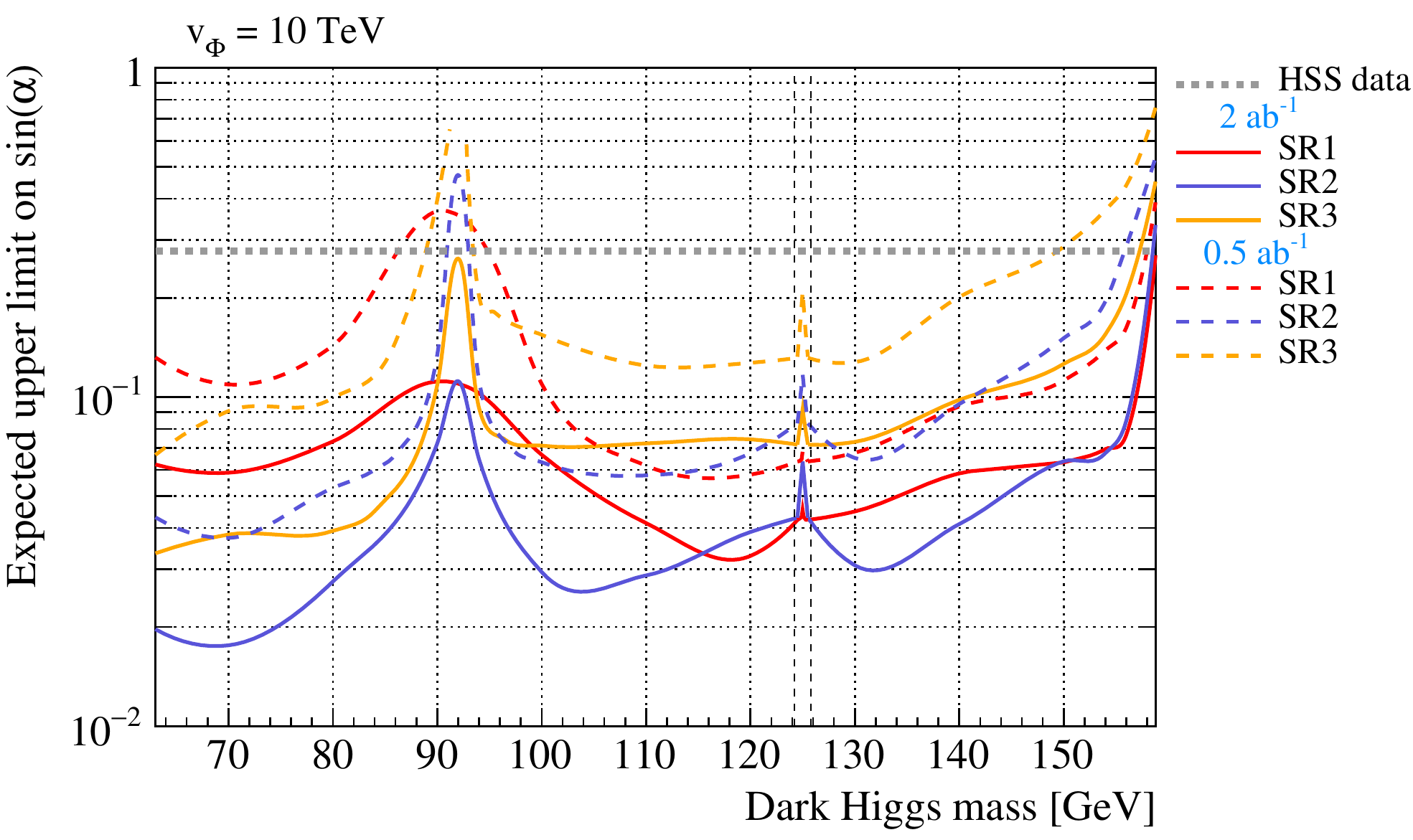}
    \caption{}
    \label{limitSys-vPhi10000}
    \end{subfigure}
    \caption{Expected $95\%$ CL upper limits on $\sin\alpha$ assuming an overall uncertainty of $10\%$ on event selection efficiencies corresponding to different signal regions obtained for the $v_\Phi$ values a) 0.5 TeV, b) 1 TeV and c) 10 TeV and the integrated luminosities of 0.5 and 2 $\mathrm{ab}^{-1}$. {Upper limits at $95\%$ CL derived from Higgs signal strength data taken from Ref. \cite{Biekotter:2022ckj} are also shown.}}
  \label{limitSys}
\end{figure}

Using the BDT response, expected $95\%$ CL upper limits on the parameter $\alpha$ are computed for all the assumed signal regions for the dark Higgs mass range $63\mhyphen\mhyphen158.8$ GeV and are presented in the $\sin\alpha\mhyphen m_s$ plane. Limits are computed using the significance formula $N_s/\sqrt{N_s+N_b}$, where $N_s$ and $N_b$ are, respectively, the number of signal and total background events in a window determined by optimizing the signal significance. The integrated luminosities assumed for computing the limits are 0.5 and 2 $\mathrm{ab}^{-1}$, which are approved by the Linear Collider Board {\cite{Bambade:2019fyw,Barklow:2015tja}. The total target integrated luminosity of 2 $\mathrm{ab}^{-1}$ at $\sqrt{s}=250$ GeV is foreseen in the ILC running scenario H20 which has been the reference scenario for ILC physics projections since 2015.}
Fig. \ref{limit} shows the limits obtained for different signal regions corresponding to the $v_{\Phi}$ values $0.5,1$ and $10$ TeV. This analysis can be affected by uncertainties arising from jet reconstruction, tau-tagging, measurement of the energy and momentum of particles, etc. To take into account potential systematic uncertainties, we consider an overall uncertainty of $10\%$ on event selection efficiencies obtained for signal and background processes. Fig. \ref{limitSys} shows the limits including the uncertainty of $10\%$ corresponding to different signal regions and $v_{\Phi}$ values. We note that in some figures, some reaches disappear {because the theoretical cross section for a given setup of $v_\F$ and $m_s$ is maximized with $\a\sim \pi/4$ (see also Sec.~\ref{prediction} for the reason why we do not show $\alpha>\pi/4$).} {The indirect $95\%$ CL upper bounds derived from the Higgs signal strength measurements \cite{Biekotter:2022ckj,ATLAS:2019nkf, CMS:2020gsy} have also been shown in Figs. \ref{limit},\ref{limitSys}. These bounds have been taken from Ref. \cite{Biekotter:2022ckj} and are obtained by performing $\chi^2$-fits to the Higgs signal strength data.}

As seen in Figs. \ref{limit} and \ref{limitSys}, in almost the whole of the considered parameter space at $v_{\Phi}$ values 0.5 and 1 TeV, the strongest limits are obtained for the signal region SR1. This signal region corresponds to the di-jet final state and has the highest signal selection efficiency among the examined signal regions. {In particular, the $h$-involved processes dominate in most of the $m_s$ range, which is the reason why the sensitivity only mildly depends on $m_s$.}
At $v_\Phi=10$ TeV, the signal region SR2, which corresponds to the di-muon final state, gives the most stringent limits for most dark Higgs masses. Although event selection efficiency corresponding to {the signal region SR2} is much smaller than that of the signal region SR1{,  the} much better signal-background discrimination {at the SR2} at high $v_\Phi$ values fully compensates for the small signal selection efficiency, resulting in stronger limits.\footnote{One notes that the reach of the SR2 signal region in $v_\F=10\TEV$ is slightly better than 
that of $v_\F=1\TEV$ at some mass ranges. We expect this is due to the cleaner signal distribution with a single peak
with the larger $v_\F$. }
 As $v_\Phi$ increases, the contribution of the $h$-involved processes to the signal decreases, and, as a result, the signal-background discrimination in the signal region SR1 gets degraded. This degradation mainly results from the uncertainties originating from the presence of jets in this signal region. The improvement (degradation) in the signal-background discrimination in the signal region SR2 (SR1) is 
significant{, especially} for dark Higgs masses lower than $m_h$.

Among the examined signal regions, the weakest one is SR7, which requires exactly one electron and one muon with opposite charges in the final state. In most regions of the parameter space, no sensitivity is obtained for this signal region. Only at the integrated luminosity of 2 $\mathrm{ab}^{-1}$~\cite{Bambade:2019fyw,Barklow:2015tja} and $v_\Phi=0.5$ TeV and when no systematic uncertainty is considered, very loose limits are obtained for this signal region for dark Higgs masses below $\sim115$ GeV as seen in Fig. \ref{limit-vPhi500}. The main contribution of the signal to this signal region comes from events with the $Z$ decay into a tau pair when both of the tau leptons decay leptonically. The smallness of the branching fraction of this decay mode and neutrinos produced as by-products in the decay of the tau leptons degrade the sensitivity in this signal region. {Similarly, at} higher $v_\Phi$ values, the {sensitivities of SR4-SR6 also get degraded rapidly} due to the reduction in the signal cross section {and due to the background dominated behavior of significance, {$N_s/\sqrt{N_s+N_b}$}. As seen, {at $v_\Phi=10\TEV$,} no sensitivity is obtained for the signal regions SR4 to SR7. It is also seen that the limits at higher $v_\Phi$ values are more sensitive to the dark Higgs mass. The reason behind this observation is {again} that the purity of the signal $s$-involved events increases as $v_\Phi$ grows, and as a result, the kinematics of the signal becomes more sensitive to the dark Higgs mass. 

As the dark Higgs mass increases, the decrease in the signal cross section degrades the sensitivity. This degrading effect can, however, be partially compensated by the improvement in the signal-background discrimination at higher dark Higgs masses. As the dark Higgs mass increases, the discrimination power of some variables like $M_{s/h\,\,candidate}$ and $\Omega_{V_1V_2}$ can be improved, leading to a better signal-background discrimination at higher dark Higgs masses (see Figs. \ref{dist-SHmass},\ref{dist-ZobjectsAngle}). The effect of this compensation is more obvious for the signal regions SR1, SR2, and SR3 at $v_\Phi$ values 0.5 and 1 TeV since the $h$-involved processes significantly contribute to the signal at these $v_\Phi$ values and the signal cross section does not drop too low at large dark Higgs masses. 

As mentioned before, for dark Higgs masses around the $Z$ boson mass, the signal peak corresponding to the $s$-involved processes coincides with the peak of the $ZZ$ background for some variables, giving rise to the degradation of the signal-background discrimination. The obtained limits for dark Higgs masses in this region, therefore, show a tendency to be weaker. This effect is more obvious for signal regions in which the $Z$ boson is reconstructed more accurately, namely SR1, SR2, and SR3. It is also seen that the sensitivity degradation around the $Z$ boson mass is more significant at higher $v_\Phi$ values (because of the higher purity of the s-involved events).

The signal cross section experiences a rapid decrease at {$|m_s- m_h|\lesssim \Gamma (s\to aa)$} due to the destructive interference term (see Fig. \ref{xsec-Ms-All}). The limits are consequently degraded at dark Higgs masses in the vicinity of $m_h$. It is seen that for some signal regions, the sensitivity is even totally lost in this mass region, e.g., for the signal regions SR4, SR5 and SR6 at $v_\Phi=0.5$ and 1 TeV. At $v_\Phi=1$ TeV, the contributions of the $h$-involved and $s$-involved processes to the signal are comparable in contrast to the cases of $v_\Phi=0.5$ and 10 TeV. A larger interference is, therefore, induced at $v_\Phi=1$ TeV, giving rise to a more serious sensitivity degradation at $m_s\approx m_h$ at this $v_\Phi$ value. {For this region, the dark Higgs boson decouples due to a quantum effect, and, by neglecting the existence of the dark Higgs boson with a certain redefinition of the mixing parameter, the invisible (SM-like) Higgs boson decay can be studied as in the usual case~\cite{Sakurai:2022cki}({cf.} \cite{Asner:2013psa}). {The signal strength bound is alleviated as well.}
 Thus we do not perform the analysis in detail in this paper.}

The obtained limits indicate that a large region of the considered parameter space can be probed with the present analysis. At $v_\Phi$ values of 0.5 and 1 TeV, the sensitivity reach doesn't change significantly for different dark Higgs masses, and the dark sector can be probed for $\alpha$ values above $\sim0.01$ and $\sim0.02$, respectively. At $v_\Phi=10$ TeV, the best-obtained limit ranges from $\sim0.02$, for dark Higgs masses near $m_h/2$, to a few tenths, for dark Higgs masses near the threshold mass. {Comparing the obtained limits with the limits from the Higgs signal strength measurements, it is seen that the limit on the $\alpha$ parameter can be improved by about one order of magnitude with the present results.}
{For a wide range of $v_\F$ values,} the invisible decay is dominated by the $s$-involved processes whose cross-section does not depend much on $v_\F$~(Fig.\,\ref{xsec-vPhi-All}) and thus the reach does not change much with $v_\Phi=\O(1-10)\TEV$. However, $s$ will also dominantly decay into SM particles via the mixing if $v_\F\gg 10\TEV$ and the invisible signal efficiency gets suppressed. This can be seen in the {$m_s\gtrsim 150\GEV$ where the reach of $\a$ increases significantly in the $v_\F=10\TEV$ case {in Figs.~\ref{limit} and} ~\ref{limitSys} ({cf.} Fig.~\ref{sWidth-Ms-All}). This is due to not only the phase space suppression but also the enhanced $s$  decay widths to $W$ and $Z$ bosons, suppressing the branching ratio for the $s$ invisible decay. 
%This observation shows that there is only a very narrow parameter region for $v_\F\approx 10\TEV \OR v_\F \ll 1\TEV$ that the Higgs boson invisible decay can be discussed separately. 
} Probing this very weakly coupled region at the ILC can be made by searching for the visible particles {\cite{Robens:2015gla,Robens:2016xkb}}{\footnote{{Apart from the searches for SM-like and dark Higgs boson invisible decays, the ILC precisely measures the Higgs boson couplings, e.g., 0.38\% for the $hZZ$ coupling \cite{Fujii:2017vwa}. Thus, indirect searches of $s$ via the deviations of the SM-like Higgs boson couplings also have the potential to probe the region of $m_{s}\gtrsim 150{\rm GeV}$ (see, e.g.,~\cite{Kanemura:2015fra,Kanemura:2016lkz,Altenkamp:2018bcs,Kanemura:2018yai,Kanemura:2019kjg}). See also the signal strength bounds of the model \cite{Biekotter:2022ckj,ATLAS:2019nkf,CMS:2020gsy}.}}}.

\begin{figure}[!h]
  \centering  
    \begin{subfigure}[b]{0.521\textwidth} 
    \centering
    \includegraphics[width=\textwidth]{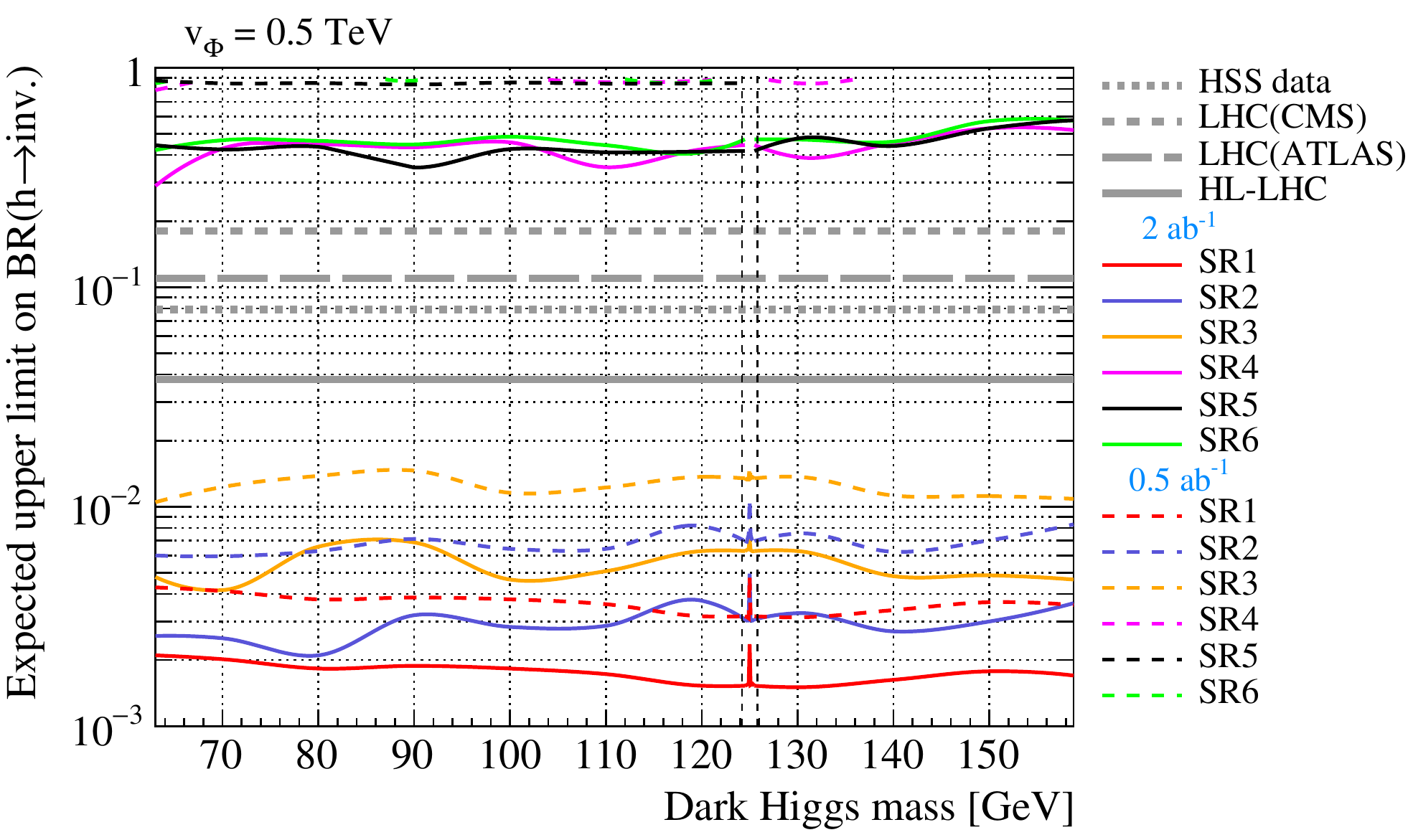}
    \caption{}
    \label{limitSysBRh-vPhi500}
    \end{subfigure} 
    \begin{subfigure}[b]{0.521\textwidth}
    \centering
    \includegraphics[width=\textwidth]{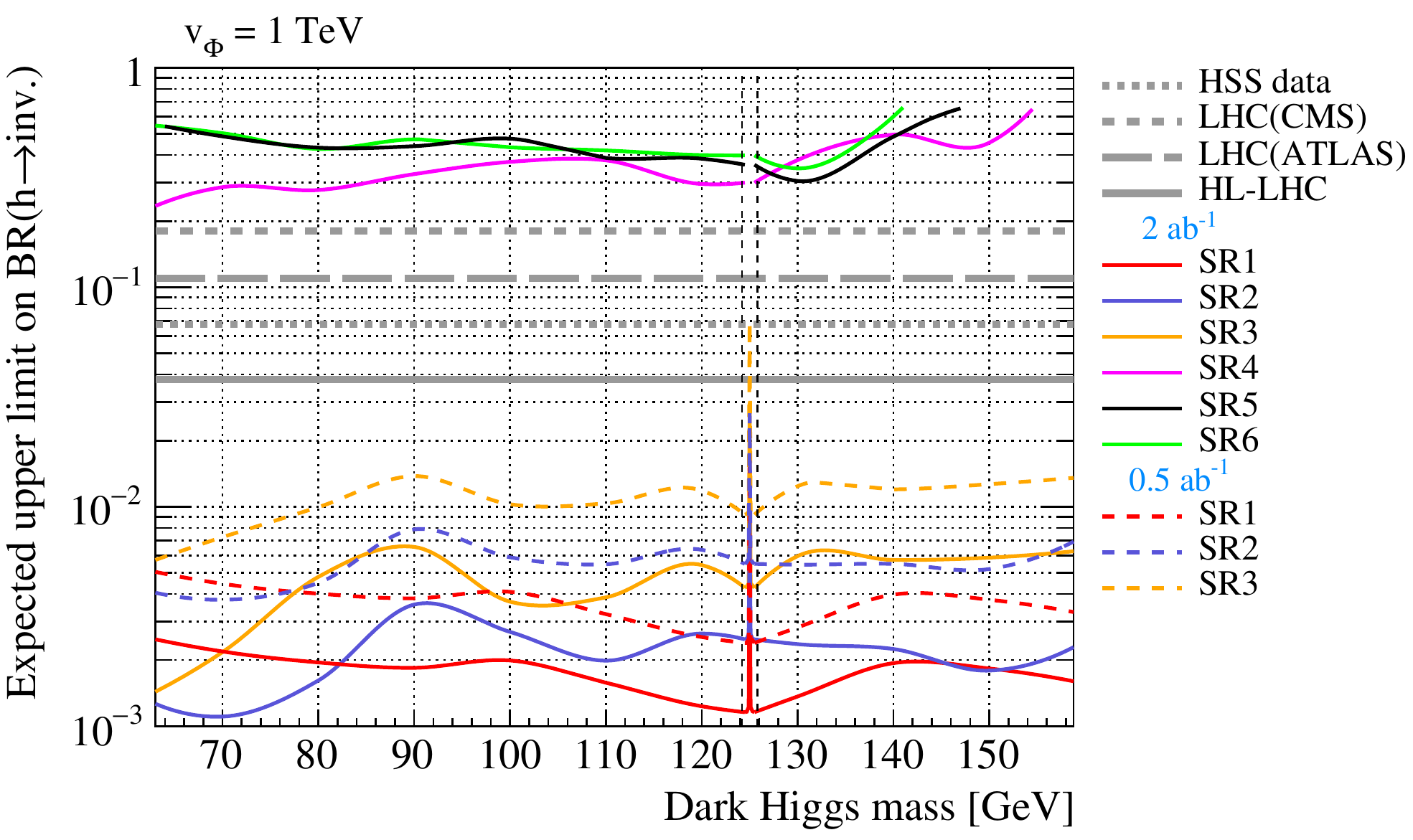}
    \caption{}
    \label{limitSysBRh-vPhi1000}
    \end{subfigure}
    \begin{subfigure}[b]{0.521\textwidth}
    \centering
    \includegraphics[width=\textwidth]{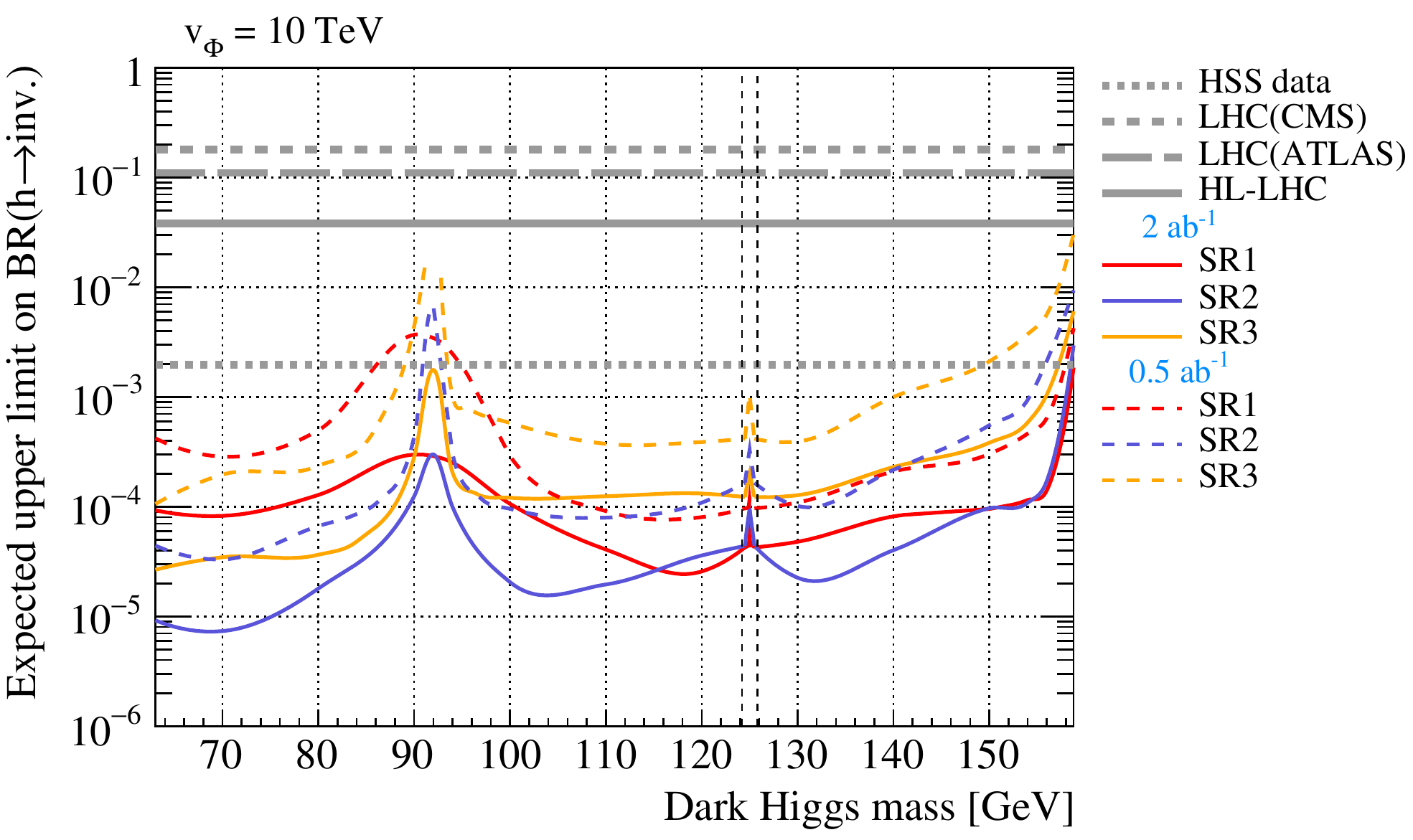}
    \caption{}
    \label{limitSysBRh-vPhi10000}
    \end{subfigure}
    \caption{{Expected $95\%$ CL upper limits on BR($h\rightarrow \mathrm{invisible}$) corresponding to different signal regions obtained for the $v_\Phi$ values a) 0.5 TeV, b) 1 TeV and c) 10 TeV and the integrated luminosities of 0.5 and 2 $\mathrm{ab}^{-1}$. 
{BR($h\rightarrow \mathrm{invisible}$) is translated as the function of $\sin{\alpha}$, $m_s$ and $v_\Phi$ (See Fig.\ref{limitSys} c.f. Fig.\ref{hWidth-Alpha-All}).} The limits derived from the Higgs signal strength data taken from \cite{Biekotter:2022ckj}, the HL-LHC projected limit \cite{Cepeda:2019klc}, and the current limits derived by the ATLAS \cite{ATLAS:2020kdi} and CMS \cite{CMS:2022qva} collaborations at $95\%$ CL are also shown.}}
  \label{limitSysBRh}
\end{figure}
{The obtained limits shown in Fig. \ref{limitSys} can be translated to the limits on the invisible decay of the SM-like Higgs boson. Fig. \ref{limitSysBRh} shows the translated upper limits on the invisible branching fraction of the SM-like Higgs boson, BR($h\rightarrow \mathrm{invisible}$), at $95\%$ CL for different $v_\Phi$ values and integrated luminosities. To make a comparison, indirect limits derived from the Higgs signal strength data \cite{Biekotter:2022ckj,ATLAS:2019nkf, CMS:2020gsy}, current limits obtained by searching for invisible decays of the SM-like Higgs boson derived by the ATLAS \cite{ATLAS:2020kdi} and CMS \cite{CMS:2022qva} collaborations, and the projected limit from the HL-LHC \cite{Cepeda:2019klc} on the invisible branching fraction of the SM-like Higgs boson at $95\%$ CL are also shown. As seen in this figure, the best limits obtained in this study are, at least, about two orders (one order) of magnitude stronger than the current (projected) limits from the LHC (HL-LHC). The difference between the obtained limits and the LHC limits increases for higher $v_\Phi$ values and reaches to about three orders of magnitude at $v_\Phi=10$ TeV. As mentioned before, at large $v_\Phi$ values, the $h$-involved processes get suppressed and thus the search for the invisible decay of the SM-like Higgs boson loses its sensitivity quickly as $v_\Phi$ grows. In this analysis, however, by searching for the invisible decays of both the SM-like and dark Higgs bosons, a large unprobed parameter region becomes accessible. A comparison also shows that the limits derived from the Higgs signal strength measurements can be improved by about one to two orders of magnitude with the present results. As emphasized in \cite{Biekotter:2022ckj}, and seen in Fig. \ref{limitSysBRh}, the limits from the Higgs signal strength data can be stronger than the direct limits derived from direct searches for invisible decays of the SM-like Higgs boson at the LHC. For a discussion on the reasons see \cite{Biekotter:2022ckj}.}

It is worth mentioning that the limits obtained in this analysis can be further improved by using the polarized electron-positron beams. ILC plans to provide polarized beams with a polarization degree of about $80\%$ ($30\%$) for electron (positron) beams \cite{Asner:2013psa}. The Higgs production is enhanced with polarized beams, providing the possibility of a deeper probe of the dark sector. 

\section{Summary and conclusions}
In a large class of models for ALPs and dark photons, light dark sector couples to the Higgs doublet field {via the doublet-singlet Higgs portal coupling, leading to the mixing of the SM-like boson $h$ with the dark Higgs boson $s$ since both the electroweak symmetry and dark symmetry are spontaneously broken. These models predict light ALPs and dark photons, which can be viable candidates for DM. The light DM scenarios exhibit similar properties, and thus a search for the light dark sector can serve as a tool for a generic probe of all these scenarios. The decay of the SM-like boson into DMs, which is a universal prediction of these models, can be used to probe the dark sector at collider experiments. This channel, which has been widely studied, can probe the dark sector to a certain extent that is allowed by the sensitivity reach of the collider. 
{Since the existence of the dark Higgs boson is also a prediction of those models, and thus it may be also a further probe of the dark sector, 
} it is very beneficial to take account of both the SM and dark Higgs bosons in the search for invisible decays.

{In this work,} we presented the first search for the invisible decays of both the SM and dark Higgs bosons. These Higgs bosons are produced via similar Feynman diagrams at the ILC in the process $e^-e^+ \rightarrow h/s  + Z$, and {$h/s$ decays into light generic dark particles}. 
The analysis is performed for different signal regions, which cover all decay modes of the $Z$ boson. ILC is assumed to operate at the center-of-mass energy of 250 GeV 
with integrated luminosities of 0.5 and 2 $\mathrm{ab}^{-1}$. Signal and background events are generated considering a realistic simulation of the ILD detector effects. Using a number of discriminating variables and with the help of a multivariate technique, the signal is well discriminated from the background. 
The dark Higgs masses larger than half of the SM-like boson mass and below the threshold of the on-shell dark Higgs production are considered in the analysis.
The obtained limits indicate that, at the $v_\Phi$ values 0.5 and 1 TeV, the dark sector can be probed for $\alpha$ values down to around $0.01\mhyphen\mhyphen0.03$. At $v_\Phi=10$ TeV, the obtained limits substantially change for different dark Higgs masses and range from $\sim0.02$, for the lowest dark Higgs masses, to a few tenths, for dark Higgs masses near the threshold mass, {since the signal events are dominated by the dark Higgs boson decay.} {A comparison shows that the present study can improve the current (projected) limits on the invisible branching fraction of the SM-like Higgs boson from the LHC (HL-LHC) by at least two orders (one order) of magnitude. The limits derived from the Higgs signal strength measurements can also be improved by about one to two orders of magnitude. The power of the search channel in this study and the importance of including the dark Higgs boson in the search are more obvious at high $v_\Phi$ values. At $v_\Phi=10$ TeV, for example, we obtained limits of $\O(10^{-5})$ on the invisible branching fraction of the SM-like Higgs boson, which is about three (two) orders of magnitude below the sensitivity reach of the HL-LHC (ILC) for the SM-like Higgs invisible decay.} In the presented analysis, the SM-like and dark Higgs bosons can be simultaneously reconstructed, and their masses can be measured. {For instance, we can measure the mass of the dark Higgs boson by setting a proper fit function, e.g., a Gaussian function, to fit to the signal Higgs candidate mass distribution.} Observation of the dark Higgs boson with the possibility of mass measurement offers a golden opportunity to probe the dark sector. 

{Again,} the presented study may serve as a tool to explore a broad class of models predicting light ALPs and dark photons. {The light dark particles may even behave as the dark radiation produced via the portal interaction alleviating the Hubble tension (see appendix.~\ref{app:2}). The analysis can be easily applied to other future $e^+e^-$ colliders such as C$^3$, CEPC, CLIC, and FCC-ee. } 

\section*{Acknowledgments}
This work was supported by JSPS KAKENHI Grant Nos.  20H01894 (K.S.), 20H05851 (W.Y.), 21K20363 (K.S.), 21K20364 (W.Y.), 22K14029 (W.Y.), and 22H01215 (W.Y.).

\appendix 
\section{Theoretical and cosmological aspects. }
So far, we have performed a collider study of the renormalizable model defined by the potential~\eqref{V}. We find that a certain parameter region with $\lambda_P,\l \sim 0.0001-1$, and $\m_H\sim m_\F \sim 100\GEV$ can be probed via the invisible decays.
Here let us comment on some theoretical and cosmological motivations of the parameter region and argue the powerfulness of our analysis. 

\subsection{Naturalness}
\label{app:1}
The parameter region that we have discussed is favored from the so-called 't Hooft naturalness~\cite{tHooft:1979rat} with a cutoff scale not too larger than the weak scale. 
Since  $\l_P$ in Eq.\,\eqref{V} is not forbidden by any symmetry, $\l_P$ may not be much smaller than 1 according to naturalness. So do $\lambda$ and $\l_H$. 
Then $\m_H^2$ ($m_\F^2$)  would acquire a collection of $-\l_P v_\f^2$ ($-\l_P v^2$). 
 Avoiding the fine-tuning of $m_\F \sim v_\F$ or $ \m_H\sim v$, we obtain the natural region $m_\F^2\sim \m_H^2$.

 We also mention that there are several loopholes in the discussion. Even if the model is not natural in the context of the 't Hooft argument, the model may become natural by assuming certain UV completion or further assumptions. 
  For instance, we may assume that there is supersymmetry above a typical scale of $\L$. Since the renormalizable coupling between $\F$ and $H$ super multiplets is forbidden due to the $\SU(2)\times \U(1)\times G_{\rm dark}$ symmetry, the resulting portal coupling in the low energy effective theory is suppressed. 
On the other hand, the dark sector may be weakly coupled to have a small portal coupling~(e.g. \cite{Nakayama:2021avl}). 

We may say that the parameter region that can be probed in our proposal belongs to the region favored by naturalness without further assumptions.

\subsection{Cosmology} \label{app:2}
In this setup, the DM (or the dark radiation relaxing the $H_0$-tension) can be produced successfully, thanks to the portal coupling. 
Whether the DM is a massive ALP\footnote{Strictly speaking, ALP may regard an axion coupled to a photon. In this paper, we denote ALP as generic axions. } or hidden photon depends on the nature of $G_{\rm dark}$ and the quality (if it is a global symmetry). 
Before discussing the $G_{\rm dark}$ specific phenomena, let us provide a $G_{\rm dark}$ independent discussion by assuming that $a$ is either a NGB or a would-be NGB. 

\paragraph{A generic discussion}
Since we are focusing on the region where $\l_P$ is not too small, the (would-be) NGB can be thermally produced via the portal coupling in the early universe. To discuss the interaction around the decoupling of NGB, we can integrate out $h$ and $s$. 
By noting the effective dark Higgs-NGB coupling
${\cal L}_{\rm int} \sim \frac{s}{\sqrt{2}v_\F} (\partial a)^2$,
 we obtain  
 the dimension-7 term 
\beq
\laq{HDO}
{\d {\cal L}} =   -\frac{\sqrt{2}{\lambda_P m_\p}}{(m_h^2-m_s^2) m_h^2} \partial a\partial a  \bar{\psi}\psi\equiv  -\frac{\sqrt{2}{m_\p}}{\L_H^2 m_h^2} \partial a\partial a  \bar{\psi}\psi,
\eeq
with $\p$ being a SM fermion.  
Until this interaction becomes irrelevant, (would-be) NGB, $a$, is produced in the early universe. Given that $\lambda_P$ is not too small and $m_h\sim m_s$, the interaction rate via this term, which scales as $\frac{m^2_\p}{\L_H^4 m_h^4} T^7 $, is slower than the Hubble expansion rate when $ T\ll 100\GEV$ in the parameter range of interest. 

The light ALP/hidden photon discussed in this part {\it cannot} be produced through the usual non-thermal production mechanisms unless the relevant temperature is very low.\footnote{A simple non-thermal production scenario with the low reheating temperature may be that the light DM is produced via reheating~\cite{Moroi:2020has, Moroi:2020bkq}. Here, the reheating temperature is required to be larger than the moduli (any other mother particle for the reheating) mass. Thus, the non-thermal sphaleron may not be active~\cite{Jaeckel:2022osh}, and we need some other possible source for the baryon number violation for the baryogenesis.  
} 
 This is because otherwise, the non-thermally produced DM would be thermalized due to the fast reaction originating from the portal interaction. Conversely, a simple production mechanism is possible via the portal interaction due to which the DM is produced thermally. 
The abundance can be calculated straightforwardly given a reheating temperature $T_R\lesssim \min{(m_h, m_s)}$ and \Eq{HDO}:
\begin{align}
\left.\frac{n_a}{s}\right|_{T=T_R} &\sim\left. \frac{\G_{\rm th}}{H} \frac{n_\p}{s}\right|_{ T=T_{R}}, \\
\Omega_a &\sim m_a \times \left. \frac{n_a}{s}\right|_{T=T_R} \times \frac{s_0}{\rho_{c}}   \notag\\
 &\label{abuandance} \sim  0.35   \frac{m_a}{20\KEV}\(\frac{m_\p}{\GEV}\)^2\(\frac{T_R}{ 2\GEV}\)^5\(\frac{3\TEV}{\L_H}\)^4.
\end{align}
Here, $s$ ($s_0$) is the entropy density soon after reheating (at the present universe), $n_a$ ($n_\psi$ ) the number density of the ALP (fermion), $\G_{\rm th}$ the thermal production rate, $\r_c$ the critical density today. 
We note that if $m_a$ is larger than keV, we need a reheating temperature lower than the weak scale. 
This kind of low reheating temperature may be predicted in a more fundamental UV model like a string theory by the decay of the moduli, string axion, gravitino, etc. Alternatively, the inflation models, which would not cause too large radiative corrections to the Higgs boson masses, also predict low reheating temperatures~\cite{Jaeckel:2020oet}. 
In the following, we consider that the reheating is caused by the moduli decay, but our discussion can apply to other possibilities as well.  
Recently, it was discussed that the out-of-equilibrium sphaleron could work efficiently during the thermalization of the reheating~\cite{Jaeckel:2022osh}, satisfying two of the Sakharov conditions for the baryogenesis. This is because the moduli for the reheating decays into the energetic SM particles, whose interaction with the ambient plasma has a center-of-mass scale much larger than the sphaleron scale.\footnote{Alternatively, given the sub weak scale reheating, we also have the possible baryogenesis scenarios with extensions of the model for baryon number violation~\cite{Dimopoulos:1987rk, Babu:2006xc, Grojean:2018fus, Pierce:2019ozl, Asaka:2019ocw, Azatov:2021irb} . } 

The thermally produced DM tends to be too warm and the mass should satisfy~\cite{Sakurai:2021ipp}
 $m_a\gtrsim 20\KEV$ 
 for a dominant DM. The parameter region $\lesssim 200\KEV$ can be tested in the future observations of the 21cm line  and milky-way sub-halo count~\cite{Sakurai:2021ipp}.

Interestingly, the same portal interaction would lead to the collider signals discussed in the main part.  
Also, it is interesting to notice that neither the DM production nor the prediction of the invisible decays depend much on the detail of what $G_{\rm dark}$ is thanks to the equivalence theorem. On the other hand, the nature of $G_{\rm dark}$ affects later cosmology and astrophysics observations. This will be discussed in the following by assuming $G_{\rm dark}=\U(1)_{\rm global},\U(1)_{\rm local}, \OR \SU(N)_{\rm local}$.

\paragraph{Case with $G_{\rm dark}=\U(1)_{\rm global}$ global symmetry: (CP-even) ALP}

Let us discuss the case when $G_{\rm dark}$ is a global abelian symmetry. Then, the resulting NG boson can be seen as an ALP, which will be considered a cold DM. 
We may assume that $\U(1)$ is the global symmetry that is explicitly broken, which may be expected from quantum gravity. 
As discussed in the main part around Eq.\,\eqref{Vexp}, we obtain a CP-even ALP mixing with the SM-like Higgs boson. 
By taking account of the effect of mixing, one can see that the ALP lifetime can be much longer than the age of the universe if $m_a\lesssim 1\MEV$, i.e., twice the electron mass.  This is because it dominantly decays into a pair of photons via a width (See \Eq{Defc_h})
\beq
\laq{agg}
\G_{a\to \g\g}\propto \frac{m_a^7}{v^4 m_\F^2},
\eeq 
which is highly suppressed by the 7th power of the small $m_a$. It was shown that a slightly smaller $m_a$ than $\MEV$ can evade the very severe bound from the $X$,$\gamma$-ray searches. 
In addition, in this regime, the ``direct detection" bound via the $a$ absorption ({cf.}~\cite{Aprile:2022vux}) or the cooling bound of stars via the mixing is also irrelevant due to the smallness of the interaction. For instance, the electron coupling, $\theta_{a h}  \frac{m_e}{v} a \bar{e} e,$ has an effective ``coupling scale" $\sim v /\theta_{ah} \sim 10^{12}\GEV \(\frac{\MEV}{m_a}\)^2 \frac{\(100\GEV\)^2}{c_h^{-1} m_h m_\F}$ which is several orders of magnitude larger than the scale that is sensitive. 
This scenario can be tested in the X-ray$\AND \gamma$-ray observations, i.e., indirect detection, rather than the direct detection.

 We also comment that the conventional ALP, CP-odd coupled to SM particles (see \cite{Mimasu:2014nea,Bauer:2017ris,Calibbi:2020jvd,Haghighat:2020nuh,Haghighat:2021djz} for related searches and experimental bounds), has tension to be the DM in this regime. This is because the ALP has a too large coupling to electrons, even via radiative corrections. 
We note that the ALP discussed in this paper has a typical decay constant 
 \beq
 f_a \sim v_\F\lesssim 10\TEV.
 \eeq
If it couples to photons anomalously, the decay rate with $m_a\gtrsim 20\KEV$ will exceed the X-ray, $\gamma$-ray extremely. We are led to consider the so-called anomaly-free ALP~\cite{Nakayama:2014cza} (see also earlier work \cite{Pospelov:2008jk, Arias:2012az}.).\footnote{{The relaxed bound of X, $\gamma$-rays in the context of the direct detection experiment via the electron coupling was first derived in \cite{Takahashi:2020bpq}. The anomaly-free ALP in the context of an EFT was studied in \cite{Takahashi:2020uio}. The above discussions are confirmed in the context of direct detection in \cite{Ferreira:2022egk}. As the UV model, a {two-Higgs-doublet} model was built~\cite{Takahashi:2020bpq} (see also \cite{Takahashi:2019qmh}).
A detailed study of light three-Higgs-doublet models (see also \cite{Nakayama:2014cza}) was discussed, and the photon coupling is found to be enhanced in a certain parameter region~\cite{Sakurai:2022roq}. In the present discussion, we do not need the tree-level ALP coupling to the electron, meaning that the photon coupling can be even more suppressed. Even in this case, the electron/light quark couplings would be induced radiatively, and the scenario is severely constrained with $f_a \lesssim 10\TEV$.}} 
 However, as long as the ALP couples to a particle that is charged under the $\SU(3)_c\times \SU(2)_L\times \U(1)_Y$, a radiative correction would induce a derivative coupling of $a$ to the electron or light quarks whose couplings are significantly constrained. 
A naively expected interaction has the form of 
\beq
\laq{dim5}
\(\frac{1}{16\pi^2}\)^{n} \frac{\partial_\mu a}{f_a}  \bar \psi \gamma^\mu \gamma_5 \psi.
\eeq
Here, $n$ is a model-dependent number of loops for the interaction to appear {($n=0$ for the tree-level coupling.)} For instance, if ALP has a photon coupling at some scale (that is canceled by a fermion loop at a lower scale), we have $n\leq 2$~\cite{Chakraborty:2021wda}. 
Given that $f_a\lesssim 10\TEV$, avoiding the bounds from direct detections, astrophysics, and $X$, $\gamma$-ray observations, we need $n\gtrsim 3$, and therefore, the even safest anomaly-free CP-odd ALP scenario is in tension with those observations.\footnote{{If $\F$ is the B-L Higgs field that couples to the right-handed neutrinos around the TeV scale, the ALP (or Majoron) decays into the SM neutrinos with a lifetime shorter than the age of the universe.} }

\paragraph{Case with $G_{\rm dark}=\U(1)_{\rm local}$ gauge symmetry: hidden photon}
If $\U(1)_{\rm local}$ is a gauge symmetry, we have a hidden photon. This hidden photon eats the NGB and gets massive. The mass is 
\beq
m_{\g'}= \sqrt{2}  g' v_\F,
\eeq
with $g'$ being the $\U(1)_{\rm local}$ gauge coupling. 
Thus the hidden photon can also be the DM.

The formula of Eq.\,\eqref{abuandance} applies to the production of the hidden photon by simply replacing $m_a$ with $m_{\g'}$ thanks to the equivalence theorem. 
The difference from the CP-even ALP is the DM phenomenology in the present universe. 
The hidden photon couples to the SM particles via the renormalizable interaction
\beq
 \delta{\cal L} =  \frac{\e}{2} F'_{\m \n}F_Y^{\m \n} 
\eeq
where $F'$ ($F_Y$) is the field strength of the hidden photon  (hyper-charge gauge boson), and $\e$ is the mixing parameter. 
Even if $m_{\g'} <1$MeV, the decay into the photons are highly suppressed. The hidden photon can only decay into three photons due to the CP nature (or we may call it Furry's theorem), and it decays via the dimension-8 terms. 
Instead of the X-ray/gamma-ray bound, the ``direct detection" as well as star cooling bounds are more important~\cite{Jaeckel:2010ni,Ringwald:2012hr,Aprile:2022vux} requiring $\e\lesssim 10^{-13}$. 
Satisfying this, we can have the thermally produced hidden photon, whose mass should be larger than $20$keV to be consistent with the 
structure formation as discussed previously.

\paragraph{Case with $G_{\rm dark}=\SU(N)_{\rm local}$ gauge symmetry: accidental light DM and dark radiation} 

Let us consider $G_{\rm dark}$ as a non-abelian gauge symmetry, $\SU(N)_{\rm local}$. 
We will show that this solves the quality problem of the $\U(1)_{\rm global}$ discussed in the previous part, and may relax the Hubble tension. Here, we do not impose any global symmetries but only the local symmetry. 
To this end, we assume that 
$\F=\F_{ij}$ is a complex symmetric tensor under this symmetry. % $\F \to g^ \F $
In a large parameter region  of the renormalizable potential of $\F$, we can get 
\beq
\vev{\F}= \frac{v_\F}{\sqrt{N}} \delta_{ij},
\eeq
which means the spontaneous breaking of
\beq
\SU(N)_{\rm local}\to \SO(N)_{\rm local}.
\eeq
This generates $N^2-1-(N-1)N/2= (N^2+N-2)/2$ would-be NGBs eaten by the gauge bosons.  
The charged massive gauge bosons have  the mass of 
\beq
m_{G}\simeq 2g_{\rm \SU(N) }v_\F,
\eeq
with $g_{\rm \SU(N)}$ being the gauge coupling. 
We also have $(N-1)N/2$ massless gauge bosons corresponding to the number of the generators of $\SO(N)_{\rm local}.$ 
Those gauge bosons, if produced sufficiently in the early universe and not confine, can play the role of the self-interacting dark radiation~\cite{Jeong:2013eza} that alleviates the Hubble tension~\cite{Planck:2018vyg,Riess:2011yx, Bonvin:2016crt,Riess:2016jrr, Riess:2018byc,Birrer:2018vtm,Riess:2021jrx}. 
In addition, we have bosons in a $(N+1) N/2-1$ representation (the symmetric tensor of $\SO(N)_{\rm local}$) and $2$ neutral (CP odd and even) Higgs bosons. The CP-even neutral boson, $\rho=\frac{\sqrt{2}}{\sqrt{N}}\tr \Re\F $, corresponds to the dark Higgs boson discussed previously.

Interestingly, there is an accidental symmetry of $\U(1)_{\rm global}$ when $N>4$. This is also spontaneously broken by the VEV of $\F$~\cite{Lee:2018yak, Ardu:2020qmo,Yin:2020dfn} (see also \cite{Randall:1992ut,DiLuzio:2017tjx, Lillard:2018fdt,Lee:2018yak} for other representations of matter for accidental symmetries). 
Thus the CP-odd neutral boson residing in $\F$ corresponds to the ALP, $a$. 
Giving a cutoff scale, $\L$, of the theory, $\U(1)_{\rm global}$ is expected to be broken by the term of 
\beq
\label{break}
\delta {\cal L} = -\frac{\det \F }{\L^{N-4}}+\O(\det \F |H|^2)+h.c..
\eeq
This gives a mass to the ALP, % satisfying
\beq
m_a \sim v_\F\( \frac{v_\F}{\L}\)^{N/2-2}.
\eeq
 In fact, taking $N=5$, we obtain 
\beq
m_a = 20\KEV \(\frac{v_\F}{1\TEV}\)^{3/2}\(\frac{M_{\rm pl}}{\L}\)^{1/2},
\eeq
consistent with the DM with $\L$ being the reduced Planck scale $M_{\rm pl}$. 
This ALP is difficult to mix with the SM-like Higgs boson if $\L\gg v_\F$. This is because there is an accidental CP symmetry in the leading term in Eq.\,\eqref{break}. 

To sum up, we have $\SU(N)_{\rm local}\times \U(1)_{\rm global}$, the latter of which is accidental when $N>4$. 
The resulting spectrum with generic parameters of the Higgs potential is as follows:\footnote{In a certain choice of the parameters, the mass scales of the dark Higgs bosons and charged bosons may be different from what we list.}
\begin{description}
\item[Scale of $v_\F$:] Dark Higgs boson, $s$, $(N+1)N/2-1$ charged bosons, 
\item[Scale of $m_G$:] $(N+1)N/2-1$ charged vector bosons. 
\item[Scale $\ll m_G$:] $(N-1)N/2$ massless gauge bosons (unless confinement), ALP $a$.  
\end{description}

By focusing on the dark Higgs component (with certain redefinition), $v_\F+s/\sqrt{2 N},$ and the SM-like field system, our previous discussion does not change much. For instance, there is a mixing term between the SM-like Higgs and dark Higgs bosons:
\beq
V\supset \lambda_P \tr(|\F|^2) |H|^2 \supset 2 \sqrt{N}v_\F v \f_r \r.
\eeq
We can also find the interaction between $s$ and NGB, $a$, as well as the would-be NGBs, $a_i:$ %{\bf check coefficient}
\beq
{\cal L}_{\rm int} \simeq \frac{s}{\sqrt{2 N}v_\F} \((\partial a)^2 +\sum_{i=1}^{N(N+1)/2-1}(\partial a_i)^2\).
\eeq
By integrating out $s$, we get the higher dimensional term of the form 
\begin{align}
{\d {\cal L}} &=   -\frac{\sqrt{2}{\lambda_P m_\p}}{(m_h^2-m_s^2) m_h^2} \(\partial a\partial a +\sum_{i=1}^{N(N+1)/2-1}(\partial a_i)^2\)  \bar{\psi}\psi\\
&\equiv  -\frac{\sqrt{2}{m_\p}}{\L_H^2 m_h^2} \(\partial a\partial a +\sum_{i=1}^{N(N+1)/2-1}(\partial a_i)^2\) \bar{\psi}\psi.
\end{align}
Thus the DM, $a,$ can be produced, again, via the thermal scattering with a low reheating temperature. 
In addition, the charged vector bosons in the early universe are also produced. We note that there are also loop-induced couplings by integrating out the charged bosons. We do not consider them since the contribution to the production of the dark particles should be subdominant.

By neglecting the annihilation of the charged vector boson soon after the reheating, we find 
\beq
\laq{relation}
 \frac{n_{\rm MVB}}{s}=\(\frac{(N+1)N}{2}-1\)\frac{n_a}{s},
\eeq
with $n_{\rm MVB}$ being the total number density of the massive charged gauge bosons, respectively. The dark sector is composed of two sectors after the reheating: the ALP sector and the $\SO(N)$ sector, where the charged particles under the $\SO(N)$ reside in. 
In addition, we have the visible SM sector. The three sectors rarely interact with each other after the reheating.  The evolutions of the SM and ALP sectors are the same as discussed previously, and we do not repeat them here. 

Let us focus on the  $\SO(N)$ sector. 
Soon after the reheating, the heavy gauge boson would annihilate into the $\SO(N)$ gauge bosons, and the thermalization takes place. According to the energy conservation, $\SO(N)$ sector carries the entropy density of 
\beq s_{\rm dark} \sim \frac{2\pi^2}{45} \(3\(\frac{(N+1)N}{2}-1\)+2\frac{(N-1)N}{2}\)^{-1/3} \(\frac{30n_{\rm MVB} T_R}{\pi^2}\)^{3/4}. \eeq

When the massive gauge boson becomes non-relativistic, they annihilate like WIMP. The annihilation cross section scales as 
\beq
\sigma\sim  \frac{g_{\SU(N)}^2 N}{4\pi v_\F^2}. 
\eeq
Thus $v_\F/(g_{\SU(N)} \sqrt{N})\ll 10\TEV$ is required so that the ``lightest charged particles" is not overproduced, while $g_{\SU(N)} N \lesssim1$
is required so that confinement is absent in cosmology for simplicity. 
 By taking account of the Sommer-felt enhancement \cite{sommerfeld, Feng:2009hw} (see also Refs \cite{Hisano:2002fk,Hisano:2003ec, Cirelli:2007xd, ArkaniHamed:2008qn}), the bound on $v_\F$ can be relaxed but it may not be much larger than $10\TEV$. This motivates the parameter region of our main focus.  After the annihilation, the $\SO(N)$ gauge bosons behave as self-interacting dark radiation. 
 Given that the relic of the charged massive gauge bosons is subdominant,\footnote{It is beyond our scope whether the charged vector bosons can play the role of the dominant DM. To clarify the possibility, the bullet cluster bound should be carefully studied by taking into account the long-range force by the $\SO(N)$ gauge bosons. } the comoving entropy would be carried by the $\SO(N)$ massless gauge bosons.

By assuming that $a$ contributes to the dominant DM via Eq.\,\eqref{abuandance}, we find that the thermally produced $\SO(N)$ sector contributes to the dark radiation represented as the deviation of the effective neutrino number,
\beq
\Delta N_{\rm eff}\sim 10^{-3}\(\frac{N}{5}\)^2 \(\frac{80}{g_{\star S}[T_R]}\)^{1/3}\frac{20\KEV}{m_a}.
\eeq
Thus satisfying $m_a>20\KEV$ for the DM small-scale structure bound, the thermally produced dark radiation via \Eq{relation} contributes to the deviation of the effective neutrino number $\D N_{\rm eff}\ll 0.1$ for $N\lesssim 10,$ satisfying the cosmological bound.
On the other hand, during the reheating, the self-interacting dark radiation can be also produced via the decay of the moduli.  We consider this production naturally happens since we can write the interaction of $|\F|^2 \f $ with $\f$ being the moduli. 
In this case the dark radiation can easily contribute to $\D N_{\rm eff} =\O(0.1)$ alleviating the $H_0$ tension. The thermally produced $a$ contributes to the DM.\footnote{There is also a component of the dark radiation/hot DM of $a$. It is safe from cosmology if the energy component is not too much. For instance, this can be satisfied if the mass of the moduli is heavy as usually assumed, and thus the produced number density of $a$ as well as the charged massive gauge bosons is small. 
}  
\\

In any of the above light DM scenarios (We can also consider dark radiation production by taking the ALP, hidden photon or non-abelian gauge boson mass to be small enough and consider a high enough reheating temperature.), 
the $s$-$h$ system as well as the invisible decays of $s, h$, do not change much.
The analysis in the main part gives a generic probe of those scenarios.

%\bibliography{ALP}
%\bibliographystyle{JHEP}  

\providecommand{\href}[2]{#2}\begingroup\raggedright\endgroup

\end{document}